\documentclass[10pt]{bmc_article}

\usepackage{cite} 
\usepackage{url}  
\usepackage{ifthen}  
\usepackage{multicol}   
\usepackage[utf8]{inputenc} 
\usepackage{amsmath} 
\usepackage{amssymb}
\usepackage{graphicx}
\usepackage{esint}
\usepackage{color}
\newboolean{publ}

\newenvironment{bmcformat}{\begin{raggedright}\baselineskip20pt\sloppy\setboolean{publ}{false}}{\end{raggedright}\baselineskip20pt\sloppy}

\begin{document}
\begin{bmcformat}


\title{Coverage, Continuity and Visual Cortical Architecture}
 

\author{Wolfgang Keil\correspondingauthor$^{1,2,3,4}$ %
       \email{Wolfgang Keil\correspondingauthor - wolfgang@nld.ds.mpg.de}%
      and
         Fred Wolf$^{1,2,3,4}$%
         \email{Fred Wolf - fred@nld.ds.mpg.de}
      }


\address{%
    \iid(1)Max-Planck-Institute for Dynamics and Self-organization, Am Fassberg 17, D-37077 G\"ottingen\\
    \iid(2)Bernstein Center for Computational Neuroscience, Am Fassberg 17, D-37077  G\"ottingen\\
     \iid(3)Georg-August-University, G\"ottingen, Faculty of Physics, Friedrich-Hund-Platz 1, D-37077 G\"ottingen\\
     \iid(4)Kavli Institute for Theoretical Physics,  Santa Barbara CA 93106-4030, USA
}%
\maketitle


\begin{abstract}
\paragraph*{Background}
The primary visual cortex of many mammals contains a continuous representation of visual space, with a
roughly repetitive aperiodic map of orientation preferences superimposed. It was recently found that orientation preference maps (OPMs) obey statistical laws which are apparently invariant among species widely separated in eutherian evolution. Here, we examine whether one of the most prominent models for the optimization of cortical maps, the elastic net (EN) model, can reproduce this common design. The EN model generates representations which optimally trade of stimulus space coverage and map continuity. While this model has been used in numerous studies, no analytical results about the precise layout of the predicted OPMs have been obtained so far.
\paragraph*{Results}
We present a mathematical approach to analytically calculate the cortical representations predicted by the EN model for the joint mapping of stimulus position and orientation. We find that in all previously studied regimes, predicted OPM layouts are perfectly periodic. An unbiased search through the EN parameter space identifies a novel regime of aperiodic OPMs with pinwheel densities lower than found in experiments. In an extreme limit, aperiodic OPMs quantitatively resembling experimental observations emerge. Stabilization of these layouts results from strong nonlocal interactions rather than from a coverage-continuity-compromise.
\paragraph*{Conclusions}
Our results demonstrate that optimization models for stimulus representations dominated by nonlocal suppressive interactions are in principle capable of correctly predicting the common OPM design. They question that visual cortical feature representations can be explained by a coverage-continuity-compromise.
\end{abstract}
\ifthenelse{\boolean{publ}}{\begin{multicols}{2}}{}
%
%
\section*{Introduction}
The pattern of orientation columns in the primary visual cortex (V1) of carnivores, primates and their close relatives are among the most intensely-studied structures in the cerebral cortex and a large body of experimental (e.g. \cite{hubel_63,Hubel:1977p5543,Grinvald:1986p5554, Blasdel1986p5555, Bonhoeffer:1991p5561, blasdel_92, bonhoeffer_93, Bartfeld1992:p11905, Blasdel:1995p1879, Bosking:1997p5557,White:2001p6452,Ohki:2006p7461,White:2007p220}) and theoretical work (e.g. \cite{vonderMalsburg:1973p6457,swindale_82,Braitenberg:1985p8092, Linsker:1986p7572, Linsker:1986p7571, Obermayer:1990p1202, Tanaka:1990p7997,Durbin:1990p1196, Worgotter:1991p7704, Miller:1994p7651, Stetter:1994p7689,Wolf:1998p1199, Wimbauer:1998p8094, Koulakov:2001p842,Ernst:2001p8090,Mayer:2002p8061, Kang:2003p7683, Lee:2003p5834, Wolf:2003p210, Thomas:2004p6132,bressloff_05,Wolf:2005p190,schnabel_2007,Reichl2009:p208101,Kaschube_2010,schnabel_2011}) has been dedicated to uncovering its organization principles and the circuit level mechanisms that underlie its development and operation. 

Orientation preference maps (OPM) exhibit a roughly repetitive arrangement of preferred orientations in which adjacent columns preferring the same orientation are separated by a typical distance in the millimeter range \cite{Hubel:1977p5543,Blasdel1986p5555, Grinvald:1986p5554,Bonhoeffer:1991p5561, Bosking:1997p5557}. This range seems to be set by cortical mechanisms both intrinsic to a particular area \cite{Hensch:2004p1678} but potentially also involving interactions between different cortical regions \cite{Kaschube:2009p6537}.
The pattern of orientation columns is however not strictly periodic because the precise local arrangement of preferred orientation never exactly repeats. Instead, orientation preference maps appear as organized by a spatially complex aperiodic array of pinwheel centers, around which columns activated by different stimulus orientations are radially arranged like the spokes of a wheel \cite{Hubel:1977p5543,Blasdel1986p5555, Grinvald:1986p5554,Bonhoeffer:1991p5561, Bosking:1997p5557}. 
The arrangement of these pinwheel centers, although spatially irregular, is statistically distinct from a pattern of randomly positioned points \cite{Kaschube_2010} as well as from patterns of phase singularities in a random pattern of preferred orientations \cite{Erwin:1995p1206, Wolf:2003p210, schnabel_2007, Kaschube_2010} with spatial correlations identical to experimental observations \cite{Erwin:1995p1206, Kaschube_2010}.
This suggests that the layout of orientation columns and pinwheels although spatially aperiodic follows a definite system of layout rules. Cortical columns can in principle exhibit almost perfectly repetitive order as exemplified by ocular dominance (OD) bands in the macaque monkey primary visual cortex \cite{Horton:1996p2865, Horton:1998p8126}. It is thus a fundamental question for understanding visual cortical architecture, whether there are layout principles that prohibit a spatially exactly periodic organization of orientation columns and instead enforce complex arrangements of these columns.

Recent comparative data has raised the urgency of answering this question and of dissecting what is constitutive of such complex layout principles. Kaschube et al. quantitatively compared pinwheel arrangements in a large data set from three species widely separated in the evolution of eutherian mammals \cite{Kaschube_2010}. 
These authors found that the spatial statistics of pinwheels are surprisingly invariant. In particular, the overall pinwheel density and the variability of pinwheel densities in regions from the scale of a single hypercolumn to substantial fractions of the entire primary visual cortex were found to be virtually identical.
Characterizing pinwheel layout on the scale of individual hypercolumns, they found the distributions of nearest-neighbor pinwheel distances to be almost indistinguishable.
Further supporting common layout rules for orientation columns in carnivores and primates, the spatial configuration of the superficial patch system \cite{Lund:2003p15} and the responses to drifting grating stimuli were recently found to very similar in cat and macaque monkey primary visual cortex \cite{Muir:2011p1}.

From an evolutionary perspective, the occurrence of quantitatively similar layouts for OPMs in primate tree shrews and carnivorous species appears highly informative. The evolutionary lineages of these taxa diverged more than 65 million years ago during the basal radiation of eutherian mammals \cite{Kaas:2006p303, Kriegs:2006:e91,Kriegs2007:p158}. According to the fossil record and cladistic reconstructions, their last common ancestors (called the boreo-eutherial ancestors) were small-brained, nocturnal, squirrel-like animals of reduced visual abilities with a telencephalon containing only a minor neocortical fraction  \cite{Kaas:2006p303,Kielian-Jaworowsk:2004}. For instance, endocast analysis of a representative stem eutherian from the late cretaceous, indicates a total anterior-posterior extent of 4mm for its entire neocortex \cite{Kaas:2006p303,Kielian-Jaworowsk:2004}. Similarly, the tenrec (\textit{Echinops telfari}), one of the closest living relatives of the boreoeutherian ancestor \cite{Bininda-Emonds:2007p507,Wible:2007p1003}, has a neocortex of essentially the same size and a visual cortex that totals only 2mm${}^{2}$  \cite{Kaas:2006p303}. Since the neocortex of early mammals was subdivided into several cortical areas \cite{Kaas:2006p303} and orientation hypercolumns measure between 0.4mm${}^{2}$ and 1.4mm${}^{2}$ \cite{Kaschube_2010}, it is difficult to envision ancestral eutherians with a system of orientation columns. In fact, no extant mammal with a visual cortex of such size is known to possess orientation columns \cite{Hooser:2007p639}.
It is therefore conceivable that systems of orientation columns independently evolved in laurasiatheria (such as carnivores) and in euarchonta (such as tree shrews and primates). Because galagos, tree-shrews, and ferrets strongly differ in habitat and ecologically relevant visual behaviors, it is not obvious that the quantitative similarity of pinwheel layout rules in their lineages evolved  driven by specific functional selection pressures (see \cite{Kaschube_2010_SOM} for an extended discussion).
Kaschube and coworkers instead demonstrated that an independent emergence of identical layout rules for pinwheels and orientation columns can be explained by mathematically universal properties of a wide class of models for neural circuit self-organization.

According to the self-organization scenario, the common design would result from developmental constraints robustly imposed by adopting a particular kind of self-organization mechanism for constructing visual cortical circuitry. Even if this scenario is correct, one question still remains: What drove the different lineages to adopt a similar self-organization mechanism?
As pointed out above, it is not easy to conceive that this adoption was favored by the specific demands of their particular visual habitats. It is, however, conceivable that general requirements for a versatile and representationally powerful cortical circuit architecture are realized by the common design. If this was true, the evolutionary benefit of meeting these requirements might have driven the adoption of large-scale self-organization and the emergence of the common design over evolutionary times.

The most prominent candidate for such a general requirement is the hypothesis of a coverage - continuity - compromise (e.g. \cite{Durbin:1990p1196, Obermayer:1990p1202,Swindale:2000p4894, Swindale:2000p6286}). It states that the columnar organization is shaped to achieve an optimal tradeoff between the coverage of the space of visual stimulus features and the continuity of their cortical representation. On the one hand, each combination of stimulus features should be well-represented in a cortical map to avoid ``blindness" to stimuli with particular feature combinations. On the other hand, the wiring cost to establish connections within the map of orientation preference should be kept low. This can be achieved if neurons that are physically close in the cortex tend to have similar stimulus preferences. These two design goals generally compete with each other. The better a cortical representation covers the stimulus space, the more discontinuous it has to be. The tradeoff between the two aspects can be modeled in what is called a dimension reduction framework in which cortical maps are viewed as two-dimensional sheets which fold and twist in a higher-dimensional stimulus space (see Fig. \ref{Keil_Wolf_figure_1}) to cover it as uniformly as possible while minimizing some measure of continuity \cite{Durbin:1990p1196, Swindale1998p827, Goodhill_2007}.   
\begin{figure}
\centering
\includegraphics[width=16cm]{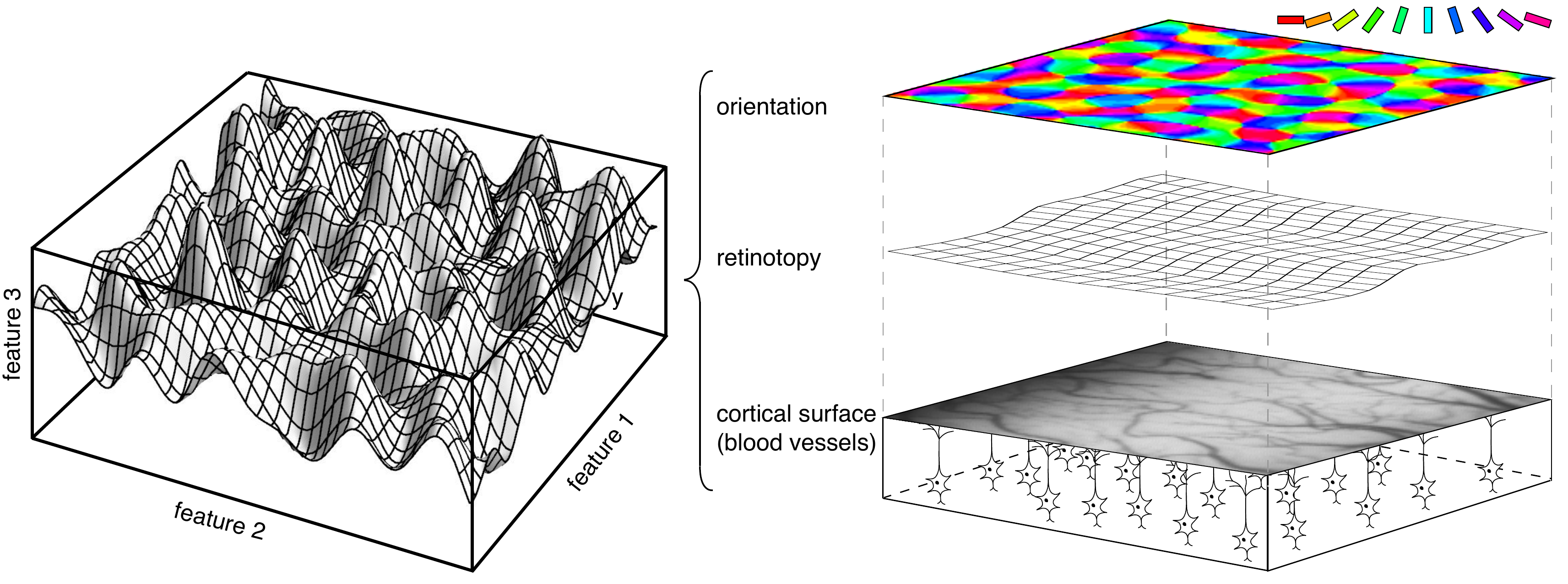}
\caption{\textbf{The dimension reduction framework.} In the dimension reduction framework, the visual cortex is modeled as a two-dimensional sheet that twists in a higher-dimensional stimulus (or feature) space to cover it as uniformly as possible while minimizing some measure of continuity (left). In this way, it represents a mapping from the cortical surface to the manifold of visual stimulus features such as orientation and retinotopy (right). \label{Keil_Wolf_figure_1}}
\end{figure}

From prior work, the coverage continuity compromise appears to be a promising candidate for a principle to explain visual cortical functional architecture.
Firstly, many studies have reported good qualitative agreement between the layout of numerically obtained dimension reducing maps and experimental observations \cite{Kohonen1982p6163, Durbin:1990p1196, Obermayer:1990p1202, Obermayer:1992pE75, Kohonen:1995p6543, Erwin:1995p1206, Swindale1998p827,Swindale:2000p4894, Swindale:2000p6286, Cimponeriu:2000p5167, Goodhill:2000p2141, CarreiraPerpinan:2004p6297, CarreiraPerpinan:2005p6295, Swindale:2004p6287, Yu:2005p5850, Farley:2007p5840, Keil:2010p6536, Giacomantonio_2007, Giacomantonio_2010}. Secondly, geometric relationships between the representations of different visual features such as orientation, spatial frequency and ocular dominance have been reproduced by dimension reduction models \cite{Hoffsummer95, Wolf:1998p1199, Goodhill:2000p2141, Swindale:2000p6286, CarreiraPerpinan:2004p6297, CarreiraPerpinan:2005p6295, Yu:2005p6302, Farley:2007p5840}.

Mathematically, the dimension reduction hypothesis implies that the layouts of cortical maps can be understood as optima or near optima (global or local minima) of a free energy functional which penalizes both ``stimulus scotomas" and map discontinuity. Unfortunately, there is currently no dimension reduction model for which the precise layouts of optimal or nearly optimal solutions have been analytically established. To justify the conclusion that the tradeoff between coverage and continuity favors the common rules of OPM design found in experiment, knowledge of optimal dimension-reducing mappings however appears essential. 

Precise knowledge of the spatial organization of optimal and nearly optimal mappings is also critical for distinguishing between optimization theories and frozen noise scenarios of visual cortical development. In a frozen noise scenario, essentially random factors such as haphazard wiring \cite{ringach_04}, the impact of spontaneous activity patterns \cite{Chiu:2001p8906} or an idiosyncratic set of visual experiences \cite{Spinelli:1979p75} determine the emerging pattern of preferred orientations. This pattern is then assumed to be ``frozen" by an unknown mechanism, capable of preventing further modification of preferred orientations by ongoing synaptic turnover and activity-dependent plasticity.
Conceptually, a frozen noise scenario is diametrically opposed to any kind of optimization theory. Even if the reorganization of the pattern prior to freezing was to follow a gradient descent with respect to some cost function, the early stopping implies that the layout is neither a local nor a global minimum of this functional. Importantly, the layout of transient states is known to exhibit universal properties that can be completely independent of model details \cite{Wolf:1998p1199,Wolf:2003p210}. As a consequence, an infinite set of distinct optimization principles will generate the same spatial structure of transient states. This implies in turn that the frozen transient layout is not specifically shaped by any particular optimization principle. Map layout will thus in principle only be informative about design or optimization principles of cortical processing architectures if maps are not just frozen transients. 

In practice, however, the predictions of frozen noise and optimization theories might be hard to distinguish. Ambiguity between these mutually exclusive theories would result in particular, if the energy landscape of the optimization principle is so ``rugged" that there is essentially a local energy minimum next to any relevant random arrangement. Dimension reduction models are conceptually related to combinatorial optimization problems like the Traveling Salesman Problem and many such problems are believed to exhibit rugged energy landscapes \cite{ritter_92, durbin_87, Applegate2006}. It is therefore essential to clarify, whether paradigmatic dimension reduction models are characterized by a rugged or a smooth energy landscape. If their energy landscapes were smooth with a small number of well-separated local minima, their predictions would be easy to distinguish from those of a frozen noise scenario.

In this study, we examine the classical example of a dimension reduction model, the elastic network (EN) model. Since the seminal work of Durbin \& Mitchison \cite{Durbin:1990p1196}, the EN model has been widely used to study visual cortical representations \cite{Erwin:1995p1206, Hoffsummer95, Wolf:1998p1199, Cimponeriu:2000p5167,Goodhill:2000p2141, CarreiraPerpinan:2004p6297, CarreiraPerpinan:2005p6295, Giacomantonio_2007, Giacomantonio_2010, Keil:2010p6536,CarreiraPerpinan:2011}. The EN model possesses an explicit energy functional which trades off a matching constraint which matches cortical cells to particular stimulus features via Hebbian learning, with a continuity constraint that minimizes euclidean differences in feature space between neighboring points in the cortex \cite{Goodhill:2000p2141}. In two ways, the EN model's explicit variational structure is very appealing. Firstly, coverage and continuity appear as separate terms in the free energy which facilitates the dissection of their relative influences. Secondly, the free energy allows for the formulation of a gradient descent dynamics. The emergence of cortical selectivity patterns and their convergence towards a minimal energy state in this dynamics might serve as a model for an optimization process taking place in postnatal development. 

Following Durbin and Mitchison, we consider the EN model for the joint mapping of two visual features: (i) position in visual space, represented in a retinotopic map (RM) and (ii) line orientation, represented in an OPM. To compute optimal dimensional reducing mappings, we first develop an analytical framework for deriving closed-form expressions for fixed points, local minima, and optima of arbitrary optimization models for the spatial layout of OPMs and RMs in which predicted maps emerge by a supercritical bifurcation as well as for analyzing their stability properties. By applying this framework to different instantiations of the EN model, we systematically disentangle the effects of individual model features on the repertoire of optimal solutions. We start with the simplest possible model version, a fixed uniform retinotopy and an orientation stimulus ensemble with only a single orientation energy and then relax the uniform retinotopy assumption incorporating retinotopic distortions. An analysis for a second widely-used orientation stimulus ensemble including also unoriented stimuli is given in Appendix I. Surprisingly, in all cases, our analysis yields pinwheel-free orientation stripes or stereotypical square arrays of pinwheels as local minima or optimal orientation maps of the EN model. Numerical simulations of the EN confirm these findings. They indicate, that more complex spatially aperiodic solutions are not dominant and that the energy landscape of the EN model is rather smooth. Our results demonstrate that while aperiodic stationary states exist, they are generally unstable in the considered model versions.

To test whether the EN model is in principle capable of generating complex spatially aperiodic optimal orientation maps, we then perform a comprehensive unbiased search of the EN optima for arbitrary orientation stimulus distributions. We identify two key parameters determining pattern selection: (i) the intracortical interaction range and (ii) the fourth moment of the orientation stimulus distribution function. We derive complete phase diagrams summarizing pattern selection in the EN model for fixed as well as variable retinotopy. Small interaction ranges together with low to intermediate fourth moment values lead to pinwheel-free orientation stripes, rhombic or hexagonal crystalline orientation map layouts as optimal states. In the regime of large interaction ranges together with higher fourth moment, we find irregular aperiodic OPM layouts with low pinwheel densities as optima. 
Only in an extreme and previously unconsidered parameter regime of very large interaction ranges and stimulus ensemble distributions with an infinite fourth moment, optimal OPM layouts in the EN model resemble experimentally observed aperiodic pinwheel-rich layouts and quantitatively reproduce the recently described species-invariant pinwheel statistics. Unexpectedly, we find that the stabilization of such layouts is not achieved by an optimal tradeoff between coverage and continuity of a localized population encoding by the maps but results from effectively suppressive long-range intracortical interactions in a spatially distributed representation of localized stimuli.

We conclude our re-examination of the EN model with a comparison between different numerical schemes for the determination of optimal or nearly optimal mappings. For large numbers of stimuli, numerically determined solutions match our analytical predictions, irrespectively of the computational method used. 
%
%
%
\section*{Results and Discussion}
\subsection*{Model definition and model symmetries}
\begin{figure}
\centering
\includegraphics[width=12cm]{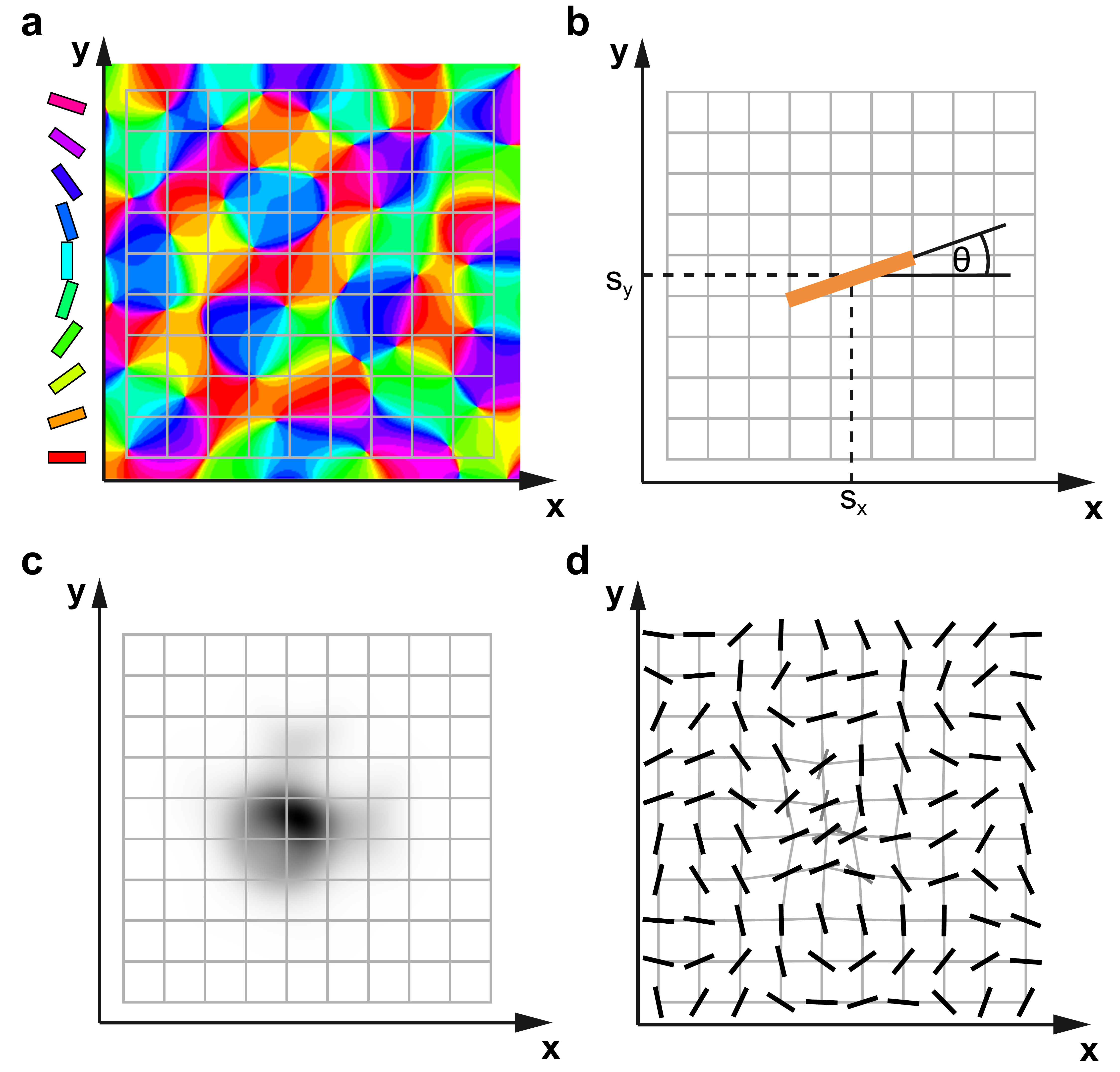}
\caption{\textbf{The elastic net (EN) model.} (\textbf{a}) Example orientation preference
map (OPM) (color code) together with a uniform map of visual space
(RM) (grid lines) (\textbf{b}) Position $\mathbf{s}_{r}=(s_{x},s_{y})$
and orientation $\theta$ of a `pointlike' stimulus. (\textbf{c})
Cortical activity, evoked by the stimulus in b for the model maps
in a. Dark regions are activated. Note, that in contrast to SOFM models, the activity pattern does not
exhibit a stereotypical Gaussian shape. (\textbf{d}) Modification of orientation
preference and retinotopy, caused by the stimulus in b. Orientation
preferences prior to stimulus presentation are indicated with grey
bars, after stimulus presentation with black bars. Most strongly modified
preferences correspond to thick black bars. Modifications of orientation preferences and retinotopy are
displayed on an exaggerated scale for illustration purposes.
\label{Keil_Wolf_figure_2}}
\end{figure}
We analyze the elastic net (EN) model for the joint optimization of position and orientation selectivity as originally introduced by Durbin \& Mitchison \cite{Durbin:1990p1196}.  In this model, the retinotopic map (RM) is represented by a mapping
$\mathbf{R}(\mathbf{x})=(R_{1}(\mathbf{x}),R_{2}(\mathbf{x}))$ which describes the receptive field center position of a neuron at cortical position $\mathbf{x}$. Any retinotopic map can be decomposed into an affine transformation $\mathbf{x} \mapsto \mathbf{X}$ from cortical to visual field coordinates, on which a vector-field of retinotopic distortions $\mathbf{r}(\mathbf{x})$ is superimposed, i.e.:
\[
\mathbf{R}(\mathbf{x})= \mathbf{X} + \mathbf{r}(\mathbf{x})
\]
with appropriately chosen units for $\mathbf{x}$ and $\mathbf{R}$. 

The orientation preference map (OPM) is represented by a second complex-valued scalar field $z(\mathbf{x})$.
The pattern of orientation preferences $\vartheta(\mathbf{x})$ is encoded by the phase of $z(\mathbf{x})$ via
\[
\vartheta(\mathbf{x})=\frac{1}{2}\arg(z(\mathbf{x}))\,.
\]
The absolute value $|z(\mathbf{x})|$ is a measure of the average
cortical selectivity at position $\mathbf{x}$. 
Solving the EN model requires to find pairs of maps $\{\mathbf{r}(\mathbf{x}), z(\mathbf{x})\}$ that represent an optimal compromise
between stimulus coverage and map continuity.  This is achieved by minimizing a free energy functional 
\begin{equation}
\mathcal{F}=\sigma^{2}\mathcal{C}+\mathcal{R}\label{eq:Elastic-Net-Energy}
\end{equation}
in which the functional $\mathcal{C}$ measures the coverage of a stimulus
space and the functional $\mathcal{R}$ the continuity of its cortical representation.
The stimulus space is defined by an ensemble \{${\mathbf{S}}$\} of idealized
pointlike stimuli, each described by two features: $s_{z}=|s_{z}|e^{2i\theta}$
and $\mathbf{s}_{r}=(s_{x,},s_{y})$ which specify the orientation $\theta$ of the stimulus and its position in
visual space $\mathbf{s}_{r}$ (Fig. \ref{Keil_Wolf_figure_2}b).
$\mathcal{C}$ and $\mathcal{R}$ are given by
\begin{eqnarray*}
\mathcal{C}[z,\mathbf{r}] & = & -\left<\ln\int d^{2}y\, e^{-\left(|s_{z}-z(\mathbf{y})|^{2}+\left| \mathbf{s}_{r}-\mathbf{X} - \mathbf{r}(\mathbf{y})\right| ^{2}\right)/2\sigma^{2}}\right>_{\mathbf{S}}\\
\mathcal{R}[z,\mathbf{r}] & = & \frac{1}{2}\int d^{2}y\,\eta\left\Vert \nabla z(\mathbf{y})\right\Vert ^{2}+\eta_{r}\sum_{j=1}^{2}\left\Vert \nabla r_{j}(\mathbf{y})\right\Vert ^{2}\,,
\end{eqnarray*}
with $\nabla=(\partial_{x},\partial_y)^{T}$, and $\eta\in[0,1]$. The ratios $\sigma^{2}/\eta$ and $\sigma^{2}/\eta_{r}$
control the relative strength of the coverage term versus the continuity
term for OPM and RM, respectively.
$\left<\cdots\right>_{\mathbf{S}}$ denotes the average over the ensemble of stimuli. 

Minima of the energy functional $\mathcal{F}$ are stable fixed points of the gradient descent dynamics
\begin{eqnarray}
\partial_{t}z(\mathbf{x}) & = & -2\frac{\delta\mathcal{F}[z,\mathbf{r}]}{\delta\bar{z}(\mathbf{x})}\equiv F^{z}[z,\mathbf{r}](\mathbf{x})\label{eq:gradient_descent}\\
\partial_{t}\mathbf{r}(\mathbf{x}) & = & -\frac{\delta\mathcal{F}[z,\mathbf{r}]}{\delta\mathbf{r}(\mathbf{x})}\equiv\mathbf{F}^{r}[z,\mathbf{r}](\mathbf{x})\nonumber
\end{eqnarray}
called the EN dynamics in the following. These dynamics read
\begin{eqnarray}
\partial_{t}z(\mathbf{x}) & = & \left\langle \left[s_{z}-z(\mathbf{x})\right]e(\mathbf{x},\mathbf{S},z,\mathbf{r})\right\rangle _{\mathbf{S}}+\eta\bigtriangleup z(\mathbf{x})\label{eq:continuous_z_dynamics}\\
\partial_{t}\mathbf{r}(\mathbf{x}) & = & \left\langle \left[\mathbf{s}_{r}-\mathbf{X} -\mathbf{r}(\mathbf{x})\right]e(\mathbf{x},\mathbf{S},z,\mathbf{r})\right\rangle _{\mathbf{S}}+\eta_{r}\bigtriangleup\mathbf{r}(\mathbf{x})\,,
\label{eq:continous_r_dynamics}
\end{eqnarray}
where $e(\mathbf{x},\mathbf{S},z,\mathbf{r})$ is the activity-pattern,
evoked by a stimulus $\mathbf{S}=(\mathbf{s}_{r},s_{z})$ in a model
cortex with retinotopic distortions $\mathbf{r}(\mathbf{x})$ and OPM $z(\mathbf{x})$.
It is given by
\begin{eqnarray*}
e(\mathbf{x},...) & = & \frac{e^{-(|\mathbf{s}_{r}-\mathbf{X} - \mathbf{r}(\mathbf{x})|^{2})/2\sigma^{2}}e^{-(|s_{z}-z(\mathbf{x})|^{2})/2\sigma^{2}}}{\int d^{2}y\, e^{-(|\mathbf{s}_{r}-\mathbf{X} - \mathbf{r}(\mathbf{y})|^{2})/2\sigma^{2}}e^{-(|s_{z}-z(\mathbf{y})|^{2})/2\sigma^{2}}}\,.
\end{eqnarray*}
Figure \ref{Keil_Wolf_figure_2} illustrates the general features of the EN
dynamics using the example of a single stimulus. Fig. \ref{Keil_Wolf_figure_2}a
shows a model orientation map with a superimposed uniform representation
of visual space. A single pointlike, oriented stimulus $\mathbf{S}=(\mathbf{s}_{r},s_{z})$
with position $\mathbf{s}_{r}=(s_{x},s_{y})$ and orientation $\theta=1/2\arg(s_{z})$
(Fig. \ref{Keil_Wolf_figure_2}b) evokes a cortical activity pattern
$e(\mathbf{x},\mathbf{S},z,\mathbf{r})$ (Fig. \ref{Keil_Wolf_figure_2}c). The activity-pattern in this example is of roughly Gaussian shape and is centered at the point, where the location $\mathbf{s}_{r}$ of the stimulus is represented in cortical space. However, depending on the model parameters and the stimulus, the cortical activity pattern may assume as well a more complex form (see also Discussion). 
According to Eqs. (\ref{eq:continuous_z_dynamics}, \ref{eq:continous_r_dynamics}), each stimulus and the evoked activity pattern induce a modification of
the orientation map and retinotopic map, shown in Fig. \ref{Keil_Wolf_figure_2}d.
Orientation preference in the activated regions is shifted towards
the orientation of the stimulus. The representation of visual space
in the activated regions is locally contracted towards the position of the stimulus.
Modifications of cortical selectivities occur due to randomly
chosen stimuli and are set proportional to a very small learning rate.
Substantial changes of cortical representations occur slowly through
the cumulative effect of a large number of activity patterns and stimuli.
These effective changes are described by the two deterministic equations for the rearrangement of cortical selectivities Eqs. (\ref{eq:continuous_z_dynamics}, \ref{eq:continous_r_dynamics}) which are obtained by stimulus-averaging
the modifications due to single activity patterns in the discrete stimulus model \cite{Wolf:1998p1199}.
One thus expects that the optimal selectivity patterns and also the way in which cortical selectivities change
over time are determined by the statistical properties of the stimulus ensemble. 
In the following, we assume that the stimulus ensemble satisfies three properties:
(i) The stimulus locations $\mathbf{s}_{r}$ are uniformly
distributed across visual space. (ii) For the distribution
of stimulus orientations, $|s_{z}|$ and $\theta$ are independent.
(iii) Orientations $\theta$ are distributed uniformly in $[0,\pi]$.

These conditions are fulfilled by stimulus ensembles used in virtually all prior studies of dimension reduction models for visual cortical architecture
(e.g. \cite{Hoffsummer95,Ritter:1988p4720, Durbin:1990p1196,Obermayer:1990p1202,Obermayer:1992p1200, Wolf:1998p1199,CarreiraPerpinan:2004p6297,CarreiraPerpinan:2005p6295,Giacomantonio_2010,CarreiraPerpinan:2011}).  
They imply several symmetries of the model dynamics Eqs. (\ref{eq:continuous_z_dynamics}, \ref{eq:continous_r_dynamics}). 
Due to the first property, the EN dynamics are equivariant under translations
\begin{eqnarray*}
\hat{T}_{\mathbf{y}}z(\mathbf{x}) & = & z(\mathbf{x}+\mathbf{y})\\
\hat{T}_{\mathbf{y}}\mathbf{r}(\mathbf{x}) & = & \mathbf{r}(\mathbf{x}+\mathbf{y})\,,
\end{eqnarray*}
rotations
\begin{eqnarray*}
\hat{R}_{\beta}z(\mathbf{x}) & = & e^{2i\beta}z(\Omega_{-\beta}\mathbf{x})\\
\hat{R}_{\beta}\mathbf{r}(\mathbf{x}) & = & \Omega_{\beta}\,\mathbf{r}(\Omega_{-\beta}\mathbf{x})
\end{eqnarray*}
with 2$\times$2 rotation matrix \[
\Omega_{\beta}=\left(\begin{array}{cc}
\cos\beta & -\sin\beta\\
\sin\beta & \cos\beta\end{array}\right)\,,
\]
and reflections
\begin{eqnarray*}
\hat{P}z(\mathbf{x}) & = & \bar{z}(\Psi\mathbf{x})\\
\hat{P}\mathbf{r}(\mathbf{x}) & = & \Psi\mathbf{r}(\Psi\mathbf{x})\,,
\end{eqnarray*}
where $\Psi = \textnormal{diag}(-1,1)$ is the 2$\times$2 reflection matrix. 
Equivariance means that 
\begin{eqnarray}
\hat{T}_{\mathbf{y}}F^{z}[z,\mathbf{r}]&= & F^{z}[\hat{T}_{\mathbf{y}}z,\hat{T}_{\mathbf{y}}\mathbf{r}]
\label{eq:translation_symmetry}\\
\hat{R}_{\beta}F^{z}[z,\mathbf{r}]& = & F^{z}[\hat{R}_{\beta}z,\hat{R}_{\beta}\mathbf{r}]\label{eq:rotation_symmetry}\\
\hat{P}F^{z}[z,\mathbf{r}]& = & F^{z}[\hat{P}z,\hat{P}\mathbf{r}]\,,\label{eq:reflection_symmetry}
\end{eqnarray}
with mutatis mutandis the same equations fulfilled by the vector-field $\mathbf{F}^r[z,\mathbf{r}]$. 

As a consequence, patterns that can be converted into one another by translation,
rotation or reflection of the cortical layers represent equivalent solutions of
the model Eqs. (\ref{eq:continuous_z_dynamics}, \ref{eq:continous_r_dynamics}),
by construction. Due to the second assumption, the dynamics is also equivariant
with respect to shifts in orientation $S_{\phi} z(\mathbf{x}) = e^{i\phi}z(\mathbf{x})$, i.e.
\begin{eqnarray}
e^{i\phi}F^{z}[z,\mathbf{r}] & = & F^{z}[e^{i\phi}z,\mathbf{r}] \label{eq:Shift-Symmetry-1}\\
\mathbf{F}^{r}[z,\mathbf{r}] & = & \mathbf{F}^{r}[e^{i\phi}z,\mathbf{r}]\,.
\label{eq:Shift-Symmetry-2}
\end{eqnarray}
Thus, two patterns are also equivalent solutions of the model, if their layout
of orientation domains and retinotopic distortions is identical, but
the preferred orientations differ everywhere by the same constant
angle. 

Without loss of generality, we normalize the ensembles of orientation stimuli such that $\left<|s_{z}|^{2}\right>_{\mathbf{S}}=\left<|s_{z}|^{2}\right> =2$ throughout this paper. This normalization can always be restored by a rescaling of $z(\mathbf{x})$ (see \cite{Wolf:1998p1199,Keil:2010p6536}).

Our formulation of the dimension-reduction problem in the EN model utilizes a continuum description, both for cortical space and the set of visual stimuli. This facilitates mathematical treatment and appears appropriate, given the high number of cortical neurons under one square millimeter of cortical surface (e.g. roughly 70000 in cat V1 \cite{Beaulieu:1989p8208}). Even an hypothesized neuronal mono-layer would consist of more than 20x20 neurons per hypercolumn area $\Lambda^2$, constituting a quite dense sampling of the spatial periodicity.
Treating the feature space as a continuum implements the concept that the cortical representation has to cover as good as possible the infinite multiplicity of conceivable stimulus feature combinations.
\subsection*{The orientation unselective fixed point}
\begin{figure}
\centering
\includegraphics[width=12cm]{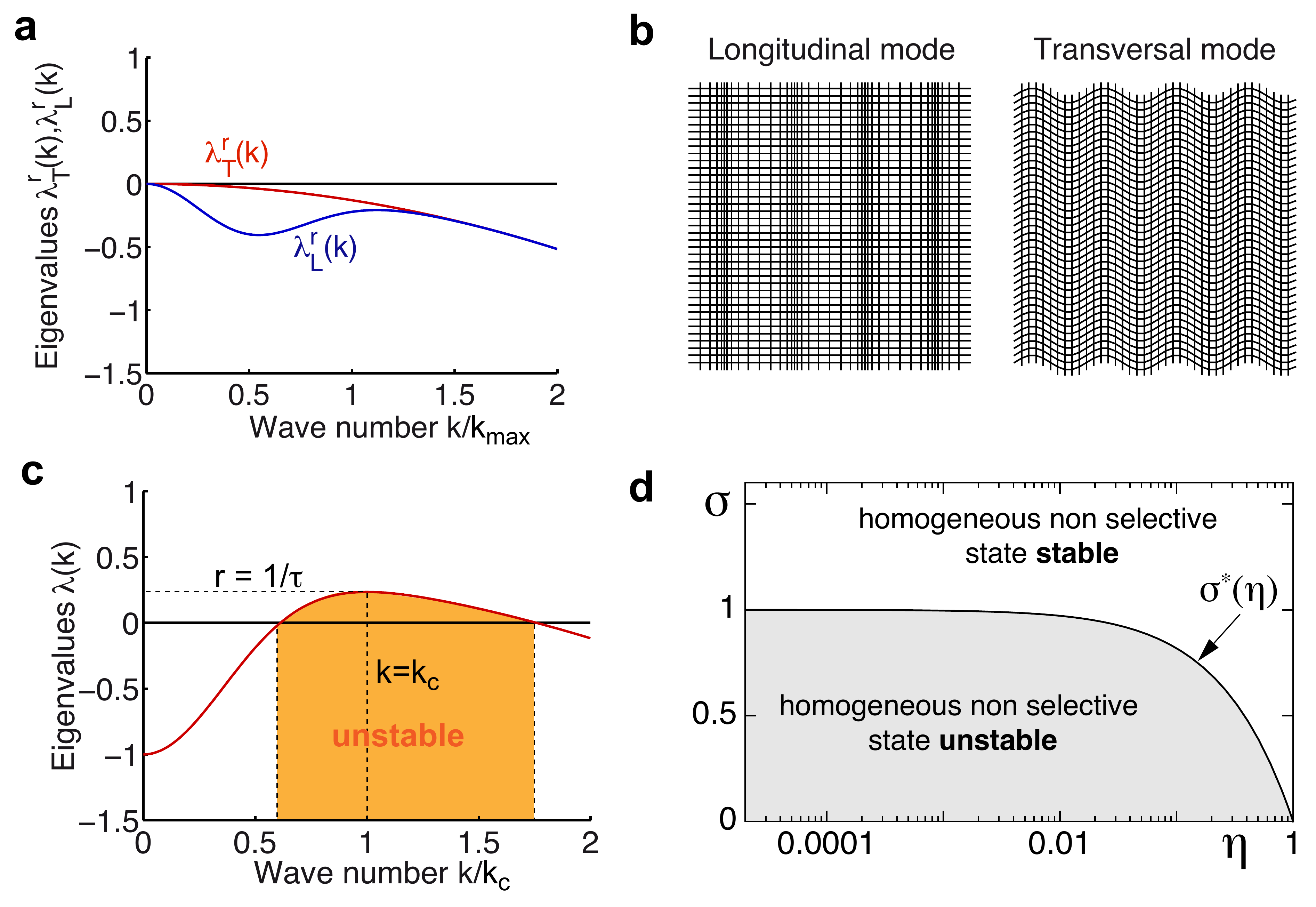}
\caption{\textbf{The linearization of the EN model dynamics around the unselective fixed point.} (\textbf{a}) Eigenvalue spectra of the linearized retinotopy dynamics for longitudinal mode ($\lambda_L^r(k)$, blue trace) and transversal mode ($\lambda_T^r(k)$, red trace). 
(\textbf{b}) Longitudinal mode $\sim\mathbf{k}_{\phi}e^{i\mathbf{k}_{\phi}\mathbf{x}}+c.c.$ (left) and transversal mode $\sim\mathbf{k}_{\phi+\pi/2}e^{i\mathbf{k}_{\phi}\mathbf{x}}+c.c.$ (right). 
 (\textbf{c}) Spectrum of eigenvalues of the linearized OPM dynamics (red trace) for $\sigma<\sigma^*(\eta)$. Orange region marks
the unstable annulus of Fourier modes (critical circle).
(\textbf{d}) Stability regions of the nonselective state in the EN model. The stability border is given by $\sigma^*(\eta)$ (Eq. \eqref{eq:sigma_star}). 
\label{Keil_Wolf_figure_3}}
\end{figure}
Two stationary solutions of the model can be established from symmetry. The simplest of these is the orientation unselective state with $z(\mathbf{x})=0$ and uniform mapping of visual space $\mathbf{r}(\mathbf{x})=\mathbf{0}$.
Firstly, by the shift symmetry Eq. (\ref{eq:Shift-Symmetry-1}), we find that $z(\mathbf{x})=0$ is a fixed point
of Eq. \eqref{eq:continuous_z_dynamics}. Secondly, by reflectional and rotational symmetry Eqs. (\ref{eq:translation_symmetry}, \ref{eq:reflection_symmetry}), we see that the right-hand side of Eq. \eqref{eq:continous_r_dynamics} has to vanish and hence the orientation unselective state with uniform mapping of visual space is a fixed point of Eqs. (\ref{eq:continuous_z_dynamics}, \ref{eq:continous_r_dynamics}).

This homogeneous unselective state thus minimizes the EN energy functional if it is a \textit{stable} solution of Eqs. (\ref{eq:continuous_z_dynamics}, \ref{eq:continous_r_dynamics}). 
The stability can be determined by considering the linearized dynamics of small deviations $\{\mathbf{r}(\mathbf{x}) , z(\mathbf{x})\}$ around this state. 
These linearized dynamics read
\begin{eqnarray}
\partial_{t}\mathbf{r}(\mathbf{x}) & \simeq & \mathbf{L}_{r}[\mathbf{r}]=\frac{1}{16\pi\sigma^{4}}\int d^{2}y\, e^{-\frac{\left(\mathbf{x}-\mathbf{y}\right)^{2}}{4\sigma^{2}}}\hat{\mathbf{A}}\,\mathbf{r}(\mathbf{y})+\eta_{r}\bigtriangleup\mathbf{r}(\mathbf{x})\label{eq:linearized-R-dynamics}\\
\partial_{t}z(\mathbf{x})& \simeq & L_{z}[z]=\left(\frac{1}{\sigma^{2}}-1\right)z(\mathbf{x})+\eta\bigtriangleup z(\mathbf{x})-\frac{1}{4\pi\sigma^{4}}\int d^{2}y\, e^{-\frac{\left(\mathbf{x}-\mathbf{y}\right)^{2}}{4\sigma^{2}}}z(\mathbf{y})\,.
\label{eq:linearized-z-dynamics}
\end{eqnarray}
where $(\hat{\mathbf{A}})_{ij}=(x_{i}-y_{i})(x_{j}-y_{j})-2\sigma^{2}\delta_{ij}$ with $\delta_{ij}$ being Kronecker's delta.
We first note that the linearized dynamics of retinotopic distortions and orientation preference decouple. Thus, up to linear order and near the homogeneous fixed point, both cortical representation evolve independently and the stability properties of the unselective state can be obtained by a separate examination of the stability properties of both cortical representations. 

The eigenfunctions of the linearized retinotopy dynamics $\mathbf{L}_{r}[\mathbf{r}]$ can be calculated by Fourier-transforming Eq. \eqref{eq:linearized-R-dynamics}:
\[
\partial_{t}\tilde{r}_{i}(\mathbf{k})=-\sum_{j=1}^{2}\left(\sigma^{2}e^{-k^{2}\sigma^{2}}k_{i}k_{j}+\eta_{r}k^{2}\delta_{ij}\right)\tilde{r}_{j}(\mathbf{k})\,,
\]
where $k=|\mathbf{k}|$ and $i = 1,2$. A diagonalization of this matrix equation yields the eigenvalues 
$$
\lambda_{L}^{r}=-k^{2}(\eta_{r}+e^{-\sigma^{2}k^{2}}\sigma^{2}),\,\lambda_{T}^{r}=-\eta_{r}k^{2}\,
$$
with corresponding eigenfunctions (in real space)
\begin{eqnarray*}
\mathbf{r}_{L}(\mathbf{x}) & = & \mathbf{k}_{\phi}e^{i\mathbf{k}_{\phi}\mathbf{x}}+\textnormal{c.c. }\\
\mathbf{r}_{T}(\mathbf{x}) & = & \mathbf{k}_{\phi+\pi/2}e^{i\mathbf{k}_{\phi}\mathbf{x}}+\textnormal{c.c.}\,,
\end{eqnarray*}
where $\mathbf{k}_{\phi} = |\mathbf{k}|(\cos\phi,\sin\phi)^T$. These eigenfunctions are longitudinal (L) or transversal (T) wave
patterns. In the longitudinal wave, the retinotopic distortion vector $\mathbf{r}(\mathbf{x})$ lies parallel to $\mathbf{k}$ which leads to a ``compression wave" (Fig. \ref{Keil_Wolf_figure_3}b, left). In the transversal wave pattern (Fig. \ref{Keil_Wolf_figure_3}b, right), the retinotopic distortion vector is orthogonal to $\mathbf{k}$. We note that the linearized Kohonen model \cite{Kohonen:1995p6543} was previously found to
exhibit the same set of eigenfunctions \cite{Ritter:1988p4720}. Because both
spectra of eigenvalues $\lambda_{T}^{r},\,\lambda_{L}^{r}$ are smaller
than zero for every $\sigma>0$, $\eta_{r}>0$, and $k>0$ (Fig. \ref{Keil_Wolf_figure_3}a),
the uniform retinotopy $\mathbf{r(x)}=\mathbf{0}$ is a stable fixed
point of Eq. \eqref{eq:continous_r_dynamics} irrespective of parameter choice.

The eigenfunctions of the linearized OPM dynamics
$L_{z}[z]$ are Fourier modes $\sim e^{i\mathbf{kx}}$ by translational symmetry. By rotational
symmetry, their eigenvalues only depend on the wave number $k$
and are given by
\[
\lambda^{z}(k)=-1+\frac{1}{\sigma^2}\left(1-e^{-k^{2}\sigma^{2}}\right)-\eta k^2\,.
\]
(see \cite{Wolf:1998p1199}). This spectrum of eigenvalues is depicted in Fig. \ref{Keil_Wolf_figure_3}c.
For $\eta>0$, $\lambda^{z}(k)$ has a single maximum at
$k_{c}=\frac{1}{\sigma}\sqrt{\ln\left(1/\eta\right)}$. For
\begin{equation}
\sigma>\sigma^{*}(\eta)=\sqrt{1+\eta\ln\eta-\eta}
\label{eq:sigma_star}
\end{equation}
this maximal eigenvalue $r=\lambda^{z}(k_{c})$ is negative. Hence, 
the unselective state with uniform retinotopy is a stable fixed point of Eqs. (\ref{eq:continuous_z_dynamics}, \ref{eq:continous_r_dynamics})
and the only known solution of the EN model in this parameter range.

For $\sigma<\sigma^{*}(\eta)$,
the maximal eigenvalue $r$ is positive, and the nonselective state
is unstable with respect to a band of Fourier modes $\sim e^{i\mathbf{kx}}$
with wave numbers around $|\mathbf{k}|\approx k_{c}$ (see Fig. \ref{Keil_Wolf_figure_3}c). This annulus
of unstable Fourier modes is called the critical circle. The finite
wavelength instability \cite{Cross:1993p922,manneville_90,greenside_09}
(or Turing instability \cite{Turing:1952p983}) leads to the
emergence of a pattern of orientation preference
with characteristic spacing $\Lambda=2\pi/k_{c}$ from the nonselective
state on a characteristic timescale $\tau=1/r$. 

One should note that as in other models for the self-organization of orientation columns,
e.g. \cite{Swindale1998p827,swindale_82}, the characteristic spatial scale $\Lambda$ arises
from effective intracortical interactions of `Mexican-hat' structure
(short-range facilitation, longer-ranged suppression). The short-range facilitation in the linearized EN dynamics is represented by the first two terms on the right hand side of Eq. \eqref{eq:linearized-z-dynamics}. Since $\sigma<1$ in the pattern forming regime, the prefactor in front of the first term is positive. Due to the second, Laplacian term, it is favored that neighboring units share selectivity properties, a process mediated by short-range facilitation.
Longer-ranged suppression is represented by the convolution term in Eq. \eqref{eq:linearized-z-dynamics}. Mathematically, this term directly results from the soft-competition in the ``activity-dependent" coverage term of Eq. \eqref{eq:Elastic-Net-Energy}. The local facilitation is jointly mediated by coverage (first term) and continuity (second term) contributions.

Fig. \ref{Keil_Wolf_figure_3}d summarizes the result of the linear stability analysis of the nonselective state. For $\sigma>\sigma^*(\eta)$, the orientation unselective state with uniform retinotopy is a minimum of the EN free energy and also the global minimum. For $\sigma<\sigma^*(\eta)$ this state represents a maximum of the energy functional and the minima must thus exhibit a space-dependent pattern of orientation selectivities. 
\subsection*{Orientation stripes}
\begin{figure}
\centering
\includegraphics[width=12cm]{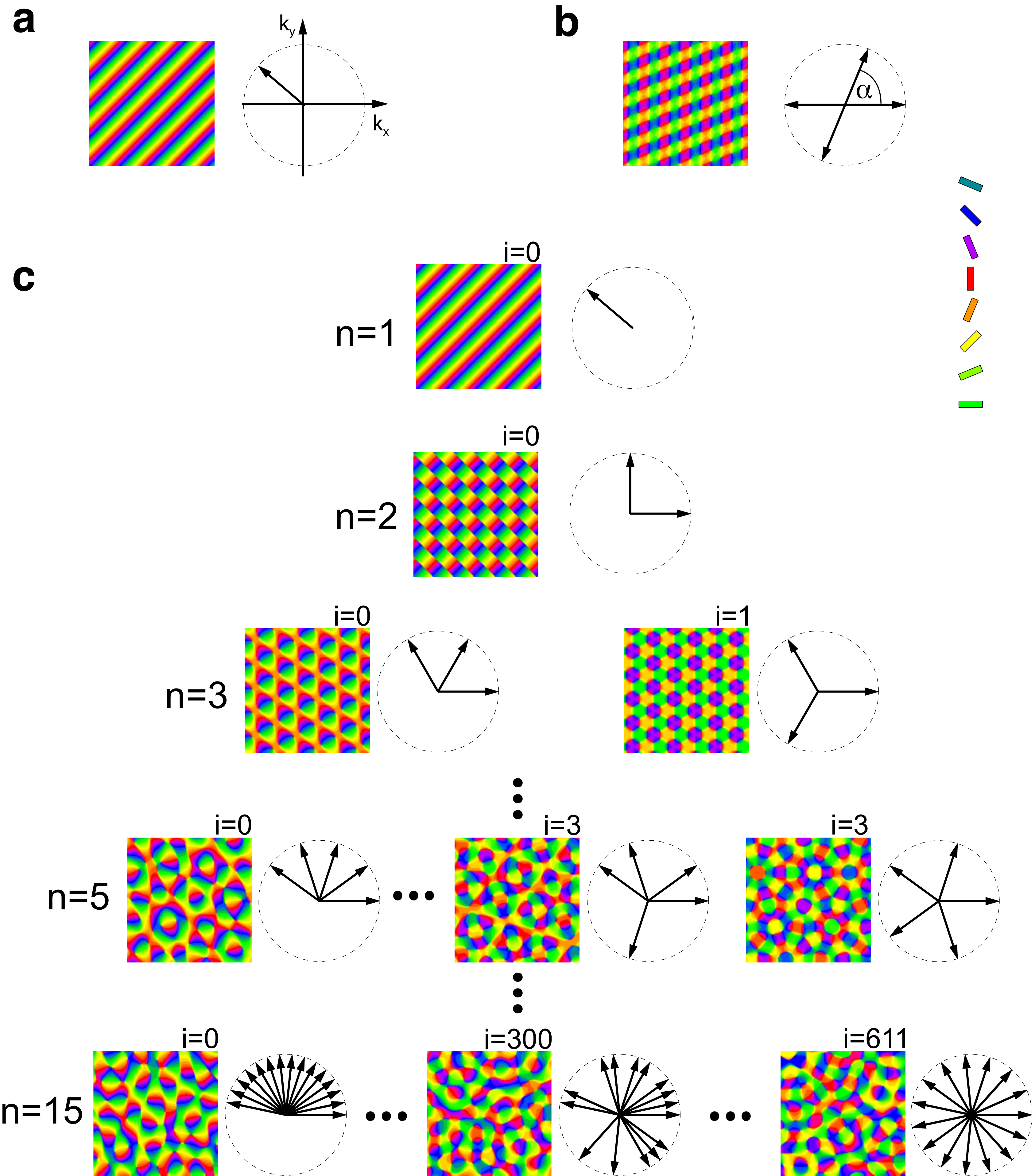}
\caption{\textbf{Exact and approximate orientation selective fixed points of OPM optimization models} 
(\textbf{a}) Pinwheel free orientation stripe (OS) pattern. Diagram shows the position
of the wave vector in Fourier space.
(\textbf{b}) Rhombic Pinwheel crystal (rPWC) with four nonzero wave vectors.
(\textbf{c}) Essentially complex planforms (ECPs). The index n indicates the number of nonzero wave vectors.
The index i enumerates nonequivalent configurations of wave vectors
with the same n, starting with i = 0 for the most anisotropic planform.
For n = 3, 5, and 15, there are 2, 4, and 612 different ECPs, respectively.
OPM layouts become more irregular with increasing n.
\label{Keil_Wolf_figure_4}}
\end{figure}
Within the potentially infinite set of orientation selective fixed points of the model, one class of solutions can be established from symmetry:
$\{\mathbf{r}(\mathbf{x}) = \mathbf{0}, z(\mathbf{x}) =A_{0}e^{i\mathbf{kx}}\}$.
In these pinwheel-free states, orientation preference is constant along one axis in cortex (perpendicular to the vector $\mathbf{k}$), and each orientation is represented in equal proportion (see Fig. \ref{Keil_Wolf_figure_4}a). Retinotopy is perfectly uniform.
Although this state may appear too simple to be biologically relevant, we will see that it plays a fundamental role in the state space of the EN model. It is therefore useful to establish its existence and basic characteristics. The existence of orientation stripe (OS) solutions follows directly from the model's symmetries Eqs. (\ref{eq:translation_symmetry}-\ref{eq:Shift-Symmetry-2}).
Computing
\[
T_{\mathbf{y}}[F^{z}[e^{i\mathbf{kx}}, \mathbf{0}]]=F^{z}[T_{\mathbf{y}}[e^{i\mathbf{kx}}], T_{\mathbf{y}}[\mathbf{0}]]=F^{z}[e^{i\mathbf{ky}}e^{i\mathbf{kx}}, \mathbf{0}]=e^{i\mathbf{ky}}F^{z}[e^{i\mathbf{kx}}, \mathbf{0}]
\]
demonstrates that $F^{z}[e^{i\mathbf{kx}}, \mathbf{0}] $ is proportional to $e^{i\mathbf{kx}}$.
This establishes that the subspace of functions $\sim e^{i\mathbf{kx}}$ is invariant under the dynamics Eq. \eqref{eq:continuous_z_dynamics}.
For $A_{0}=0$, we recover the trivial fixed point of the EN dynamics
by construction, as shown above. This means that within this subspace $A_{0}=0$ is either a
minimum or a maximum of the EN energy functional Eq. \eqref{eq:Elastic-Net-Energy}.
Furthermore, for $A_{0}\rightarrow\infty$ the EN energy tends to
infinity. If the trivial fixed point is unstable, it corresponds to
a maximum of the EN energy functional. Therefore, there must exist at
least one minimum with $A_{0}\neq0$ in the subspace of functions $\sim e^{i\mathbf{kx}}$
which then corresponds to a stationary state of the EN dynamics. 

Regarding the dynamics of retinotopic deviations, the model's symmetries Eqs. (\ref{eq:translation_symmetry}-\ref{eq:Shift-Symmetry-2}) can be invoked to show that for the state $\{\mathbf{0}, A_{0}e^{i\mathbf{kx}}\}$, the right-hand side of Eq.  \eqref{eq:continous_r_dynamics} has to be constant in space:
\[
T_{\mathbf{y}}[\mathbf{F}^{r}[e^{i\mathbf{kx}}, \mathbf{0}]]=\mathbf{F}^{r} [T_{\mathbf{y}}[e^{i\mathbf{kx}}], T_{\mathbf{y}}[\mathbf{0}]]=\mathbf{F}^{r}[e^{i\mathbf{ky}}e^{i\mathbf{kx}}, \mathbf{0}]=\mathbf{F}^{r}[e^{i\mathbf{kx}}, \mathbf{0}]
\]
If this constant was non-zero the retinotopic map would drift with constant velocity. This, however, is impossible in a variational dynamics
such that this constant must vanish. The orientation stripe solution (Fig. \ref{Keil_Wolf_figure_4}a) is to the best of our knowledge the only exact nontrivial stationary solution of Eqs. (\ref{eq:continuous_z_dynamics}, \ref{eq:continous_r_dynamics}) that can be established without any approximations.
\subsection*{Doubly periodic and quasiperiodic solutions}
In the EN model as considered in this study, the maps of visual space and orientation preference are jointly optimized to trade off coverage and continuity leading to mutual interactions between the two cortical representations. These mutual interactions vanish in the rigid retinotopy limit $\eta_r\rightarrow \infty$ and the perfectly uniform retinotopy becomes an optimal solution for arbitrary orientation column layout $z(\mathbf{x})$. As it is not clear how essential the mutual interactions with position specificity are in shaping the optimal orientation column layout, we continue our investigation of solution classes by considering global minima of optimization models with fixed uniform retinotopy. The mutual interactions will be taken into account in a subsequent step. 

In the rigid retinotopy limit, minima of the energy functional are stable stationary states of the dynamics of the orientation preference map Eq. \eqref{eq:continuous_z_dynamics} with $\mathbf{r}(\mathbf{x})=\mathbf{0}$.
To compute orientation selective stationary solutions of this OPM dynamics, we employ
that in the vicinity of a supercritical bifurcation where the nonselective fixed point $z(\mathbf{x})=0$ becomes unstable, the entire set of nontrivial fixed points is determined by the third-order
terms of the Volterra series representation of the operator
$F^{z}[z, \mathbf{0}]$ \cite{manneville_90,Wolf:2005p190, wolf_04, greenside_09}.
The symmetries Eqs. (\ref{eq:translation_symmetry}-\ref{eq:Shift-Symmetry-2})
restrict the general form of such a third-order approximation for
any model of OPM optimization to
\begin{equation}
\partial_{t}z(\mathbf{x})\approx L_{z}[z]+N_{3}^{z}[z,z,\bar{z}]\,,
\label{eq:third-order-approx-z-dynamics}
\end{equation}
where the cubic operator $N_{3}^{z}$ is written in trilinear form, i.e. 
$$
N_{3}^{z}\left[\sum_j \alpha_j z_j,\sum_k \beta_k z_k,\sum_l \gamma_l \bar{z}_l\right] = \sum_{j,k,l} \alpha_j \beta_k \gamma_l\, N_{3}^{z}[z_j, z_k, \bar{z}_l] \,.
$$
In particular, all even terms in the Volterra Series representation
of $F^{z}[z, \mathbf{0}]$ vanish due to the Shift-Symmetry Eqs. (\ref{eq:Shift-Symmetry-1}, \ref{eq:Shift-Symmetry-2}).
Explicit analytic computation of the cubic nonlinearities for the
EN model is cumbersome but not difficult (see Methods) and yields
a sum
\begin{equation}
N_{3}^{z}[z,z,\bar{z}]=\sum_{j=1}^{11}a_{j}N_{3}^{j}[z,z,\bar{z}]\,.\label{eq:sum-of-cubic-operators}
\end{equation}
The individual nonlinear operators $N_{3}^{j}$ are with one exception nonlocal
convolution-type operators and are given in the Methods section (Eq. \eqref{eq:cubic_z_nonlinearities}), together
with a detailed description of their derivation. Only the coefficients
$a_{j}$ depend on the properties of the ensemble of oriented stimuli.

To calculate the fixed points of Eq. \eqref{eq:third-order-approx-z-dynamics},
we use a perturbative method called weakly nonlinear analysis that
enables us to analytically examine the structure and stability of inhomogeneous
stationary solutions in the vicinity of a finite-wavelength instability. Here,
we examine the stability of so-called planforms \cite{manneville_90,Cross:1993p922,greenside_09}.
Planforms are patterns that are composed of a finite number of Fourier
components, such as
\[
z(\mathbf{x})=\sum_{j}A_{j}(t)e^{i\mathbf{k}_j\mathbf{x}}
\]
for a pattern of orientation columns. 
With the above planform ansatz, we neglect any spatial dependency of the amplitudes $A_{j}(t)$ for example due to long-wave deformations for the sake of simplicity and analytical tractability.
When the dynamics is close to a finite wavelength instability, the essential Fourier components
of the emerging pattern are located on the critical circle $|\mathbf{k}_{j}|=k_{c}$.
The dynamic equations for the amplitudes of these Fourier components
are called amplitude equations. For a discrete number of $N$ Fourier
components of $z(\mathbf{x})$ whose wave vectors lie equally spaced
on the critical circle, the most general system of amplitude equations
compatible with the symmetries Eqs. (\ref{eq:translation_symmetry}-\ref{eq:Shift-Symmetry-2})
has the form \cite{wolf_04,Wolf:2005p190}
\begin{equation}
\dot{A}_{i}=rA_{i}-A_{i}\sum_{j=1}^{N}g_{ij}|A_{j}|^{2}-\bar{A}_{i^{-}}\sum_{j=1}^{N}f_{ij}A_{j}A_{j^{-}}\,,
\label{eq:general-amplitude-equations}
\end{equation}
with $r>0$. Here, $g_{ij}$ and $f_{ij}$ are the real-valued coupling
coefficients between the amplitudes $A_{i}$ and $A_{j}$. They depend on the differences between indices $|i-j|$ and are entirely
determined by the nonlinearity $N_{3}^{z}[z,z,\bar{z}]$ in Eq. \eqref{eq:third-order-approx-z-dynamics}.
If the wave vectors $\mathbf{k}_{i}=(\cos\alpha_{i},\sin\alpha_{i})k_{c}$
are parametrized by the angles $\alpha_{i}$, then the coefficients
$g_{ij}$ and $f_{ij}$ are functions only of the angle $\alpha=|\alpha_{i}-\alpha_{j}|$
between the wave vectors $\mathbf{k}_{i}$ and $\mathbf{k}_{j}$.
One can thus obtain the coupling coefficients from two continuous functions $g(\alpha)$
and $f(\alpha)$ that can be obtained from the nonlinearity
$N_{3}^{z}[z,z,\bar{z}]$ (see Methods for details). In the following, these functions
are called angle-dependent interaction functions.
The amplitude equations Eq. \eqref{eq:general-amplitude-equations}
are variational if and only if $g_{ij}$ and $f_{ij}$ are real-valued. In this case they can be derived via 
$$
\dot{A}_j(t) = \frac{\partial U_A}{\partial \bar{A}_j}
$$ from an energy
\begin{equation}
U_{A}=-r\sum_{i=1}^{N}|A_{i}|^{2}+\frac{1}{2}\sum_{i,\, j=1}^{N}g_{ij}|A_{i}|^{2}|A_{j}|^{2}+\frac{1}{2}\sum_{i,\, j=1}^{N}f_{ij}\bar{A}_{i}\bar{A}_{i^{-}}A_{j}A_{j^{-}}\,.
\label{eq:potential-of-amplitude-equations}
\end{equation}

If the coefficients $g_{ij}$ and $f_{ij}$ are derived from Eq. \eqref{eq:Elastic-Net-Energy}, the energy $U_A$  for a given planform solution corresponds to the energy density of the EN energy functional considering only terms up to fourth order in $z(\mathbf{x})$. 

The amplitude equations  Eq. \eqref{eq:general-amplitude-equations} enable to calculate an infinite set of orientation selective fixed points.
For the above orientation stripe solution with one nonzero wave vector $z(\mathbf{x})=A_{0}e^{i\mathbf{kx}}$, the amplitude equations predict the so far undetermined amplitude
\begin{equation}
|A_{0}|^{2}=\frac{r}{g_{ii}}
\label{eq:amplitude-orientation-stripes}
\end{equation}
and its energy
\begin{equation}
U_{\textnormal{OS}}=-\frac{r}{2g_{ii}}\,.
\label{eq:energy-of-orientation-stripes}
\end{equation}
Since $g_{ii}>0$, this shows that OS stationary solutions only exist for $r>0$, i.e. in the symmetry breaking regime. As for all following fixed-points, $U_{\textnormal{OS}}$ specifies the energy difference to the homogeneous unselective state $z(\mathbf{x})=0$. 

A second class of stationary solutions can be found with the ansatz \[
z(\mathbf{x})=A_{1}e^{i\mathbf{k}_{1}\mathbf{x}}+A_{2}e^{i\mathbf{k}_{2}\mathbf{x}}+A_{3}e^{-i\mathbf{k}_{1}\mathbf{x}}+A_{4}e^{-i\mathbf{k}_{2}\mathbf{x}}\,,\]
with amplitudes $A_{j}=|A_{j}|e^{i\phi_{j}}$ and $\angle(\mathbf{k}_{1},\mathbf{k}_{2})=\alpha>0$.
By inserting this ansatz into Eq. \eqref{eq:general-amplitude-equations} and assuming uniform amplitude
$|A_{1}|=|A_{2}|=|A_{2}|=|A_{4}|=\mathcal{A}$, we obtain
\begin{equation}
\mathcal{A}^2=\frac{r}{g_{00}+g_{0\pi}+g_{0\alpha}+g_{0\pi-\alpha}-2f_{0\alpha}}\,.
\label{eq:amplitude-rPWC}
\end{equation}
The phase relations of the four amplitudes are given by
\begin{eqnarray*}
\phi_{1}+\phi_{3} & = & \phi_{0}\\
\phi_{2}+\phi_{4} & = & \phi_{0}+\pi\,.
\end{eqnarray*}
These solutions describe a regular rhombic lattice of pinwheels and are therefore called rhombic pinwheel crystals (rPWC) in the following. Three phases can be chosen arbitrarily according to the two above conditions, e.g. $\phi_0$, $\Delta_0 = \phi_1- \phi_3$ and $\Delta_1 = \phi_2- \phi_4$. For an rPWC parametrized by these phases, $\Delta_0$ shifts the absolute positions of the pinwheels in x-direction,  $\Delta_1$ shifts the absolute positions of the pinwheels in y-direction,  and $\phi_0$ shifts all the preferred orientations by a constant angle. 
The energy of an rPWC solution is
\begin{equation}
U_{\textnormal{rPWC}}=-\frac{2r}{g_{00}+g_{0\pi}+g_{0\alpha}+g_{0\pi-\alpha}-2f_{0\alpha}}\,.
\label{eq:energy-of-rPWC}
\end{equation}
An example of such a solution is depicted in Fig. \ref{Keil_Wolf_figure_4}b.
We note that rPWCs have been previously found in several other models for OPM
development \cite{Koulakov:2001p842, Lee:2003p5834, schnabel_diss_09 , Reichl2009:p208101, schnabel_2011}. 
The pinwheel density $\rho$ of an rPWC, i.e. the number of pinwheels
in an area of size $\Lambda^{2}$, is equal to $\rho=4\sin\alpha$ \cite{Kaschube_2010_SOM}.
The angle $\alpha$ which minimizes the energy $U_{\textnormal{rPWC}}$ can be computed
by maximizing the function
\begin{equation}
s(\alpha)=g_{0\alpha}+g_{0\pi-\alpha}-2f_{0\alpha}
\label{eq:rhombic-energy-angle}
\end{equation}
in the denominator of Eq. \eqref{eq:energy-of-rPWC}.

The two solution classes discussed so far, namely OS and rPWCs, exhibit one prominent feature, absent
in experimentally observed cortical OPMs, namely perfect spatial periodicity. Many cortical maps
including OPMs do not resemble a crystal-like grid of repeating units. Rather the
maps are characterized by roughly repetitive but aperiodic spatial arrangement of feature preferences (e.g. \cite{Bonhoeffer:1991p5561, Bosking:1997p5557}).
This does not imply that the precise layout of columns is arbitrary. It rather means that the rules of column design cannot be exhaustively characterized by mapping a ``representative" hypercolumn.

Previous studies of abstract models of OPM development introduced the family of so-called essentially complex planforms (ECPs) as stationary solutions of
Eq. \eqref{eq:general-amplitude-equations}. This solution class encompasses a large variety of realistic quasiperiodic OPM layouts and is therefore a good candidate solution class for models of OPM layout. In addition, Kaschube et al. demonstrated that models in which these are optimal solutions can reproduce all essential features of the common OPM design in ferret, tree-shrew, and galago \cite{Kaschube_2010}.

An n-ECP solution can be written as
\[
z(\mathbf{x})=\sum_{j=1}^{n}A_{j}e^{il_{j}\mathbf{k}_{j}\mathbf{x}}\,,
\]
with $n=N/2$ wave vectors $\mathbf{k}_{j}=k_{c}(\cos(\pi j/n),\sin(\pi j/n))$
distributed equidistantly on the upper half of the critical circle,
complex amplitudes $A_{j}$ and binary variables $l_{j}=\pm1$ determining whether
the mode with wave vector $\mathbf{k}_{j}$ or $-\mathbf{k}_{j}$
is active (nonzero). Because these planforms cannot realize a real-valued
function they are called essentially complex \cite{Wolf:2005p190}.
For an n-ECP, the third term on the right hand side of Eq. \eqref{eq:general-amplitude-equations}
vanishes and the amplitude equations for the active modes $A_{i}$
reduce to a system of Landau equations
\begin{eqnarray*}
\dot{A}_{i} & = & rA_{i}-A_{i}\sum_{j=1}^{n}g_{ij}|A_{j}|^{2}\,.
\label{eq:amplitudes_for_ECPs}
\end{eqnarray*}
where $g_{ij}$ is the $n\times n$-coupling matrix for the active
modes. Consequently, the stationary amplitudes obey
\begin{equation}
|A_{i}|^{2}=r\sum_{j=1}^{n}\left(\mathbf{g}^{-1}\right){}_{ij}\,.\label{eq:amplitude-of-n-ECP}
\end{equation}
The energy of an n-ECP is given by
\begin{equation}
U_{\textnormal{ECP}}=-\frac{r}{2}\sum_{i,j}\left(\mathbf{g}^{-1}\right)_{ij}\,.
\label{eq:energy-of-an-n-ECP}
\end{equation}
We note that this energy in general depends on the configuration of active modes, given by the $l_j$'s, and therefore planforms with the same number of active modes may not be energetically degenerate.

Families of n-ECP solutions are depicted in Fig. \ref{Keil_Wolf_figure_4}c.
The 1-ECP corresponds to the pinwheel-free OS pattern discussed above.
For fixed $n\geq3$, there are multiple planforms not related by symmetry
operations which considerably differ in their spatial layouts. For $n\geq4$, the
patterns are spatially quasiperiodic, and are a generalization of the so-called Newell-Pomeau turbulent crystal \cite{Mermin:1985p6463, Newell:1993p6534}. For $n\geq10$, their layouts resemble experimentally observed OPMs.
Different n-ECPs however differ considerably
in their pinwheel density. Planforms whose nonzero wave vectors are
distributed isotropically on the critical circle typically have a
high pinwheel density (see Fig. \ref{Keil_Wolf_figure_4}c,
n=15 lower right). Anisotropic planforms generally contain considerably
fewer pinwheels (see Fig. \ref{Keil_Wolf_figure_4}c,
n=15 lower left). All large n-ECPs, however, exhibit a complex quasiperiodic
spatial layout and a nonzero density of pinwheels.

In order to demonstrate that a certain planform is an optimal solution of
an optimization model for OPM layouts in which patterns emerge via a
supercritical bifurcation, we not only have to show that it is a
stationary solution of the amplitude equations but have to analyze its
stability properties with respect to the gradient descent dynamics as well
as its energy compared to all other candidate solutions.

Many stability properties can be characterized by examining the amplitude equations \eqref{eq:general-amplitude-equations}. In principle, the stability range of an n-ECPs may be bounded by two different instability
mechanisms: (i) an intrinsic instability by which stationary solutions
with $n$ active modes decay into ones with lower $n$. (ii) an 
extrinsic instability by which stationary solutions with a ``too low"
number of modes are unstable to the growth of additional active modes.
These instabilities can constrain the range of stable $n$ to a small
finite set around a typical $n$ \cite{wolf_04,Wolf:2005p190}. A mathematical evaluation of both
criteria leads to precise conditions for extrinsic and intrinsic stability
of a planform (Methods). In the following, a planform is said to be stable, if it is
both extrinsically and intrinsically stable. A planform is said to
be an optimum (or optimal solution) if it is stable and possesses the minimal energy
among all other stationary planform solutions.

Taken together, this amplitude equation approach enables to analytically compute the fixed points and
optima of arbitrary optimization models for visual cortical map layout in which the functional architecture
is completely specified by the pattern of orientation columns $z(\mathbf{x})$ and emerges via a supercritical bifurcation.
Via a third-order expansion of the energy functional together with weakly nonlinear analysis, the otherwise analytically intractable partial integro-differential 
equation for OPM layouts reduces to a much simpler system of ordinary differential equations, the amplitude equations.  Using these, several families of solutions, orientation stripes, rhombic pinwheel crystals and essentially complex planforms, can be systematically evaluated and comprehensively compared to identify sets of unstable, stable and optimal, i.e. lowest energy fixed points.

As already mentioned, the above approach is suitable for arbitrary optimization models for visual cortical
map layout in which the functional architecture is completely specified by the pattern of orientation
columns $z(\mathbf{x})$ which in the EN model is fulfilled in the rigid retinotopy limit. We now start by considering the EN optimal solutions in this limit and subsequently generalize  this approach to models in which the visual cortical architecture is jointly specified by maps of orientation and position preference that are matched to one another.
\subsection*{Representing an ensemble of ``bar"-stimuli}
\begin{figure}
\centering
\includegraphics[width=10cm]{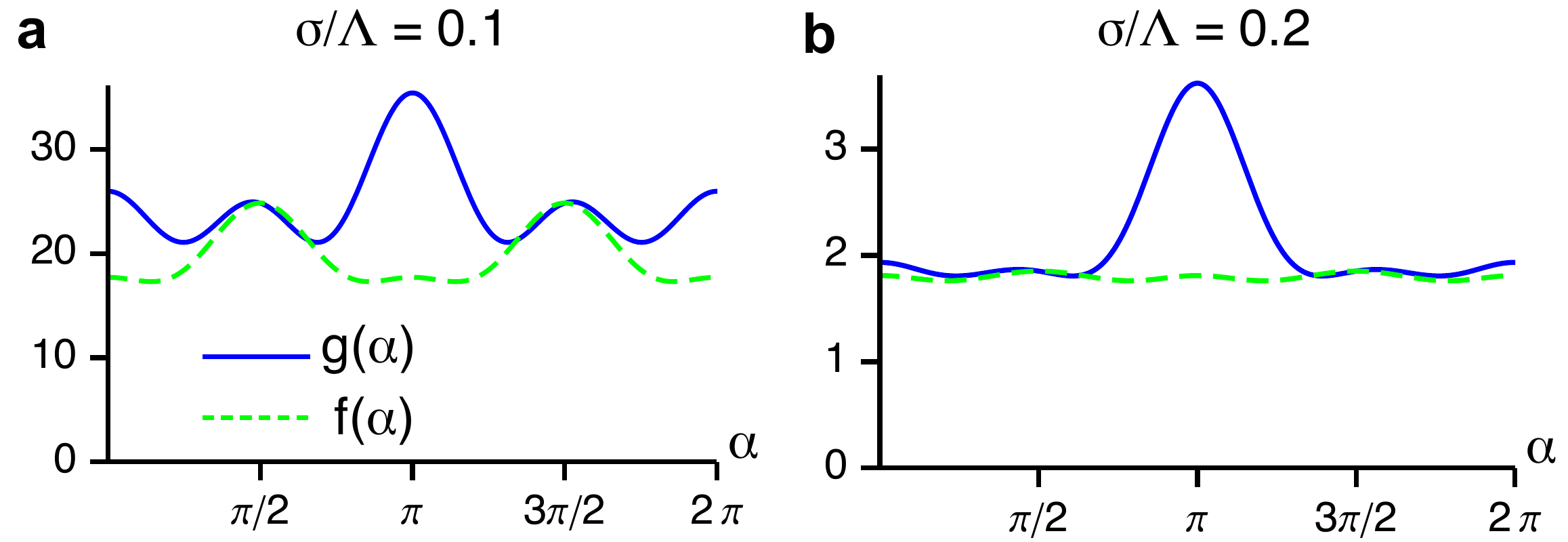}
\caption{\textbf{Angle-dependent interaction functions for the EN model with fixed
retinotopy and circular orientation stimulus ensemble.}
(\textbf{a},\textbf{b}) $g(\alpha)$ and $f(\alpha)$ for $\sigma/\Lambda=0.1$ (a) and $\sigma/\Lambda=0.2$ (b).
\label{Keil_Wolf_figure_5}}
\end{figure}
We start our investigation of optimal dimension-reducing mappings
in the EN model using the simplest and most frequently used orientation
stimulus ensemble, the distribution with $s_{z}$-values uniformly
arranged on a ring with radius $r_{s_{z}}=\sqrt{2}$ \cite{Swindale:1992p4338,Swindale1998p827,Swindale:2004p6287,CarreiraPerpinan:2004p6297,CarreiraPerpinan:2005p6295}.
We call this stimulus ensemble the \textit{circular} stimulus ensemble in
the following. According to the linear stability analysis of the nonselective
fixed point, at the point of instability, we choose $\sigma=\sigma^{*}(\eta)$
such that the linearization Eq. \eqref{eq:linearized-z-dynamics} is completely
characterized by the continuity parameter $\eta$. Equivalent to specifying $\eta$ is to fix the ratio of activation range $\sigma$ and column spacing $\Lambda$
\begin{equation}
\sigma/\Lambda=\frac{1}{2\pi}\sqrt{\log(1/\eta)}\label{eq:sig_per_lambda_definition}
\end{equation}
as a more intuitive parameter. This ratio measures the effective interaction-range relative to the expected spacing of
the orientation preference pattern. In abstract optimization models
for OPM development a similar quantity has been demonstrated to be
a crucial determinant of pattern selection \cite{wolf_04,Wolf:2005p190}.
We note, however, that due to the logarithmic dependence of $\sigma/\Lambda$
on $\eta$, a slight variation of the effective interaction range 
may correspond to a variation of the continuity parameter $\eta$ over several orders of magnitude.

In order to investigate the stability of stationary planform solutions in the EN model with a circular orientation
stimulus ensemble, we have to determine the angle-dependent
interaction functions $g(\alpha)$ and $f(\alpha)$. For the coefficients
$a_{j}$ in Eq. \eqref{eq:sum-of-cubic-operators} we obtain
\begin{equation*}
\begin{array}{lll}
a_{1}=\frac{1}{4\sigma^{6}}-\frac{1}{\sigma^{4}}+\frac{1}{2\sigma^{2}} & a_{2}=\frac{1}{4\pi\sigma^{6}}-\frac{1}{8\pi\sigma^{8}}\hspace{1em} & a_{3}=-\frac{1}{16\pi\sigma^{8}}+\frac{1}{8\pi\sigma^{6}}\\
a_{4}=-\frac{1}{8\pi\sigma^{8}}+\frac{1}{4\pi\sigma^{6}}-\frac{1}{8\pi\sigma^{4}}\hspace{1em} & a_{5}=-\frac{1}{16\pi\sigma^{8}} & a_{6}=\frac{1}{8\pi\sigma^{6}}-\frac{1}{16\pi\sigma^{8}}\\
a_{7}=\frac{1}{12\pi^{2}\sigma^{10}}-\frac{1}{12\pi^{2}\sigma^{8}} & a_{8}=\frac{1}{24\pi^{2}\sigma^{10}} & a_{9}=-\frac{3}{64\pi^{3}\sigma^{12}}\\
a_{10}=\frac{1}{12\pi^{2}\sigma^{10}}-\frac{1}{12\pi^{2}\sigma^{8}} & a_{11}=\frac{1}{24\pi^{2}\sigma^{10}}\,.
\end{array}
\end{equation*}
The angle-dependent interaction functions of the EN model with a circular
orientation stimulus ensemble are then given by
\begin{eqnarray}
g(\alpha) & = & \frac{1}{\sigma^{4}}\left(1-2e^{-k_{c}^{2}\sigma^{2}}-e^{2k_{c}^{2}\sigma^{2}(\cos\alpha-1)}\left(1-2e^{-k_{c}^{2}\sigma^{2}\cos\alpha}\right)\right)\nonumber \\
 &  & +\frac{1}{2\sigma^{2}}\left(e^{2k_{c}^{2}\sigma^{2}(\cos\alpha-1)}-1\right)+\frac{8}{\sigma^{6}}e^{-2k_{c}^{2}\sigma^{2}}\sinh^{4}\left(1/2k_{c}^{2}\sigma^{2}\cos\alpha\right)\nonumber \\
f(\alpha) & = & \frac{1}{\sigma^{4}}\left(1-e^{-2k_{c}^{2}\sigma^{2}}\left(\cosh(2k_{c}^{2}\sigma^{2}\cos\alpha)+2\cosh(k_{c}^{2}\sigma^{2}\cos\alpha)\right)+2e^{-k_{c}^{2}\sigma^{2}}\right)\nonumber \\
 &  & +\frac{1}{2\sigma^{2}}\left(e^{-2k_{c}^{2}\sigma^{2}}\cosh(2k_{c}^{2}\sigma^{2}\cos\alpha)-1\right)+\frac{4}{\sigma^{6}}e^{-2k_{c}^{2}\sigma^{2}}\sinh^{4}\left(1/2k_{c}^{2}\sigma^{2}\cos\alpha\right)\,.\label{eq:angle-dependent interaction-function-circular-ensemble}
 \end{eqnarray}
These functions are depicted in Fig. \ref{Keil_Wolf_figure_5}
for two different values of the interaction range $\sigma/\Lambda$. We note that both functions are positive for all $\sigma/\Lambda$ which is a sufficient condition for a supercritical bifurcation from the homogeneous nonselective state in the EN model. 

Finally, by minimizing the function $s(\alpha)$ in Eq. \eqref{eq:rhombic-energy-angle}, we find that
the angle $\alpha$ which minimizes the energy of rPWC fixed-point is $\alpha=\pi/2$. This corresponds to a square array of pinwheels (sPWC). Due to the orthogonal arrangement oblique and cardinal orientation columns and the maximized pinwheel density of $\rho=4$, the square array of pinwheels has the maximal coverage among all rPWC solutions. 
\subsubsection*{Optimal solutions close to the pattern formation threshold}
\begin{figure*}
\includegraphics[width=16cm]{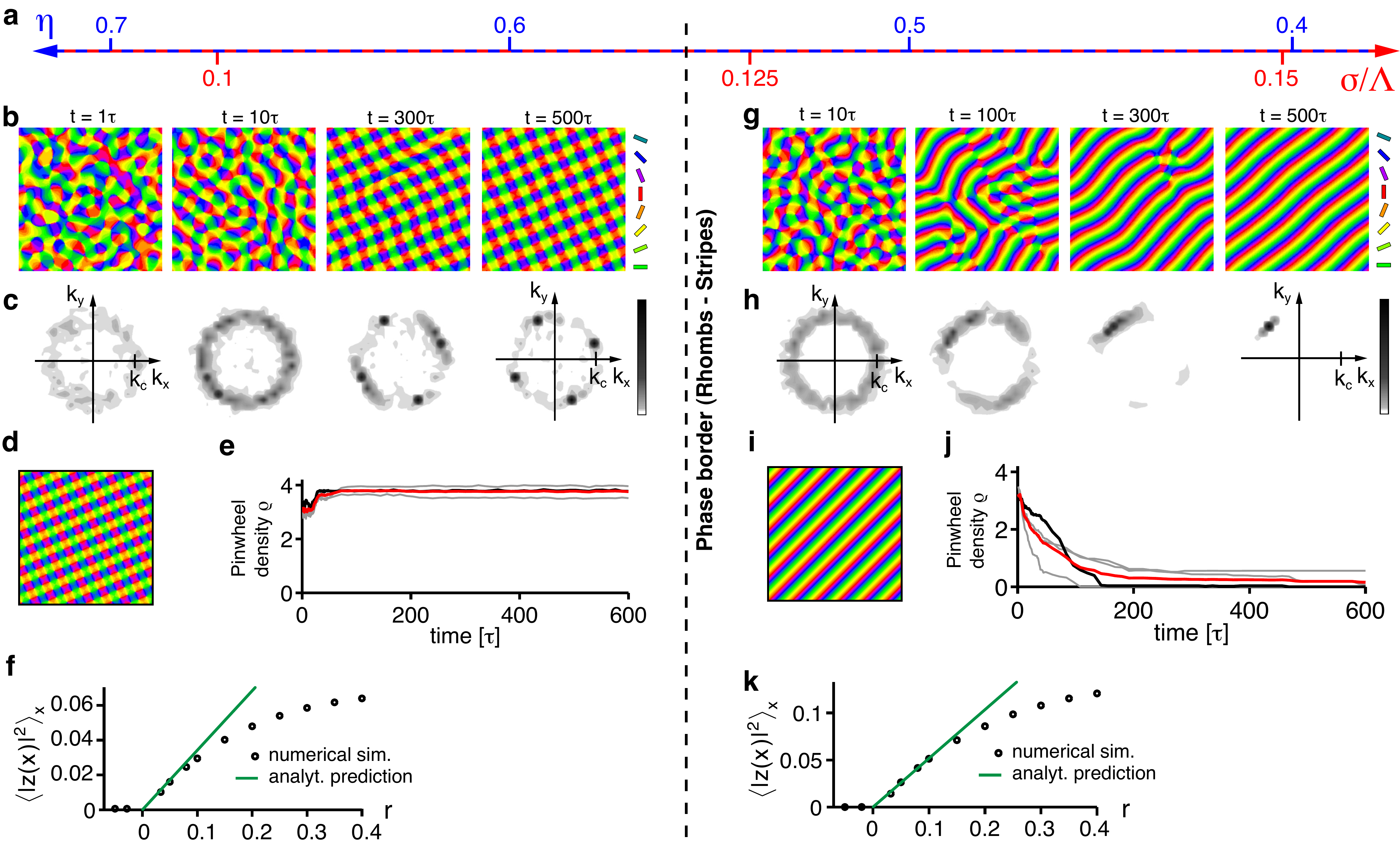}
\caption{\textbf{Optimal solutions of the EN model with a  circular orientation stimulus ensemble
\cite{Swindale:1992p4338,Swindale1998p827,Swindale:2004p6287,CarreiraPerpinan:2004p6297,CarreiraPerpinan:2005p6295}
and fixed representation of visual space.}
(\textbf{a}) At criticality, the phase space of this model is parameterized by either the continuity
parameter $\eta$ (blue labels) or the interaction range
$\sigma/\Lambda$ (red labels, see text).
(\textbf{b}-\textbf{c}) OPMs (b) and their power spectra (c)
in a simulation of Eq. \eqref{eq:continuous_z_dynamics} with $\mathbf{r}(\mathbf{x})=\mathbf{0}$ and $r=0.1$,
$\sigma/\Lambda=0.1\,(\eta=0.67)$ and circular stimulus ensemble.
(\textbf{d}) Analytically predicted optimum for $\sigma/\Lambda\lesssim0.122$
(quadratic pinwheel crystal).
(\textbf{e}) Pinwheel density time courses for four different simulations
(parameters as in b; gray traces, individual realizations; black trace,
simulation in b; red trace, mean value).
(\textbf{f})
Mean squared amplitude of the stationary pattern, obtained in simulations (parameters as in
b) for different values of the control parameter $r$ (black circles)
and analytically predicted value (solid green line).
(\textbf{g}-\textbf{h})
OPMs (g) and their power spectra (h)
in a simulation of Eq. \eqref{eq:continuous_z_dynamics} with $\mathbf{r}(\mathbf{x})=\mathbf{0}$ and $\sigma/\Lambda=0.15\,(\eta=0.41)$
(other parameters as in b).
(\textbf{i}) Analytically predicted optimum for $\sigma/\Lambda\gtrsim0.122$ (orientation stripes).
(\textbf{j}) Pinwheel density time courses for four different simulations
(parameters as in g; gray traces, individual realizations; black trace,
simulation in g; red trace, mean value).
(\textbf{k}) Mean squared amplitude of the stationary pattern, obtained in simulations
(parameters as in g) for different values of the control parameter
$r$ (black circles) and analytically predicted value (solid green line).
\label{Keil_Wolf_figure_6}}
\end{figure*}

We first tested for the stability of pinwheel-free OS solutions and the sPWCs, by analytical evaluation
of the criteria for intrinsic and extrinsic stability (Methods). 
We found \textit{both}, OS and sPWCs, to be intrinsically and extrinsically stable
for all $\sigma/\Lambda$. 
Next, we tested for the stability of n-ECP solutions with $2\leq n\leq20$. 
We found all n-ECP configurations with $2\leq n\leq20$ to be intrinsically \textit{unstable}
for all $\sigma/\Lambda$. Hence, none of these planforms represent
optimal solutions of the EN model with a circular stimulus ensemble, while both OS and sPWC are always local minima of the energy functional.

By evaluating the energy assigned to the sPWC (Eq. \eqref{eq:energy-of-rPWC})
and OS (Eq. \eqref{eq:energy-of-orientation-stripes}), we next identified
two different regimes: (i) For short interaction range $\sigma/\Lambda\lesssim0.122$ the
sPWC possesses minimal energy and is therefore the predicted global minimum.
(ii) For $\sigma/\Lambda\gtrsim0.122$ the OS is optimal. Figure \ref{Keil_Wolf_figure_6}a shows the
resulting simple phase diagram. The sPWC and OS are separated by a phase border at $\sigma/\Lambda\approx0.122$.
We numerically confirmed these analytical predictions by extensive
simulations of Eq. \eqref{eq:continuous_z_dynamics} with $\mathbf{r}(\mathbf{x})=\mathbf{0}$ and the circular stimulus ensemble (see Methods for details). Fig. \ref{Keil_Wolf_figure_6}b-c
show snapshots of a representative simulation with short interaction range ($r=0.1$, $\sigma/\Lambda=0.1\,(\eta=0.67)$).
After the phase of initial pattern emergence (symmetry breaking), the OPM layout rapidly
approaches a square array of pinwheels, the analytically predicted optimum (Fig. \ref{Keil_Wolf_figure_6}d). Pinwheel density time courses (see Methods) display a rapid convergence to a value close to the predicted density of 4 (Fig. \ref{Keil_Wolf_figure_6}e).
Fig. \ref{Keil_Wolf_figure_6}f shows the stationary mean squared amplitudes of the pattern obtained for different values of the control parameter $r$ (black circles). For small control
parameters, the pattern amplitude is perfectly predicted by Eq. \eqref{eq:amplitude-rPWC} (solid green line). Fig. \ref{Keil_Wolf_figure_6}g-h
show snapshots of a typical simulation with longer interaction range ($r=0.1$, $\sigma/\Lambda=0.15\,(\eta=0.41)$).
After the emergence of an OPM with numerous pinwheels, pinwheels undergo pairwise annihilation as previously described for various models of OPM development and optimization \cite{Wolf:1998p1199,Koulakov:2001p842, Wolf:2005p190}.
The OP pattern converges to a pinwheel-free stripe pattern, which is the analytically computed optimal solution in this parameter regime (Fig. \ref{Keil_Wolf_figure_6}i). Pinwheel densities decay towards zero over the time course of the simulations (Fig. \ref{Keil_Wolf_figure_6}j). Also in this parameter regime,
the mean squared amplitude of the pattern is well-predicted Eq. \eqref{eq:amplitude-orientation-stripes} for small $r$. (Fig.
\ref{Keil_Wolf_figure_6}k). 

In summary, the phase diagram of the EN model with a circular stimulus
ensemble close to threshold is divided into two regions: (i) for a small
interaction range (large continuity parameter) a square array of
pinwheels is the optimal dimension-reducing mapping and (ii)
for a larger interaction range (small continuity parameter)
orientation stripes are the optimal dimension-reducing mapping. Both
states are stable throughout the entire parameter range. All other planforms, in
particular quasiperiodic n-ECPs are unstable.

At first sight, this structure of the EN phase diagram may appear rather counterintuitive. A solution with many pinwheel-defects is energetically favored over a solution with no defects in a regime with large continuity parameter where discontinuity should be strongly penalized in the EN energy term. However, a large continuity parameter at pattern formation threshold inevitably leads to a short interaction range $\sigma$ compared to the characteristic spacing $\Lambda$ (see Eq. \eqref{eq:sig_per_lambda_definition}). In such a regime, the gain in coverage by representing many orientation stimuli in a small area spanning the typical interaction range, e.g. with a pinwheel, is very high. Our results show that the gain in coverage by a spatially regular positioning of pinwheels outweighs the accompanied loss in continuity above a certain value of the continuity parameter. 
\subsubsection*{EN dynamics far from pattern formation threshold}
\begin{figure}
\centering
\includegraphics[width=13cm]{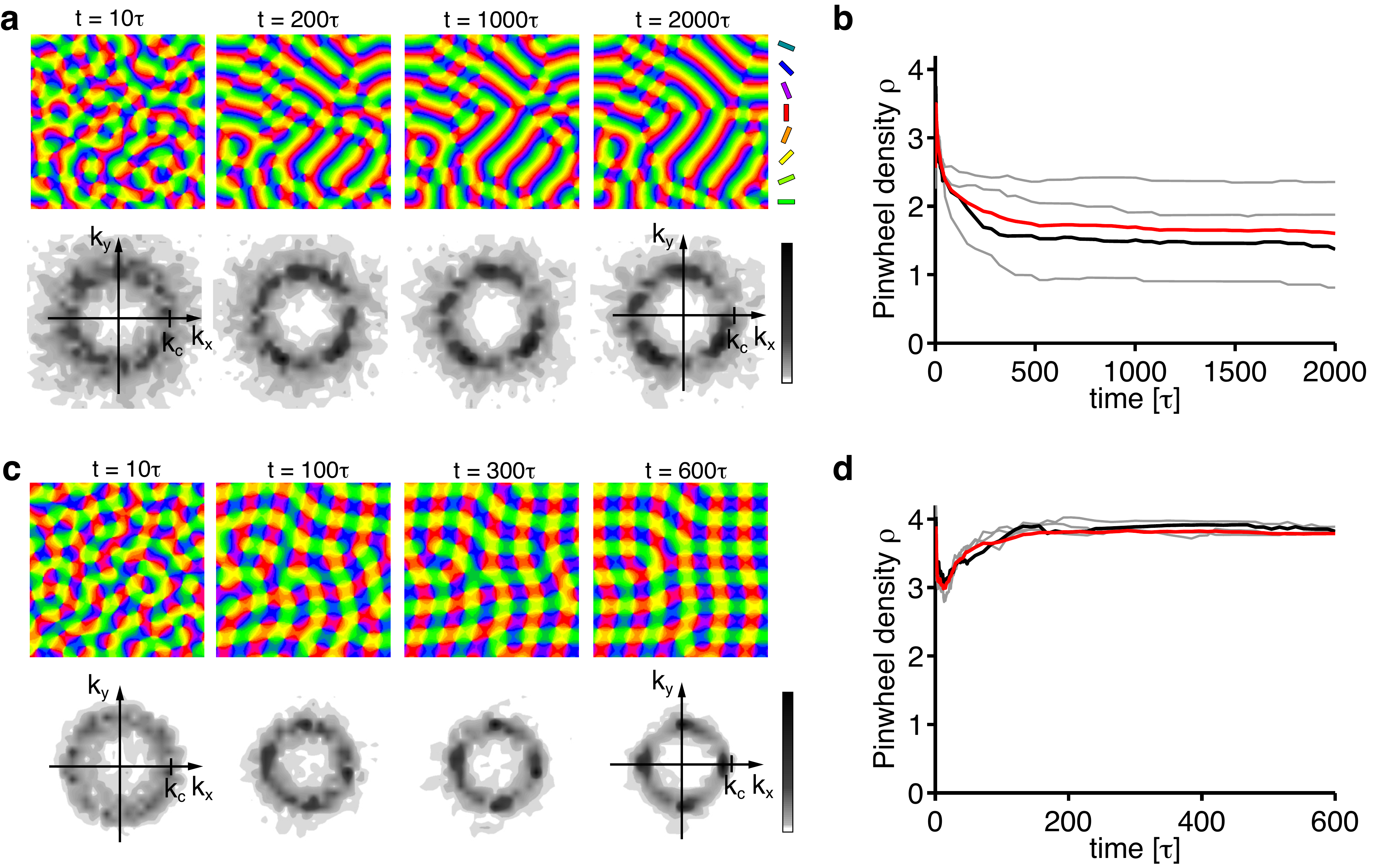}
\caption{\textbf{Numerical analysis of the EN dynamics with circular orientation stimulus
ensemble and fixed representation of visual space far from pattern formation threshold.}
(\textbf{a}) OPMs and their power spectra in a simulation of Eq. \eqref{eq:continuous_z_dynamics} with $\mathbf{r}(\mathbf{x})=\mathbf{0}$, $r=0.8$, $\sigma/\Lambda=0.3\,(\eta=0.028)$ and circular orientation
stimulus ensemble. Pinwheel density time courses for four different simulations
(parameters as in a; gray traces, individual realizations; black trace,
simulation in a; red trace, mean value) (\textbf{c},\textbf{ d}) OPMs and their power spectra in a simulation of
Eq. \eqref{eq:continuous_z_dynamics} with $\mathbf{r}(\mathbf{x})=\mathbf{0}$,  $r=0.8$, $\sigma/\Lambda=0.12\,(\eta=0.57)$
and circular orientation stimulus ensemble. (\textbf{d}) Pinwheel density
time courses for four different simulations (parameters as in c; gray traces,
individual realizations; black trace, simulation in c; red trace,
mean value)
\label{Keil_Wolf_figure_7}}
\end{figure}
Close to pattern formation threshold, we found only two stable solutions, namely OS and sPWCs. Neither of the two exhibits the characteristic aperiodic and pinwheel-rich organization of experimentally observed orientation preference maps. Furthermore, the pinwheel densities of both solutions ($\rho=0$
for OS and $\rho=4$ for sPWCs) differ considerably from experimentally
observed values \cite{Kaschube_2010} around $3.14$. One way towards more realistic
stable stationary states might be to increase the distance from pattern
formation threshold. In fact, further away from threshold, our perturbative calculations
may fail to correctly predict optimal solutions of the model due to the increasing
influence of higher order terms in the Volterra series expansion of
the right hand side in Eq. \eqref{eq:continuous_z_dynamics}.

To asses this possibility, we simulated Eq. \eqref{eq:continuous_z_dynamics} with $\mathbf{r}(\mathbf{x})=\mathbf{0}$ and
a circular stimulus ensemble for very large values of the control parameter $r$.
Fig. \ref{Keil_Wolf_figure_7} displays snapshots
of such a simulation for $r=0.8$ as well as their pinwheel density
time courses for two different values of $\sigma/\Lambda$. Pinwheel
annihilation in the case of large $\sigma/\Lambda$ is less rapid
than close to threshold (Fig. \ref{Keil_Wolf_figure_7}a, b).
The OPM nevertheless converges towards a layout with rather low pinwheel
density with pinwheel-free stripe-like domains of different directions
joined by domains with essentially rhombic crystalline pinwheel
arrangement. The linear zones increase their size over the time course
of the simulations, eventually leading to stripe-patterns for large simulation times. For smaller interaction ranges $\sigma/\Lambda$, the OPM layout
rapidly converges towards a crystal-like rhombic arrangement of pinwheels,
however containing several dislocations (Fig. \ref{Appendix_Keil_Wolf_figure_3}a) \cite{manneville_90}. Dislocations are defects of roll or square patterns, where two rolls or squares merge into one, thus increasing the local wavelength of the pattern \cite{Cross:1993p922, greenside_09}. 
Nevertheless, for all simulations, the pinwheel density rapidly reaches
a value close to 4 (Fig. \ref{Keil_Wolf_figure_7}c) and the square arrangement of pinwheels is readily
recognizable. Both features, the dislocations in the rhombic patterns and domain walls in the stripe patterns, have been frequently observed in pattern-forming systems far
from threshold \cite{manneville_90,greenside_09}.

In summary, the behavior of the EN dynamics with circular stimulus ensemble far from pattern formation threshold agrees very well with our analytical predictions close to threshold. Again, orientation stripes and quadratic pinwheel crystals are identified as the only stationary solutions. Aperiodic and pinwheel-rich patterns which resemble experimentally OPM layouts were not observed.
\subsection*{Taking retinotopic distortions into account}
So far, we have examined the optimal solutions of the EN model for the simplest and
most widely used orientation stimulus ensemble.
Somewhat unexpected from previous reports, the optimal states in this case do
not exhibit the irregular structure of experimentally observed
orientation maps. Our treatment however differs from previous approaches in
that the mapping of visual space so far was assumed to be undistorted and fixed, i.e. $\mathbf{r}(\mathbf{x})=\mathbf{0}$.
We recall that in their seminal publication, Durbin and Mitchison in particular demonstrated interesting correlations between the map of orientation preference and the map of visual space  \cite{Durbin:1990p1196}. These correlations suggest a strong coupling between the two that may completely alter the model's dynamics and optimal solutions.

It is thus essential to clarify whether the behavior of the EN model observed above changes or persists
if we relax the simplifying assumption of undistorted retinotopy and allow for retinotopic distortions.
By analyzing the complete system of equations Eqs. (\ref{eq:continuous_z_dynamics}, \ref{eq:continous_r_dynamics}), we study the EN model exactly as originally introduced by Durbin and
Mitchison \cite{Durbin:1990p1196}. 

We again employ the fact that in the vicinity of a supercritical bifurcation where the non-orientation
selective state becomes unstable, the entire set of nontrivial fixed
points of Eqs. (\ref{eq:continuous_z_dynamics}, \ref{eq:continous_r_dynamics}) is determined by the third-order terms of the Volterra series
representation of the nonlinear operators $F^{z}[z,\mathbf{r}]$ and
$\mathbf{F}^{r}[z,\mathbf{r}]$.
The model symmetries Eqs. (\ref{eq:translation_symmetry}-\ref{eq:Shift-Symmetry-2})
restrict the general form of the leading order terms for any model
for the joint optimization of OPM and RM to
\begin{eqnarray}
\partial_{t}z(\mathbf{x}) & = & L_{z}[z]+Q^{z}[\mathbf{r},z]+N_{3}^{z}[z,z,\bar{z}]\label{eq:z-approximation-with-retinotopic deviations}+\dots\\
\partial_{t}\mathbf{r}(\mathbf{x}) & = & \mathbf{L}_{r}[\mathbf{r}]+\mathbf{Q}^{r}[z,\bar{z}]+\dots\,.
\label{eq:r-approximation-with-retinotopic-deviations}
\end{eqnarray}
Because the uniform retinotopy is linearly stable, retinotopic distortions
are exclusively induced by a coupling of the RM to the OPM via the
quadratic vector-valued operator $\mathbf{Q}^{r}[z,\bar{z}]$. These retinotopic
distortions will in turn alter the dynamics of the OPM via the quadratic complex-valued
operator $Q^{z}[\mathbf{r},z]$. Close to the point of pattern onset
($r\ll1$), the timescale of OPM development, $\tau=1/r$, becomes
arbitrarily large and retinotopic deviations evolve
on a much shorter timescale. This separation of timescales allows
for an adiabatic elimination of the variable $\mathbf{r}(\mathbf{x})$,
assuming it to always be at the equilibrium point of Eq. \eqref{eq:r-approximation-with-retinotopic-deviations}:
\begin{equation}
\mathbf{r}(\mathbf{x})=-\mathbf{L}_{r}^{-1}\left[\mathbf{Q}^{r}[z,\bar{z}]\right]\,.\label{eq:slaved_retinotopy}
\end{equation}
We remark that as $\lambda_{T/L}^{r}(k)<0$ for all finite wave numbers
$k>0$, the operator $\mathbf{L}_{r}[\mathbf{r}]$ is indeed invertible
when excluding global translations in the set of possible perturbations
of the trivial fixed point. Via Eq. \eqref{eq:slaved_retinotopy},
the coupled dynamics of OPM and RM is thus reduced to a third-order
effective dynamics of the OPM:
\begin{eqnarray}
\partial_{t}z(\mathbf{x}) & \approx & L_{z}[z]\underbrace{-\, Q^{z}[\mathbf{L}_{r}^{-1}\left[\mathbf{Q}^{r}[z,\bar{z}]\right],z]}_{N_{3}^{r}[z,z,\bar{z}]}+N_{3}^{z}[z,z,\bar{z}]\nonumber \\
 & = & L_{z}[z]+N_{3}^{r}[z,z,\bar{z}]+N_{3}^{z}[z,z,\bar{z}]\,.
 \label{eq:full_z_eq_insert_r}
 \end{eqnarray}
The nonlinearity $N_{3}^{r}[z,z,\bar{z}]$ accounts for the coupling
between OPM and RM. Its explicit analytical calculation for the EN model
is rather involved and yields a sum
\[
N_{3}^{r}[z,z,\bar{z}]=\sum_{j=1}^{12}a_{r}^{j}N_{r}^{j}[z,z,\bar{z}]\,.
\]
The individual nonlinear operators $N_{r}^{j}$ are nonlinear convolution-type
operators and are presented in the Methods section together with a
detailed description of their derivation. Importantly, it turns out that the coefficients $a_{r}^{j}$
are completely \textit{independent} of the orientation stimulus ensemble.

The adiabatic elimination of the retinotopic distortions results in an equation for the OPM (Eq. \eqref{eq:full_z_eq_insert_r}) which has the same
structure as Eq. \eqref{eq:third-order-approx-z-dynamics}, the only
difference being an additional cubic nonlinearity. Due to this similarity,
its stationary solutions can be determined by the same methods as
presented for the case of a fixed retinotopy. 
Again, via weakly nonlinear analysis we obtain amplitude equations of the form Eq. \eqref{eq:general-amplitude-equations}.
The nonlinear coefficients $g_{ij}$ and $f_{ij}$
are determined from the angle-dependent interaction functions $g(\alpha)$
and $f(\alpha)$. 
For the operator $N_{3}^{r}[z,z,\bar{z}]$, these functions are given by
\begin{eqnarray*}
g_{r}(\alpha) & = & \frac{\left(\left(1-\sigma^{2}-2e^{-k_{c}^{2}\sigma^{2}}\right)e^{2k_{c}^{2}\sigma^{2}(\cos\alpha-1)}+e^{-k_{c}^{2}\sigma^{2}}\right)^{2}}{2\sigma^{4}\left(\text{\ensuremath{\eta_{r}}}+\sigma^{2}e^{-2k_{c}^{2}\sigma^{2}(\cos\alpha-1)}\right)}\\
f_{r}(\alpha) & = & \frac{1}{2}\left(g_{r}(\alpha)+g_{r}(\alpha+\pi)\right)\,,
\end{eqnarray*}
verifying that , $N_{3}^{r}[z,z,\bar{z}]$ is independent of the
orientation stimulus ensemble. 
Besides the interaction range $\sigma/\Lambda$ the continuity parameter $\eta_{r}\in[0,\infty]$
for the RM appears as an additional parameter in the angle-dependent interaction function. 
Hence, the phase diagram of the EN model will acquire one additional dimension when retinotopic distortions are
taken into account.  We note, that in the limit $\eta_{r}\rightarrow\infty$, the functions $g_{r}(\alpha)$ and $f_{r}(\alpha)$ tend to zero and 
as expected one recovers the results presented above for fixed uniform retinotopy. The functions $g_{r}(\alpha)$ and $f_{r}(\alpha)$
are depicted in Fig. \ref{Keil_Wolf_figure_8} for various interaction ranges $\sigma/\Lambda$ and retinotopic continuity parameters $\eta_{r}$.

\begin{figure}
\centering
\includegraphics[width=8.5cm]{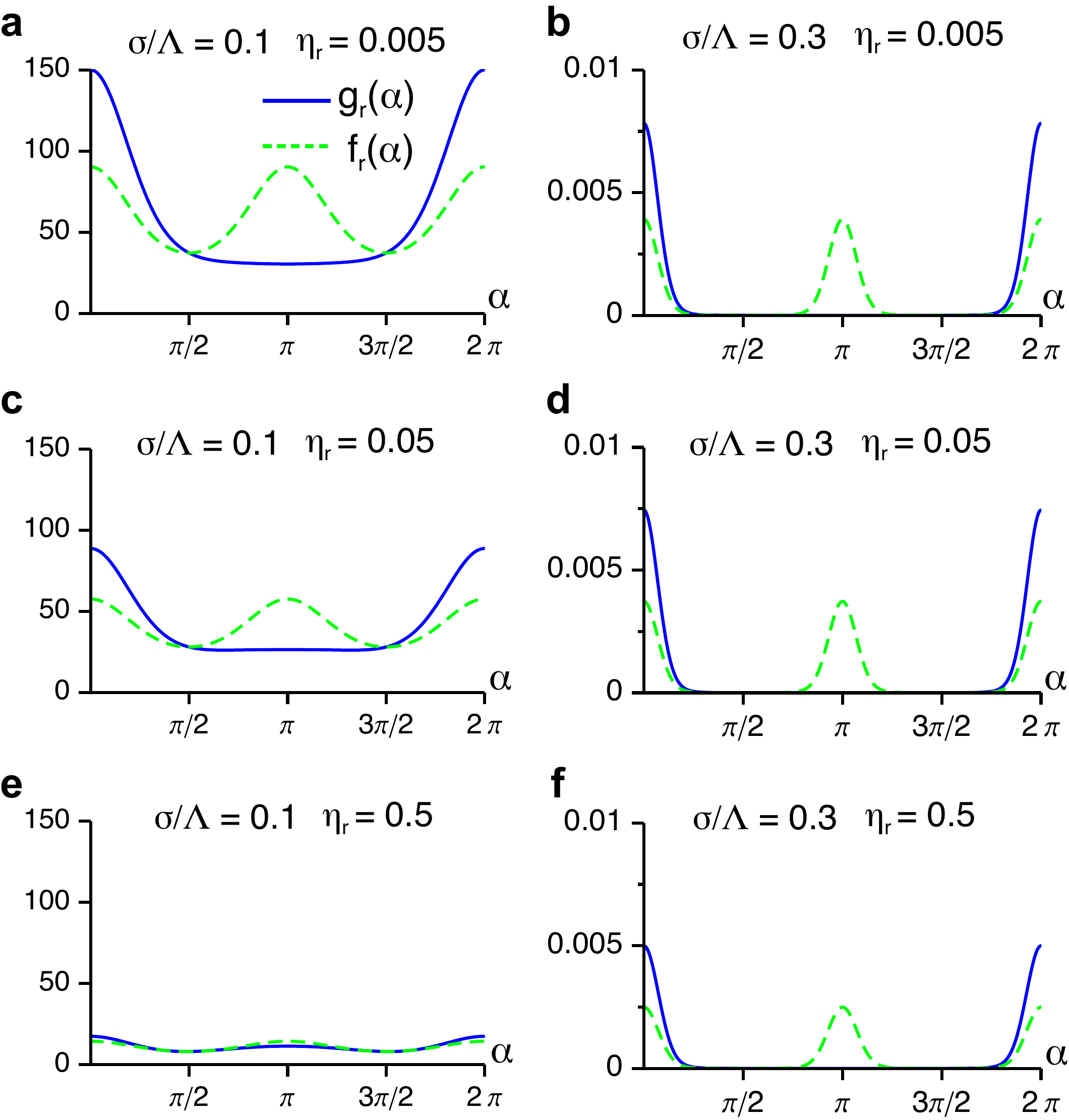}
\caption{\textbf{Angle-dependent interaction functions for the coupling between OPM and RM in the EN model.}
(\textbf{a}, \textbf{b}) $g_{r}(\alpha)$ and $f_{r}(\alpha)$ for $\eta_{r}=0.005$ and
$\sigma/\Lambda=0.3$ (a) and $0.1$ (b). 
(\textbf{c}, \textbf{ d}) $g_{r}(\alpha)$ and $f_{r}(\alpha)$ for $\eta_{r}=0.05$ and and
$\sigma/\Lambda=0.3$ (c) and $0.1$ (d). 
(\textbf{e}, \textbf{ f})
$g_{r}(\alpha)$ and $f_{r}(\alpha)$ for $\eta_{r}=0.5$ and $\sigma/\Lambda=0.3$
(e) and $0.1$ (f). 
\label{Keil_Wolf_figure_8}}
\end{figure}
%
%
%
%
\subsubsection*{Coupled Essentially Complex n-Planforms}
\begin{figure*}
\centering
\includegraphics[width=13cm]{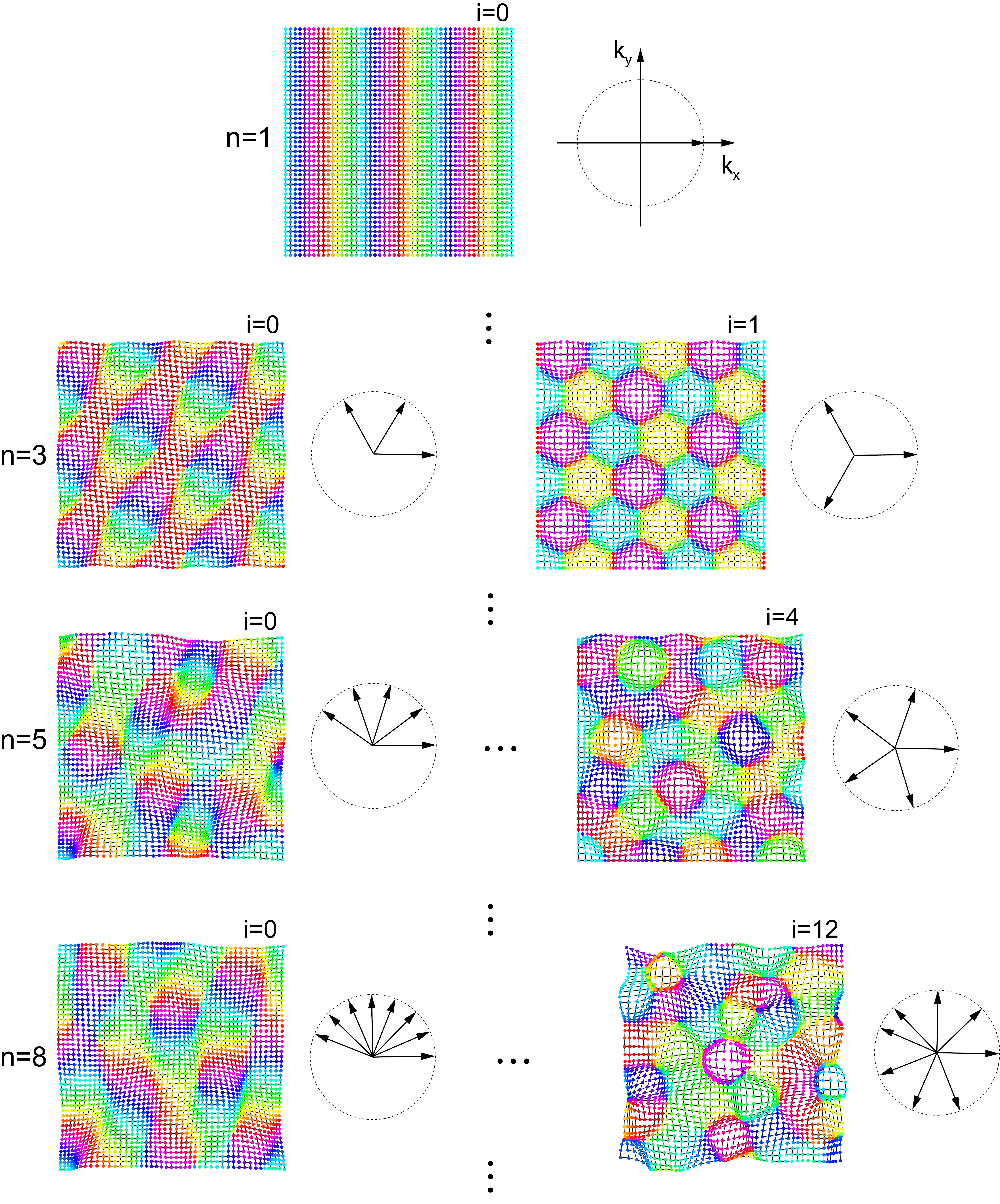}
\caption{\textbf{Coupled n-ECPs as dimension-reducing solutions of the EN model}
Coupled n-ECP are displayed in visual space showing simultaneously the distortion of the retinotopic
map and the orientation preference map ($\sigma/\Lambda=0.3\,(\eta=0.028)$,
$\eta_{r}=\eta$, circular stimulus ensemble). The distorted grid represents a
the cortical square array of cells. Each grid intersection is at the receptive
field center of the corresponding cell. Preferred stimulus orientations
are color-coded as in Fig. \ref{Keil_Wolf_figure_2}a. As in Fig. \ref{Keil_Wolf_figure_4},
n and i enumerate the number of nonzero wave vectors and non-equivalent
configurations of wave vectors with the same n, respectively. The
coupled 1-ECP is a pinwheel-free stripe pattern without retinotopic
distortion. Only the most anisotropic and the most isotropic coupled
n-ECPs are shown for each n. Note that for all ECPs, high gradients
within the orientation mapping coincide with low gradients of the
retinotopic mapping and vice versa. Retinotopic distortions are displayed
on a 5-fold magnified scale for visualization purposes. 
\label{Keil_Wolf_figure_9}}
\end{figure*}
In the previous section, we found that by an adiabatic elimination of the retinotopic distortions in the dynamics Eqs. (\ref{eq:z-approximation-with-retinotopic deviations}, \ref{eq:r-approximation-with-retinotopic-deviations}) the system of partial integro-differential equations can be reduced to a single equation for the OPM. In this case, the stationary solutions of the OPM dynamics are again planforms composed of a discrete set of Fourier modes
\begin{equation}
z(\mathbf{x})=\sum_{j}^{N}A_{j}e^{i\mathbf{k}_{j}\mathbf{x}}\,,
\label{eq:planform-ansatz}
\end{equation}
with $|\mathbf{k}|=k_{c}$. 
However, each of these stationary planform OPM solutions induces a specific pattern of retinotopic distortions via Eq. \eqref{eq:slaved_retinotopy}. The joint mapping $\{z(\mathbf{x}+\mathbf{r}(\mathbf{x}), \mathbf{x})\}$ is then an approximate stationary solution of Eqs. (\ref{eq:z-approximation-with-retinotopic deviations}, \ref{eq:r-approximation-with-retinotopic-deviations}) and will be termed \textit{coupled planform solution} in the following. In contrast to other models for the joint mapping of orientation and visual space (e.g. \cite{Lee:2003p5834,Thomas:2004p6132,Thomas:2006p119}), the coupling between the representation of visual space and orientation in the EN model is not induced by model symmetries but a mere consequence of the joint optimization of OPM and RM that requires them to be matched to one another.

For planforms Eq. \eqref{eq:planform-ansatz}, it is possible to analytically evaluate Eq. \eqref{eq:slaved_retinotopy} and compute the associated
retinotopic distortions $\mathbf{r}(\mathbf{x})$. After a somewhat lengthy calculation (see Methods), one obtains
\begin{eqnarray}
\mathbf{r}(\mathbf{x}) & = & -\sum_{k=1,j<k}^{n}\frac{\mathbf{\Delta}_{jk}}{\lambda_{L}^{r}(|\mathbf{\Delta}_{jk}|)}
\left(\frac{1}{\sigma^{2}}\left(e^{-k_{c}^{2}\sigma^{2}/2} - e^{-\mathbf{\Delta}_{jk}^{2}\sigma^{2}/2}\right)^2
-e^{-\Delta_{jk}^{2}\sigma^{2}}\right)\nonumber \\
 &  & *\left(\Im\left(A_{j}\bar{A}_{k}\right)\cos\left(\mathbf{\Delta}_{jk}\mathbf{x}\right)+\Re\left(A_{j}\bar{A}_{k}\right)\sin\left(\mathbf{\Delta}_{jk}\mathbf{x}\right)\right)\,,\label{eq:retinotopy_for_planforms}
\end{eqnarray}
with $\mathbf{\Delta}_{jk}=\mathbf{k}_{j}-\mathbf{k}_{k}$ and $\lambda_{L}^{r}(k)=-k^{2}(\eta_{r}+e^{-\sigma^{2}k^{2}}\sigma^{2})$.
The retinotopic distortions Eq. (\ref{eq:retinotopy_for_planforms}) represent superpositions of longitudinal
modes (see Fig. \ref{Keil_Wolf_figure_3}b). Hence, coupled
planform stationary solutions of the EN dynamics do not contain any transversal
mode components. According to Eq. \eqref{eq:retinotopy_for_planforms},
the pinwheel-free coupled 1-ECP state has the functional
form $\{\mathbf{r}(\mathbf{x})=\mathbf{0}, z(\mathbf{x})=A_{0}e^{i\mathbf{k}\mathbf{x}}\}$.
This means, that the orientation stripe solution does not induce
any deviations from the perfect retinotopy as shown previously from symmetry. This is not the case for the square pinwheel crystal (sPWC)
\begin{eqnarray*}
z_{\textnormal{sPWC}}(\mathbf{x}) \propto \sin(k_{c}x_{1})+ i\sin(k_{c}x_{2})\,,
\end{eqnarray*}
the second important solution for undistorted retinotopy.
Inserting this ansatz into Eq. \eqref{eq:retinotopy_for_planforms} and neglecting terms of order
$\mathcal{O}\left(\left(e^{-k_{c}^{2}\sigma^{2}}\right)^{2}\right)$
or higher, we obtain
\[
\mathbf{r}_{\textnormal{sPWC}}(\mathbf{x}) \propto \frac{e^{-k_{c}^{2}\sigma^{2}}}{\sigma^{2}\lambda_{2}^{r}(2k_{c})}\left(\begin{array}{c}
k_{c}\sin(2k_{c}x_{1})\\
k_{c}\sin(2k_{c}x_{2})\end{array}\right)\,.
\]
These retinotopic distortions are a superposition of one longitudinal
mode in x-direction and one in y-direction, both with doubled wave number $\sim2k_{c}$.
The doubled wave number implies that the form of retinotopic
distortions is independent of the topological charge of the pinwheels. Importantly,
the gradient of the retinotopic mapping $\mathbf{R}(\mathbf{x})=\mathbf{X}+\mathbf{r}_{\textnormal{sPWC}}(\mathbf{x})$
is reduced at all pinwheel locations. The coupled sPWC is therefore in two ways a high coverage mapping as expected. Firstly, the representations of cardinal and oblique stimuli (real and imaginary part of $z(\mathbf{x})$) are orthogonal to each other. Secondly, the regions of highest gradient in the orientation map correspond to low gradient regions in the retinotopic map.

In Fig. \ref{Keil_Wolf_figure_9}, the family of coupled n-ECPs is displayed, showing simultaneously the distortions of the retinotopic map and the orientation preference map. Retinotopic distortions are generally weaker for anisotropic n-ECPs and stronger for isotropic n-ECPs. However, for all stationary solutions the regions of high gradient in the orientation map coincide with low gradient regions (the folds of the grid) in the retinotopic map. This is precisely what is generally expected from a dimension-reducing mapping \cite{Durbin:1990p1196, Goodhill:2000p2141, Cimponeriu:2000p5167,Swindale:1992p4338}. In the following section, we will investigate which of these solutions become optimal depending on the two parameters $\sigma/\Lambda$ and  $\eta_r$ that parametrize the model.
\subsection*{The impact of retinotopic distortions}
\begin{figure}
\centering
\includegraphics[width=16cm]{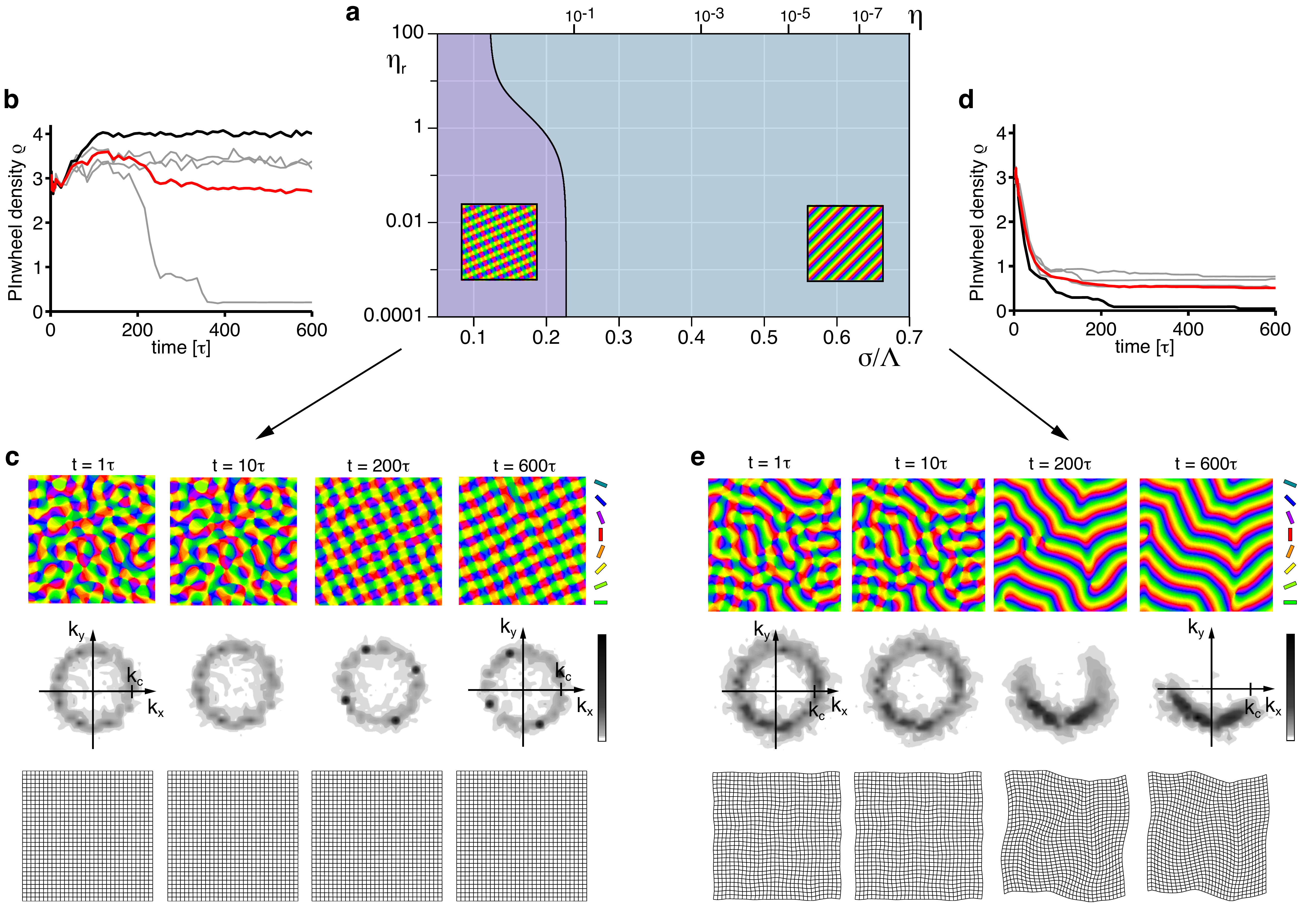}
\caption{\textbf{Phase diagram of the EN model with variable retinotopy for a circular
stimulus ensemble \cite{Swindale:1992p4338,Swindale1998p827,Swindale:2004p6287,CarreiraPerpinan:2004p6297,CarreiraPerpinan:2005p6295}}
(\textbf{a}) Regions of the $\eta_{r}$-$\sigma/\Lambda$-plane in
which n-ECPs or rPWCs have minimal energy.
(\textbf{b}) Pinwheel density time courses for four different simulations of Eqs. (\ref{eq:continuous_z_dynamics}, \ref{eq:continous_r_dynamics}) with $r=0.1$, $\sigma/\Lambda=0.13\:(\eta=0.51$),
$\eta_{r}=\eta$ (grey traces, individual realizations; red trace,
mean value; black trace, realization shown in c).
(\textbf{c}) OPMs (upper row), their power spectra (middle row),
and retinotopic maps (lower row) obtained in a simulation of Eqs. (\ref{eq:continuous_z_dynamics}, \ref{eq:continous_r_dynamics}); parameters as in b.
(\textbf{d}) Pinwheel density time courses for four different simulations of Eqs. (\ref{eq:continuous_z_dynamics}, \ref{eq:continous_r_dynamics}) with $r=0.1$, $\sigma/\Lambda=0.3\:(\eta=0.03$),
$\eta_{r}=\eta$ (grey traces, individual realizations; red trace,
mean value; black trace, realization shown in e).
(\textbf{e}) OPMs (upper row), their power spectra (middle row), and retinotopic maps
(lower row) in a simulation of Eqs. (\ref{eq:continuous_z_dynamics}, \ref{eq:continous_r_dynamics}); parameters as in d.
\label{Keil_Wolf_figure_10}}
\end{figure}
According to our analysis, at criticality, the
nontrivial stable fixed points of the EN dynamics are determined
by the continuity parameter $\eta\in(0,1)$ for the OPM or, equivalently,
the ratio $\sigma/\Lambda=\frac{1}{2\pi}\sqrt{\log(1/\eta)}$ and
the continuity parameter $\eta_{r}$ for the mapping of visual space.
We first tested for the stability of pinwheel-free orientation stripes
(OS) and rhombic pinwheel crystals (rPWCs) solutions of Eq. \eqref{eq:general-amplitude-equations},
with coupling matrices $g_{ij}$ and $f_{ij}$  as obtained
from the nonlinearities in Eq. \eqref{eq:full_z_eq_insert_r}. The angle which minimizes the energy $U_{\textnormal{rPWC}}$ (Eq. \eqref{eq:energy-of-rPWC}) is not affected by the coupling between retinotopic and orientation preference map and is thus again $\alpha=\pi/4$. 
By numerical evaluation of the criteria for intrinsic and extrinsic
stability, we found \textit{both}, OS and sPWC, to be intrinsically and extrinsically
stable for all $\sigma/\Lambda$ and $\eta_{r}$. 

Next, we tested for the stability of coupled n-ECP solutions for $2\leq n\leq20$.
We found all coupled n-ECP configurations with $n\geq2$ to be intrinsically
unstable for all $\sigma/\Lambda$ and $\eta_{r}$. Evaluating
the energy assigned to sPWCs and OS identified two different regimes:
(i) for shorter interaction range $\sigma/\Lambda$ the sPWC is the minimal energy state and
(ii) for larger interaction range $\sigma/\Lambda$ the optimum is an OS as indicated by the phase diagram in Fig.
\ref{Keil_Wolf_figure_10}a. The retinotopic
continuity parameter has little influence on the energy
of the two fixed points. The phase border separating stripes
from rhombs runs almost parallel to the $\eta_{r}$-axis. We
numerically confirmed these analytical predictions by extensive simulations
of Eq. (\ref{eq:continuous_z_dynamics}, \ref{eq:continous_r_dynamics}) (see Methods for details).
Fig. \ref{Keil_Wolf_figure_10}c
shows snapshots of a representative simulation with small interaction range ($r=0.1$, $\sigma/\Lambda=0.1\,(\eta=0.67),\,\eta_{r}=\eta$).
After the initial symmetry breaking phase, the OPM layout rapidly
converges towards a crystalline array of pinwheels, the predicted optimum in this parameter regime
(Fig. \ref{Keil_Wolf_figure_10}c). Retinotopic deviations
are barely visible. Fig. \ref{Keil_Wolf_figure_10}b
displays pinwheel density time courses for four such simulations.
Note that in one simulation, the pinwheel density drops to almost
zero. In this simulation, the OP pattern converges to a stripe-like
layout. This is in line with the finding of bistability of rhombs
and stripes in all parameter regimes. Although the sPWC represents the
global minimum in the simulated parameter regime, OS are also a stable
fixed point and, depending on the initial conditions, may arise as
the final state of a fraction of the simulations. 
In the two simulations with pinwheel densities around 3.4, patterns at later simulation stages consist of different domains of rhombic pinwheel lattices with $\alpha < \pi/2$.

Fig. \ref{Keil_Wolf_figure_10}d-e
show the corresponding analysis with parameters for larger interaction range $r=0.1$, $\sigma/\Lambda=0.15\,(\eta=0.41),\,\eta_{r}=\eta$.
Here after initial pinwheel creation, pinwheels typically annihilate
pairwisely and the OPM converges to an essentially pinwheel-free stripe
pattern, the predicted optimal solution in this parameter regime (Fig. \ref{Keil_Wolf_figure_10}e). Retinotopic deviations
are slightly larger. The behavior of the EN model for the joint optimization of RM
and OPM thus appears very similar compared to the fixed retinotopy case. Perhaps
surprisingly, the coupling of both feature maps has little effect on
the stability properties of the fixed points and the resulting optimal solutions.

As in the previous case, the structure of the phase diagram in Fig. \ref{Keil_Wolf_figure_10}a appears somewhat counterintuitive. A high coverage and pinwheel-rich solution is the optimum in a regime with large OPM continuity parameter where discontinuities in the OPM such as pinwheels should be strongly penalized. A pinwheel-free solution with low coverage and high continuity is the optimum in a regime with small continuity parameter. As explained above, a large OPM continuity parameter at pattern formation threshold implies a small interaction range $\sigma/\Lambda$ (see Eq. \eqref{eq:sig_per_lambda_definition}). In such a regime, the gain in coverage by representing many orientation stimuli in a small area spanning the typical interaction range, e.g. with a pinwheel, is very high. Apparently this gain in coverage by a regular positioning of pinwheels outweighs the accompanied loss in continuity for very large OPM continuity parameters. This counterintuitive interplay between coverage and continuity thus seems to be almost independent of the choice of retinotopic continuity parameters.

The circular orientation stimulus ensemble contains only stimuli with a fixed and finite {}``orientation energy'' or elongation $|s_{z}|$. This raises the question of whether the simple nature of the circular stimulus ensemble might restrain the dynamics of the EN model. The EN dynamics are expected to
depend on the characteristics of the activity patterns evoked by
the stimuli and these will be more diverse and complex
with ensembles containing a greater diversity of stimuli. Therefore,
we repeated the above analysis of the EN model for a richer
stimulus ensemble where orientation stimuli
are uniformly distributed on the disk $\{s_{z,}\,|s_{z}|\leq2\}$, a choice adopted by a subset of previous studies, e.g. \cite{Obermayer:1990p1202,Obermayer:1992p1200,Obermayer:1992p4756,Wolf:1998p1199}. In particular, this ensemble contains unoriented stimuli with $|s_{z}|=0$.
Intuitively, the presence of these unoriented stimuli might be expected to change the role of pinwheels in the optimal OPM layout. Pinwheels' population activity is untuned for orientation. Pinwheel centers may therefore acquire a key role for the representation of unoriented stimuli. Nevertheless, we found the behavior of the EN model when considered with this richer stimulus ensemble to be virtually indistinguishable from the circular stimulus ensemble. Details of the derivations, phase diagrams and numerically obtained solutions are given in Appendix I.
\subsection*{Are there stimulus ensembles for which realistic, aperiodic maps are optimal?}
So far, we have presented a comprehensive analysis of optimal dimension-reducing mappings of the EN model for two widely used orientation stimulus distributions (previous sections and Appendix I). In both cases optima were either regular crystalline pinwheel lattices or pinwheel-free orientation stripes. 
These results might indicate that the EN model for the joint optimization of OPM and RM is per se incapable of reproducing the structure of orientation preference maps as found in the visual cortex. Drawing such a conclusion is suggested in view of the apparent insensitivity of the model's optima to the choice of stimulus ensemble.
The two stimulus ensembles considered so far however do not exhaust the infinite space of stimulus distributions that are admissible in principle.
From the viewpoint of ``biological plausibility" it is certainly not obvious that one should strive to examine stimulus distributions very different from these, as long as the guiding hypothesis is that the functional architecture of the primary visual cortex optimizes the joint representation of the classical elementary stimulus features.
If, however, stimulus ensembles were to exist, for which optimal EN mappings truly resemble the biological architecture, their characteristics may reveal essential ingredients of alternative optimization models for visual cortical architecture.

Adopting this perspective raises the technical question of whether an unbiased search of the infinite space of stimulus ensembles only constrained by the model's symmetries Eqs. (\ref{eq:translation_symmetry}-\ref{eq:Shift-Symmetry-2}) is possible. To answer this question, we examined whether the amplitude equations Eq. \eqref{eq:general-amplitude-equations} can be obtained for an arbitrary orientation stimulus distribution. Fortunately, we found that the coefficients of the amplitude equations are completely determined by the finite set of moments of order less than 5 of the distributions. The approach developed so far can thus be used to comprehensively examine the nature of EN optima resulting for any stimulus distribution with finite fourth-order moment. While such a study does not completely exhaust the infinite space of all eligible distributions, it appears to only exclude ensembles with really exceptional properties. These are probability distributions with diverging fourth moment, i.e. ensembles that exhibit a heavy tail of essentially ``infinite" orientation energy stimuli. 

Since the coupling between OPM and RM did not have a large impact in the case of the two classical stimulus ensembles, we start the search through the space of orientation stimulus ensembles by considering the EN model with fixed retinotopy $\mathbf{r}(\mathbf{x})=\mathbf{0}$.
The coefficients $a_{i}$ for the nonlinear operators $N_{i}^{3}[z,z,\bar{z}]$ in Eq. \eqref{eq:sum-of-cubic-operators} for arbitrary stimulus ensembles are given by
\begin{equation}
\begin{array}{lll}
a_{1}=\frac{\left<|s_{z}|^{4}\right>}{16\sigma^{6}}-\frac{\left<|s_{z}|^{2}\right>}{2\sigma^{4}}+\frac{1}{2\sigma^{2}} & a_{2}=\frac{\left<|s_{z}|^{2}\right>}{8\pi\sigma^{6}}-\frac{\left<|s_{z}|^{4}\right>}{32\pi\sigma^{8}}\hspace{1em} & a_{3}=-\frac{\left<|s_{z}|^{4}\right>}{64\pi\sigma^{8}}+\frac{\left<|s_{z}|^{2}\right>}{16\pi\sigma^{6}}\\
a_{4}=-\frac{\left<|s_{z}|^{4}\right>}{32\pi\sigma^{8}}+\frac{\left<|s_{z}|^{2}\right>}{8\pi\sigma^{6}}-\frac{1}{8\pi\sigma^{4}}\hspace{1em} & a_{5}=-\frac{\left<|s_{z}|^{4}\right>}{64\pi\sigma^{8}} & a_{6}=\frac{\left<|s_{z}|^{2}\right>}{16\pi\sigma^{6}}-\frac{\left<|s_{z}|^{4}\right>}{64\pi\sigma^{8}}\\
a_{7}=\frac{\left<|s_{z}|^{4}\right>}{48\pi^{2}\sigma^{10}}-\frac{\left<|s_{z}|^{2}\right>}{24\pi^{2}\sigma^{8}} & a_{8}=\frac{\left<|s_{z}|^{4}\right>}{96\pi^{2}\sigma^{10}} & a_{9}=-\frac{3\left<|s_{z}|^{4}\right>}{256\pi^{3}\sigma^{12}}\\
a_{10}=\frac{\left<|s_{z}|^{4}\right>}{48\pi^{2}\sigma^{10}}-\frac{\left<|s_{z}|^{2}\right>}{24\pi^{2}\sigma^{8}} & a_{11}=\frac{\left<|s_{z}|^{4}\right>}{96\pi^{2}\sigma^{10}}\,.\end{array}
\label{eq:general-nonlinear_coefficients}
\end{equation}
\begin{figure}
\centering
\includegraphics[width=9.5cm]{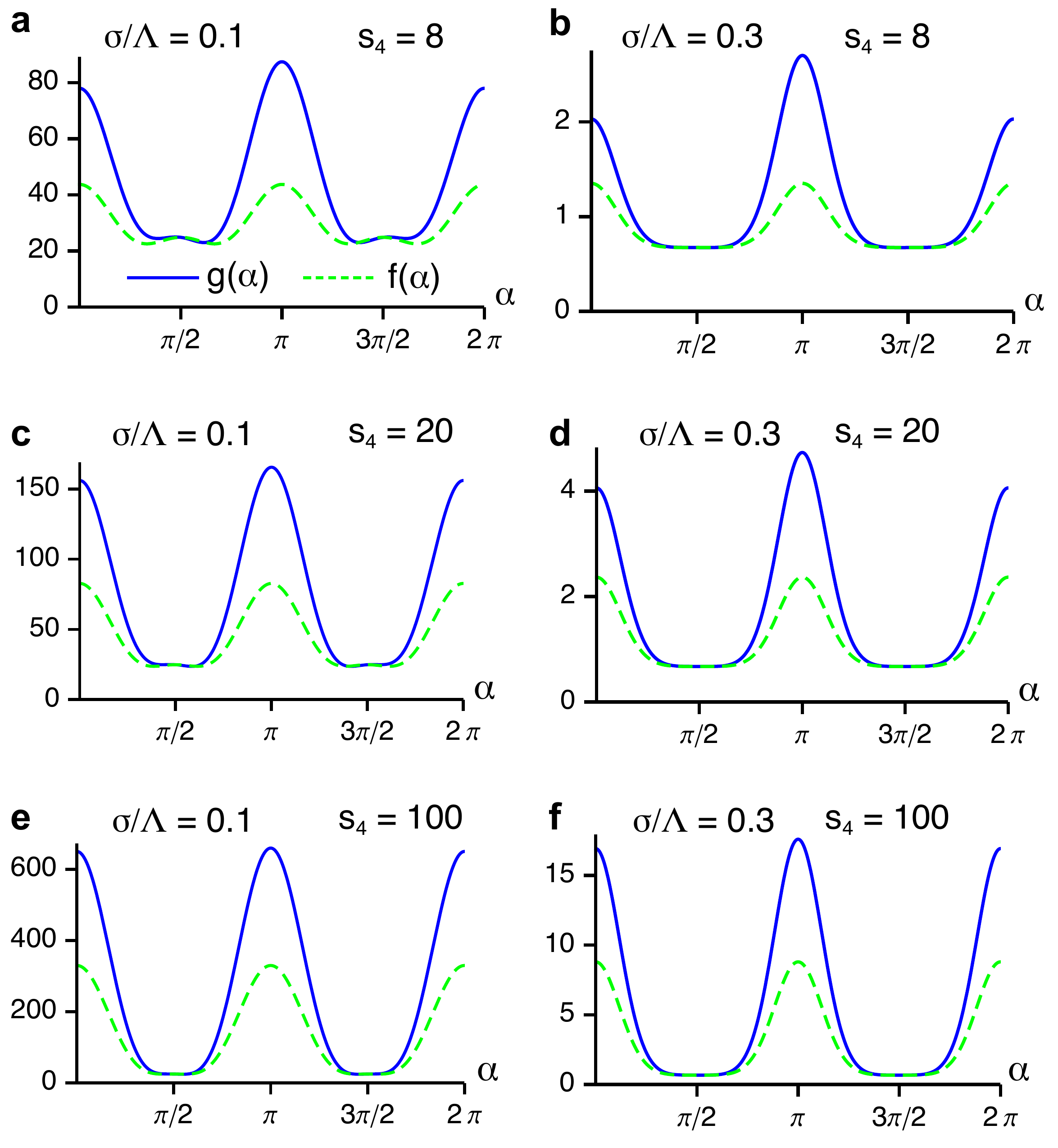}
\caption{\textbf{Angle-dependent interaction functions for the EN model with fixed
retinotopy for different fourth-moment values of the orientation stimulus distribution and effective interaction-widths.}
(\textbf{a}, \textbf{b}) $g(\alpha)$ and $f(\alpha)$ for $s_4 = 8$ and $\sigma/\Lambda=0.1$ (a) and $\sigma/\Lambda=0.3$ (b). 
(\textbf{c}, \textbf{d}) $g(\alpha)$ and $f(\alpha)$ for $s_4 = 20$ and $\sigma/\Lambda=0.1$ (c) and $\sigma/\Lambda=0.3$ (d). 
(\textbf{e}, \textbf{f}) $g(\alpha)$ and $f(\alpha)$ for $s_4 = 100$ and $\sigma/\Lambda=0.1$ (e) and $\sigma/\Lambda=0.3$ (f). 
\label{Keil_Wolf_figure_11}}
\end{figure}
The corresponding angle-dependent interaction functions are given
by (see Methods)
\begin{eqnarray}
g(\alpha) & = & \frac{\left<|s_{z}|^{2}\right>}{2\sigma^{4}}\left(1-2e^{-k_{c}^{2}\sigma^{2}}-e^{2k_{c}^{2}\sigma^{2}(\cos\alpha-1)}\left(1-2e^{-k_{c}^{2}\sigma^{2}\cos\alpha}\right)\right)\nonumber \\
 &  & +\frac{1}{2\sigma^{2}}\left(e^{2k_{c}^{2}\sigma^{2}(\cos\alpha-1)}-1\right)+\frac{2\left<|s_{z}|^{4}\right>}{\sigma^{6}}e^{-2k_{c}^{2}\sigma^{2}}\sinh^{4}\left(1/2k_{c}^{2}\sigma^{2}\cos\alpha\right)\nonumber \\
f(\alpha) & = & \frac{\left<|s_{z}|^{2}\right>}{2\sigma^{4}}\left(1-e^{-2k_{c}^{2}\sigma^{2}}\left(\cosh(2k_{c}^{2}\sigma^{2}\cos\alpha)+2\cosh(k_{c}^{2}\sigma^{2}\cos\alpha)\right)+2e^{-k_{c}^{2}\sigma^{2}}\right)\nonumber \\
 &  & +\frac{1}{2\sigma^{2}}\left(e^{-2k_{c}^{2}\sigma^{2}}\cosh(2k_{c}^{2}\sigma^{2}\cos\alpha)-1\right)+\frac{\left<|s_{z}|^{4}\right>}{\sigma^{6}}e^{-2k_{c}^{2}\sigma^{2}}\sinh^{4}\left(1/2k_{c}^{2}\sigma^{2}\cos\alpha\right)\,.
 \label{eq:general-angle-dependent-interaction-function}
 \end{eqnarray}
Again, without loss of generality, we set $\left<|s_{z}|^{2}\right>=2$.
At criticality, both functions are parametrized by the continuity
parameter $\eta\in(0,1)$ for the OPM or, equivalently, the interaction range
$\sigma/\Lambda=\frac{1}{2\pi}\sqrt{\log(1/\eta)}$ and the fourth
moment $\left<|s_{z}|^{4}\right>$ of the orientation
stimulus ensemble. The fourth moment, is a measure of the peakedness of a stimulus distribution.
High values generally indicate a strongly peaked distribution with a large fraction of non-oriented stimuli ($|s_{z}|^{4}\approx0$),
together with a large fraction of high orientation energy stimuli ($|s_{z}|^{4}$
large).

The dependence of $g(\alpha)$ on the fourth moment of the orientation stimulus distribution
and $f(\alpha)$ suggests that different stimulus distributions may
indeed lead to different optimal dimension-reducing mappings. The
circular stimulus ensemble possesses the minimal possible fourth moment, with $\left<|s_{z}|^{4}\right> = \left(\left<|s_{z}|^{2}\right>\right)^2=4$. The fourth moment of the uniform stimulus ensemble is $\left<|s_{z}|^{4}\right> = 16/3$. 
The angle-dependent interaction functions for both
ensembles (Eq. (\ref{eq:angle-dependent interaction-function-circular-ensemble}), Fig. \ref{Appendix_Keil_Wolf_figure_1} in Appendix I) are recovered, when inserting these values into Eq. \eqref{eq:general-angle-dependent-interaction-function}.

To simplify notation in the following, we define
\[
s_{4}=\left<|s_{z}|^{4}\right>-\left<|s_{z}|^{2}\right>^2 =\left<|s_{z}|^{4}\right> - 4
\]
as the parameter characterizing an orientation stimulus distribution.
This parameter ranges from zero for the circular stimulus ensemble
to infinity for ensembles with diverging fourth moments. Fig. \ref{Keil_Wolf_figure_11}
displays the angle-dependent interaction functions for different values
of $\sigma/\Lambda$ and $s_{4}$. In all parameter regimes, $g(\alpha)$
and $f(\alpha)$ are larger than zero.  The amplitude dynamics are
therefore guaranteed to converge to a stable stationary fixed point and the bifurcation from the nonselective fixed point in the EN model is predicted to be supercritical in general.

By evaluating the energy assigned to the rPWC
and n-ECPs, we investigated the structure of the two-dimensional phase space of the EN model with
an arbitrary orientation stimulus distribution. Firstly, it is not difficult to show that the angle $\alpha$ which minimizes the energy $U_{\textnormal{rPWC}}$ (Eq. \eqref{eq:energy-of-rPWC}) of a rhombic pinwheel crystal is $\alpha = \pi/4$ for all $\sigma/\Lambda$ and $s_4$. Hence, a square lattice of pinwheels is in all parameter regimes energetically favored over any other rhombic lattice configuration of pinwheels.
Fig. \ref{Keil_Wolf_figure_12} displays the phase diagram of the EN model with an arbitrary orientation stimulus distribution.
For orientation stimulus distributions with small fourth moments, optimal
mappings consist of either parallel pinwheel-free stripes or quadratic
pinwheel crystals. These distributions include the circular and the
uniform stimulus ensembles with $s_{4}=0$ and $s_{4}=4/3.$ Above
a certain value of the fourth moment around $s_{4}=6$, n-ECPs with
$n>2$ become optimal mappings. For a short interaction range
$\sigma/\Lambda$, hexagonal pinwheel crystals dominate the phase
diagram in a large region of parameter space. With increasing interaction range, we observe a sequence of phase transitions by which higher n-ECPs become optimal.
For $n>3$, these optima are spatially aperiodic. In all parameter regimes, we found that the n-ECP with the most anisotropic mode configuration (Fig. \ref{Keil_Wolf_figure_4}c, left column) is the energetically favored state for $n>3$. Pinwheel densities of these planforms are indicated in Fig. \ref{Keil_Wolf_figure_12} and are typically smaller than 2.0. We note that this is well below experimentally observed pinwheel density values \cite{Kaschube_2010}. 

Optimal mappings of orientation preference for finite fourth moment in the EN model are thus either orientation stripes, periodic arrays of
pinwheels (hexagonal, square), or aperiodic pinwheel arrangements with low pinwheel density.
\begin{figure}
\centering
\includegraphics[width=10cm]{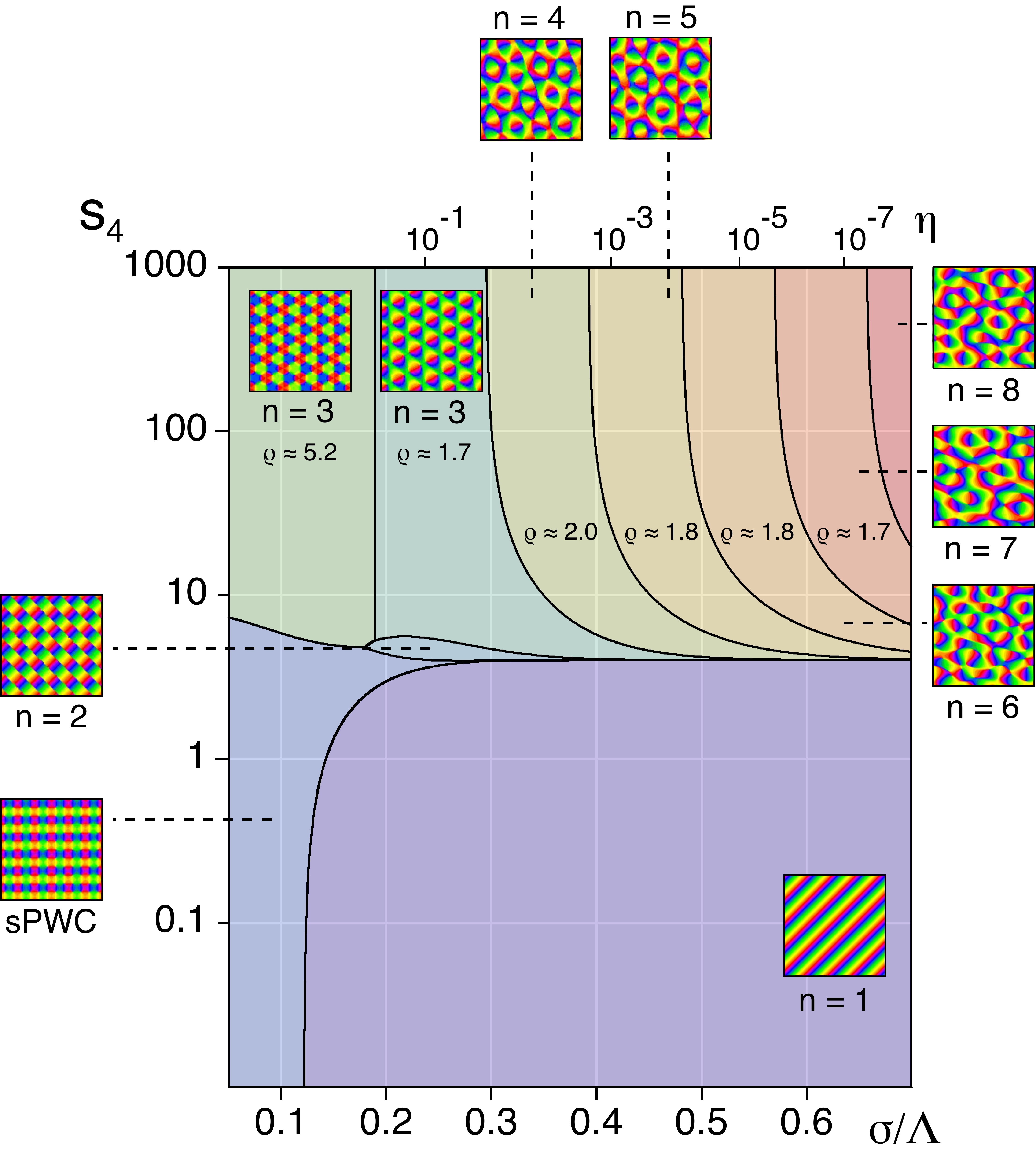}
\caption{\textbf{Stripe-like, crystalline, and quasi-crystalline cortical representations
as optimal solutions to the mapping of orientation preference with fixed uniform retinotopy in the EN model.}
The graph shows the regions of the $s_{4}$-$\sigma/\Lambda$-plane in which n-ECPs or sPWCs have
minimal energy. For $n\geq3$, pinwheel densities of the energetically
favored n-ECP configuration are indicated.
\label{Keil_Wolf_figure_12}}
\end{figure}

We numerically tested these analytical predictions by
simulations of Eq. \eqref{eq:continuous_z_dynamics} ($\mathbf{r}(\mathbf{x})=\mathbf{0}$) with two additional
stimulus ensembles with $s_{4}=6$ and $s_{4}=8$ (see Methods). 
Fig. \ref{Keil_Wolf_figure_13}a shows snapshots of a simulation with ($r=0.1$, $\sigma/\Lambda=0.2\,(\eta=0.2)$) and $s_{4}=6$.
After the initial phase of pattern emergence, the OPM layout
converges towards an arrangement of fractured stripes which resembles the 2-ECP state (Fig. \ref{Keil_Wolf_figure_13}a, most right), the optimum predicted in this regime. In the Fourier spectra, two distinct peaks of the active modes are clearly visible in the final stages of the simulation (Fig. \ref{Keil_Wolf_figure_13}a, lower row). The 2-ECP state is exotic in the sense that it is the only n-ECP containing line defects and thus the pinwheel density is not a well-defined quantity. This explains the pronounced numerical variability in the measured pinwheel densities in simulations during the convergence towards a 2-ECP state (Fig. \ref{Keil_Wolf_figure_13}b).

Fig. \ref{Keil_Wolf_figure_13}c shows snapshots of a simulation with ($r=0.1$, $\sigma/\Lambda=0.2\,(\eta=0.2)$) and $s_{4}=8$ (Gaussian stimulus ensemble). After the initial phase of pattern emergence, the OPM layout
converges towards a regular hexagonal arrangement of pinwheels which resembles the anisotropic 3-ECP (Fig. \ref{Keil_Wolf_figure_13}c, far right), the optimum predicted in this regime. In the Fourier spectra, three distinct peaks forming an angle of 60 degrees are clearly visible in the later stages of the simulation (Fig. \ref{Keil_Wolf_figure_13}c, lower row). Pinwheel densities in the simulations consistently approach the theoretically predicted value of $2 \cos(\pi/6)\simeq 1.73$ (Fig. \ref{Keil_Wolf_figure_13}d).

\begin{figure}	
\centering
\includegraphics[width=13cm]{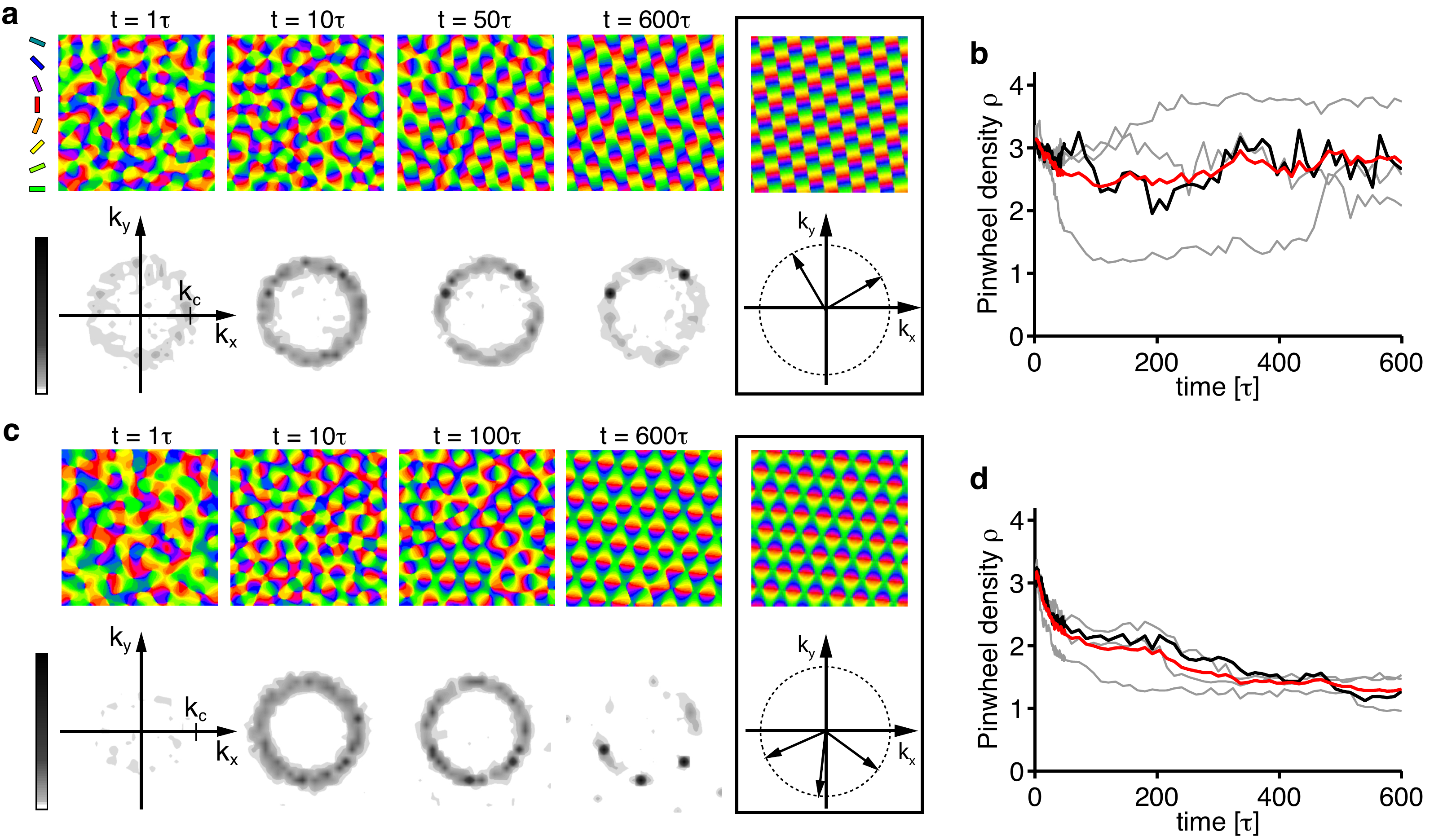}
\caption{\textbf{Approaching crystalline n-ECP optima in the EN model with fixed retinotopy.}
(\textbf{a}) OPMs (upper row) and their power spectra (lower row)
in a simulation of Eq. \eqref{eq:continuous_z_dynamics} with $\mathbf{r}(\mathbf{x})=\mathbf{0}$, $r=0.1$, $\sigma/\Lambda=0.2$ and
$s_{4}=6$. The predicted optimum is the 2-ECP (black frame).
(\textbf{b}) Pinwheel density time courses for four different
simulations (parameters as in a; gray traces, individual realizations;
black trace, simulation in a; red trace, mean value) 
(\textbf{c}) OPMs (upper row) and their power spectra (lower row)
in a simulation of Eq. \eqref{eq:continuous_z_dynamics} with $\mathbf{r}(\mathbf{x})=\mathbf{0}$, $r=0.1$, $\sigma/\Lambda=0.3$ and $s_{4}=8$. 
The predicted optimum is the anisotropic 3-ECP (black frame).
(\textbf{d}) Pinwheel density time courses for four different
simulations (parameters as in c; gray traces, individual realizations;
black trace, simulation in c; red trace, mean value).\label{Keil_Wolf_figure_13}}
\end{figure}
\subsubsection*{Permutation symmetric limit}
In the previous section, we uncovered a parameter regime for the EN model in which optimal solutions are spatially aperiodic. This can be viewed as a first step towards realistic optimal solutions. In the identified regime, however, among the family of n-ECPs only those with pinwheel densities well below experimentally observed values \cite{Kaschube_2010} are energetically favored (see Fig. \ref{Keil_Wolf_figure_12}).
In this respect, the repertoire of aperiodic optima of the EN model differs from previously considered abstract variational models for OPM development \cite{Wolf:2005p190, schnabel_2007 , Kaschube_2010, schnabel_2011}. In these models, an energetic degeneracy of aperiodic states with low and high pinwheel densities has been found which leads to a pinwheel statistics of the repertoire of optimal solutions that quantitatively reproduces experimental observations \cite{Kaschube2008, Kaschube_2010}.
What is the reason for this difference between the two models? In \cite{Wolf:2005p190}, the energetic degeneracy of aperiodic states with low and high pinwheel densities was derived from a so-called permutation symmetry
\begin{equation}
N_{3}^{z}[u,v,w]=N_{3}^{z}[w,u,v]\,,
\label{eq:permutation-symmetry}
\end{equation}
of the cubic nonlinearities of the model. It can be easily seen, that the cubic nonlinearities obtained in the third order expansion of the EN model do not exhibit this permutation symmetry (see Methods). As shown by Reichl \cite{reichl_2010}, the absence of permutation symmetry can lead to a selection of a subrange of pinwheel densities in the repertoire of optima of OPM models. Depending on the degree of permutation symmetry breaking, the family of optima of such models, albeit encompassing aperiodic OPM layouts, may consist of layouts with either unrealistically low or high pinwheel densities. Furthermore, for very strong permutation symmetry breaking, stationary solutions from solution classes other than the n-ECPs and rPWCs with low or high pinwheel densities may become optima of models for OPM development. In order to determine a regime in which the EN model optima quantitatively resemble experimentally observed OPM layouts, it is therefore important to quantify the degree of permutation symmetry breaking in the EN model and to examine whether permutation symmetric limits exist. 
\begin{figure}
\centering
\includegraphics[width=17cm]{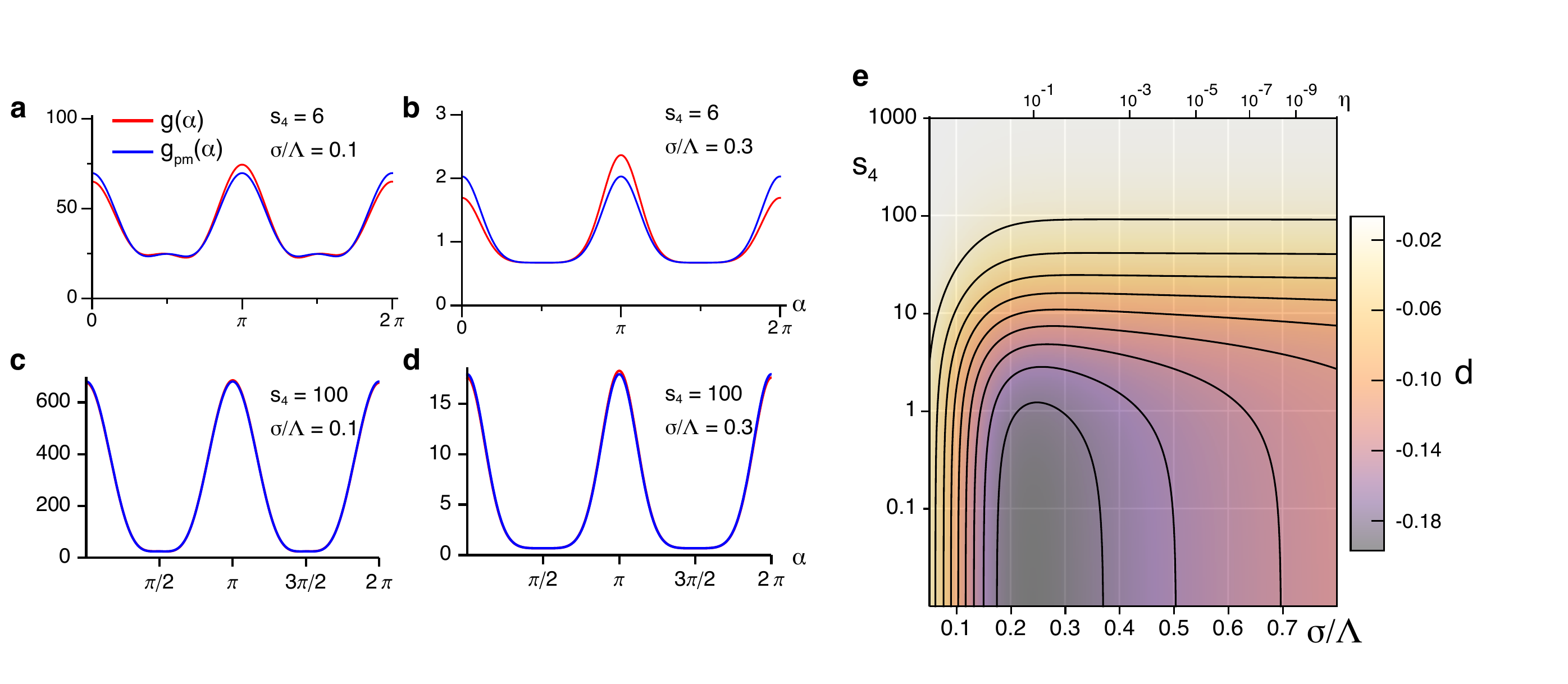}
\caption{\textbf{Quantifying permutation symmetry breaking in the EN model.}
(\textbf{a}-\textbf{d}) $g(\alpha)$ (red traces) and the {}``permutation symmetrized"  function $g_{pm}(\alpha) = 1/2(g(\alpha) + g(\alpha+\pi))$ (blue traces, see Eq. \eqref{eq:permutation-symmetrized-g-function}) for $\sigma / \Lambda = 0.1$ and $0.3$ and $s_4 = 6$ and $100$.  (\textbf{e}) Permutation symmetry parameter d (Eq. \eqref{eq:degree-of-perm-sym-breaking})
in the EN model with fixed retinotopy. Permutation symmetry breaking
is largest for $\sigma/\Lambda \approx 0.25$ and small $s_{4}$. In
the limit $s_{4} \rightarrow \infty $, permutation symmetry is restored.
\label{Keil_Wolf_figure_14}}
\end{figure}
As shown in the Methods section, any cubic nonlinearity $N_3^z[z,z,\bar{z}]$ that obeys the permutation symmetry Eq. \eqref{eq:permutation-symmetry} has a corresponding angle-dependent interaction function $g(\alpha)$ which is $\pi$-periodic. Therefore, we examine the degree of permutation symmetry breaking in the EN model by comparing the angle-dependent interaction function $g(\alpha)$ of its third order expansion (see Eq. \eqref{eq:general-angle-dependent-interaction-function} and Fig. \ref{Keil_Wolf_figure_11}) to the $\pi$-periodic function
$g_{pm}(\alpha)=1/2\left(g(\alpha)+g(\alpha+\pi)\right)$.
This {}``permutation-symmetrized'' part of the angle-dependent interaction function
of the EN model for general orientation stimulus ensembles reads
\begin{eqnarray}
g_{pm}(\alpha) & = & \frac{2\left<|s_{z}|^{4}\right>}{\sigma^{6}}e^{-2k_{c}^{2}\sigma^{2}}\sinh^{4}\left(1/2k_{c}^{2}\sigma^{2}\cos\alpha\right)\nonumber\\
 &  & -\frac{\left<|s_{z}|^{2}\right>}{2\sigma^{4}}e^{-2k_{c}^{2}\sigma^{2}}\left(\left(\cosh\left(2k_{c}^{2}\sigma^{2}\cos\alpha\right)-2\cosh\left(k_{c}^{2}\sigma^{2}\cos\alpha\right)\right)-2e^{k_{c}^{2}\sigma^{2}}-e^{2k_{c}^{2}\sigma^{2}}\right)\nonumber\\
 &  & +\frac{1}{2\sigma^{2}}\left(1+e^{-2k_{c}^{2}\sigma^{2}}\cosh(2k_{c}^{2}\sigma^{2}\cos\alpha)\right)\,.
 \label{eq:permutation-symmetrized-g-function}
 \end{eqnarray}
A comparison between $g_{pm}(\alpha)$ and $g(\alpha)$ is depicted
in Fig.  \ref{Keil_Wolf_figure_14}a-d. It shows that essentially insensitive to the
interaction range $\sigma/\Lambda$, at large values of the fourth moment original
and permutation symmetrized angle-dependent interaction functions converge. We quantified the degree of permutation symmetry breaking
with the parameter 
\begin{equation}
d=\frac{\|g-g_{pm}\|_{2}}{\|g\|_{2}}\textnormal{sgn}(g(0)-g(\pi))\,.\label{eq:degree-of-perm-sym-breaking}
\end{equation}
This parameter is zero in the case of a permutation symmetric cubic nonlinearity.
In the case of a g-function completely antisymmetric around $\alpha=\pi/2$,
the parameter is either plus or minus one, depending on whether the
maximum of $g_{pm}$ is at zero or $\pi$. 
If $d$ is smaller than zero, low pinwheel densities are expected to be energetically
favored and vice versa. The values of $d$ in parameter space
is depicted in Fig. \ref{Keil_Wolf_figure_14}e.
It is smaller than zero in the entire phase space, implying a tendency
for low pinwheel density optimal states, in agreement with the phase diagram
in Fig. \ref{Keil_Wolf_figure_12}. Permutation symmetry breaking is largest
for $\sigma/\Lambda$ around 0.25 and small fourth moment values of the orientation stimulus distribution.
It decays to zero for large fourth moments proportionally to $1/ s_4$ as can be seen by inserting Eqs. \eqref{eq:general-angle-dependent-interaction-function} and \eqref{eq:permutation-symmetrized-g-function} into Eq. \eqref{eq:degree-of-perm-sym-breaking}  . In the infinite fourth moment limit $s_4\rightarrow \infty$, the cubic nonlinearities of the third order expansion of the EN model become permutation symmetric. 

In this case, the EN model is parametrized by only one parameter, the effective intracortical interaction range $\sigma/\Lambda$ and we obtain a rather simple phase diagram (Fig. \ref{Keil_Wolf_figure_15}). Optimal solutions are n-ECPs for increasing $\sigma/\Lambda$ and we observe a sequence of phase transitions towards a higher number of active modes and therefore more complex spatially aperiodic OPM layouts. Importantly, for a subregion in the phase diagram with given number of active modes, all possible n-ECP mode configurations are energetically degenerate. It is precisely this degeneracy that has been previously shown to result in a pinwheel statistics of the repertoire of aperiodic optima which quantitatively agrees with experimental observations \cite{Kaschube_2010}. 
Therefore, our unbiased search in fact identified a regime, namely a very large effective interaction range and infinite fourth moment of the orientation stimulus ensemble, in which the EN model formally predicts which quantitatively reproduce  the experimentally observed  V1 architecture.

Unexpectedly, however, this regime coincides with the limit of applicability of our approach. Permutation symmetry is exactly obtained by approaching stimulus distribution with diverging fourth moment for which the amplitude equations may become meaningless. We would generally expect that the EN for very large but finite fourth moment can closely resemble a permutation symmetric model. However, to consolidate the relevance of this regime, it appears crucial to establish the robustness of the limiting behavior to inclusion of retinotopic distortions.
\begin{figure}
\centering
\includegraphics[width=10cm]{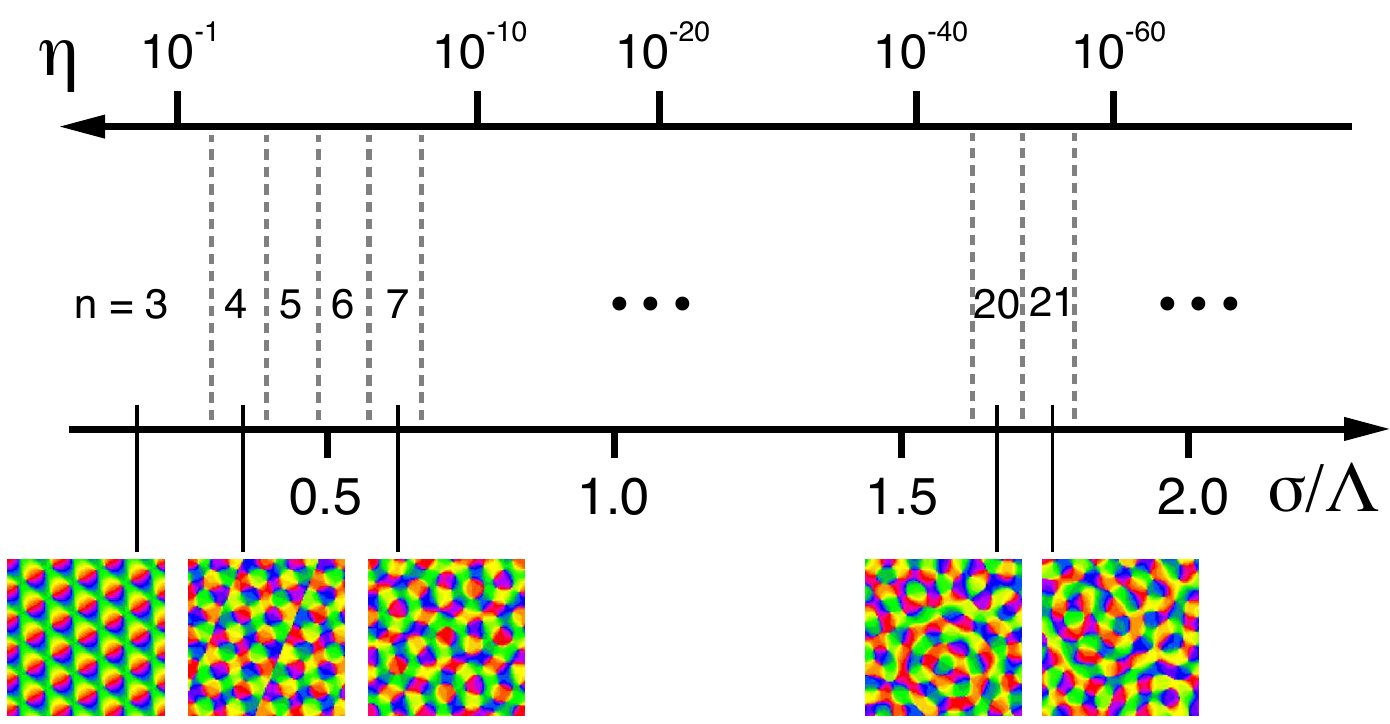}
\caption{\textbf{Phase diagram of the EN model with fixed retinotopy in the permutation symmetric limit $s_{4}\rightarrow\infty$.}
The graphs show the regions on the $\sigma/\Lambda$-axis
(lower axis) and the corresponding $\eta$-axis (upper axis), where
n-ECPs or sPWCs have minimal energy. High n-ECPs ($n\gtrsim10$) exhibit universal
pinwheel statistics. Note however the extremely small $\eta$-values
for large $\sigma/\Lambda$.
\label{Keil_Wolf_figure_15}}
\end{figure}
\subsection*{Optimal solutions of the EN model with variable retinotopy and arbitrary orientation stimulus ensembles}
\begin{figure*}
\centering
\includegraphics[width=15cm]{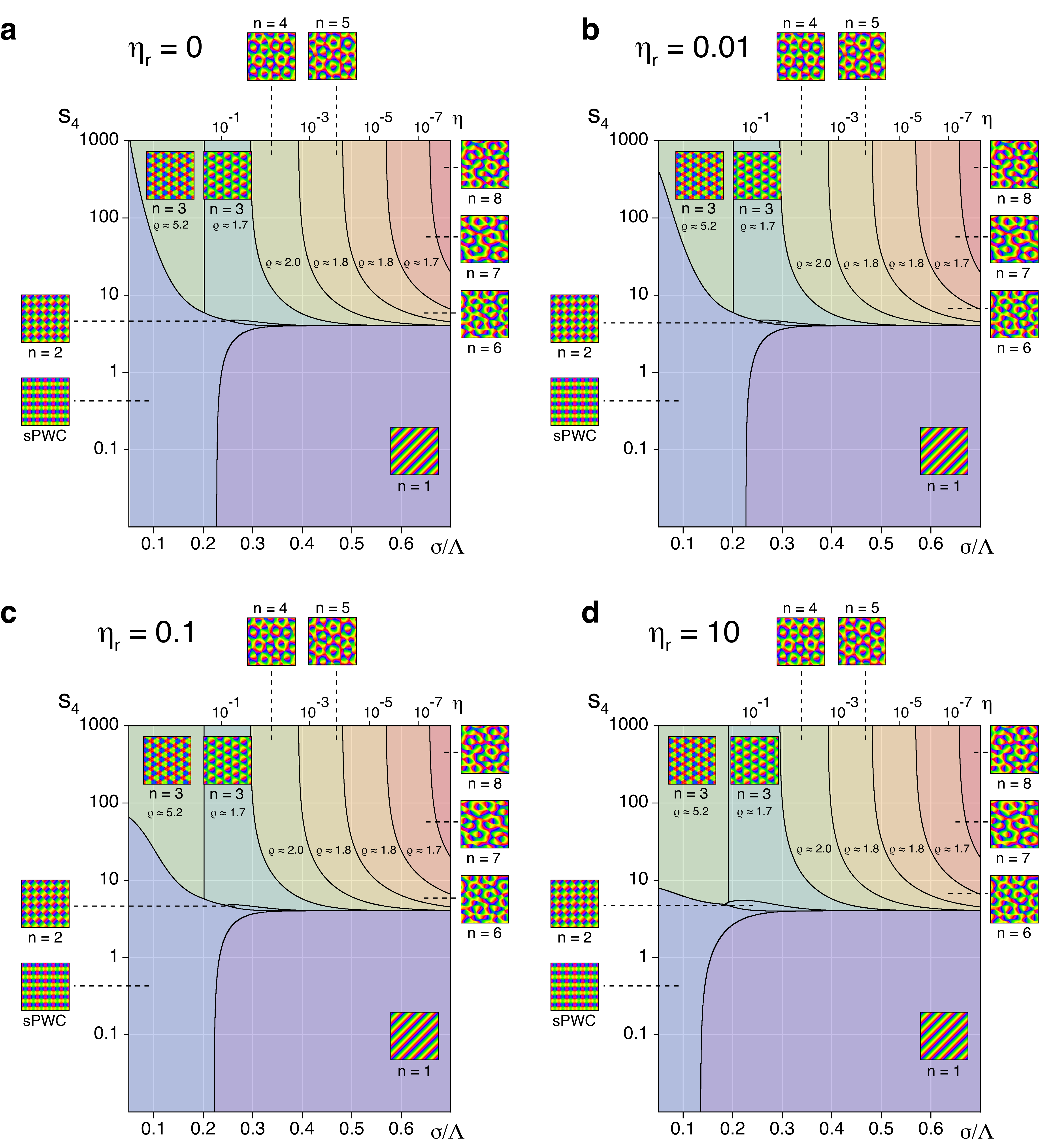}
\caption{\textbf{Stripe-like, crystalline, and quasi-crystalline cortical representations
as optimal solutions to the joint mapping problem of visual space
and orientation preference in the EN.}
(\textbf{a-d}) Phase diagrams for the joint mapping of visual space and orientation preference in
the EN near criticality for $\eta_{r}=0$ (a), $\eta_{r}=0.01$ (b), $\eta_{r}=0.1$ (c), and $\eta_{r}=10$ (d). The graphs show the regions
of the $s_{4}$-$\sigma/\Lambda$-plane in which coupled n-ECPs or sPWCs have minimal energy. For $n\geq3$, pinwheel densities
of the energetically favored n-ECP configuration are indicated.
Note the strong similarity between the phase diagrams and the phase diagrams in the fixed retinotopy
case (Fig. \ref{Keil_Wolf_figure_12}).
\label{Keil_Wolf_figure_16}}
\end{figure*}
In the EN model for the joint mapping of visual space and orientation
preferences, the angle-dependent interaction functions depend on four
parameters: $\eta$, $\sigma$, the fourth moment $\left<|s_{z}|^{4}\right>$
of the stimulus ensemble and $\eta_{r}$. By setting $\sigma=\sigma^{*}(\eta)$,
we are left with three free parameters at criticality. Therefore,
a three-dimensional phase diagram now completely describes pattern
selection in the EN model. For better visualization, in Fig. \ref{Keil_Wolf_figure_16}
we show representative cross sections through this three-dimensional
parameter space for fixed $\eta_{r}$. Firstly, we note the strong similarity
between the phase diagram for fixed retinotopy (Fig. \ref{Keil_Wolf_figure_12})
and the cross sections through the phase diagrams for the joint mappings
shown in Fig. \ref{Keil_Wolf_figure_16}. This expresses the
fact that retinotopic mapping and OPM are only weakly coupled or mathematically,
$g_{r}(\alpha)\ll g(\alpha)$ in all parameter regimes (see Appendix III). Again, for
distributions with small fourth moment, optimal mappings consist of either
pinwheel-free orientation stripes or sPWCs. Above a certain fourth moment value around $s_4 = 6$,  higher coupled n-ECPs are optimal. 
For small interaction range $\sigma/\Lambda$, hexagonal pinwheel crystals (coupled 3-ECPs) represent
optimal mappings in a large fraction of parameter space. With increasing $\sigma/\Lambda$,
we observe a sequence of phase transitions by which higher n-ECPs become optimal.
Anisotropic planforms at the lower end of the spectrum of pinwheel
densities are always energetically favored over high pinwheel density layouts. 
The only difference between the cross-sections is that the region covered by sPWCs increases for decreasing $\eta_{r}$. The phase diagram for large $\eta_r=10$ is virtually indistinguishable from the phase diagram in Fig. \ref{Keil_Wolf_figure_16}.

Optimal mappings of orientation preference are thus either orientation stripes, periodic
arrays of pinwheels (hexagonal, quadratic) or quasiperiodic pinwheel arrays
with low pinwheel density. Retinotopic distortions lead to lower gradients
of the retinotopic mapping at high gradient regions of the OPM. This
is in line with some of the experimental evidence \cite{Yu:2008p6292,Swindale:2000p4894}
but contradicts others \cite{Das:1997p5979}. 

Most importantly, we note that the results on permutation symmetry breaking in the fixed retinotopy case are not altered by allowing for retinotopic distortions. Since $g_r(\alpha)$ does not depend on the fourth moment of the orientation stimulus distribution, non-permutation symmetric terms decay as $1/s_4$ for large $s_4$. Hence, in the limit $s_4 \rightarrow\infty$, permutation symmetry is restored and we recover the phase diagram Fig. \ref{Keil_Wolf_figure_15} also for the EN model with variable retinotopy independent of $\eta_r$. As the energy contribution of retinotopic deviations $\mathbf{r}(\mathbf{x})$ becomes negligible in the infinite fourth moment limit, the optima are then simply the corresponding \textit{coupled} n-ECPs and these states are energetically degenerate for fixed n. For very large effective interaction range and infinite fourth moment of the orientation stimulus ensemble, the EN model with variable retinotopy is able to quantitatively reproduce the experimentally observed pinwheel statistics in OPMs. It furthermore predicts reduced gradients of the visual space mapping at high gradient regions of the OPM.
\subsection*{Finite stimulus samples and discrete stimulus ensembles}
\begin{figure*}
\centering
\includegraphics[width=14cm]{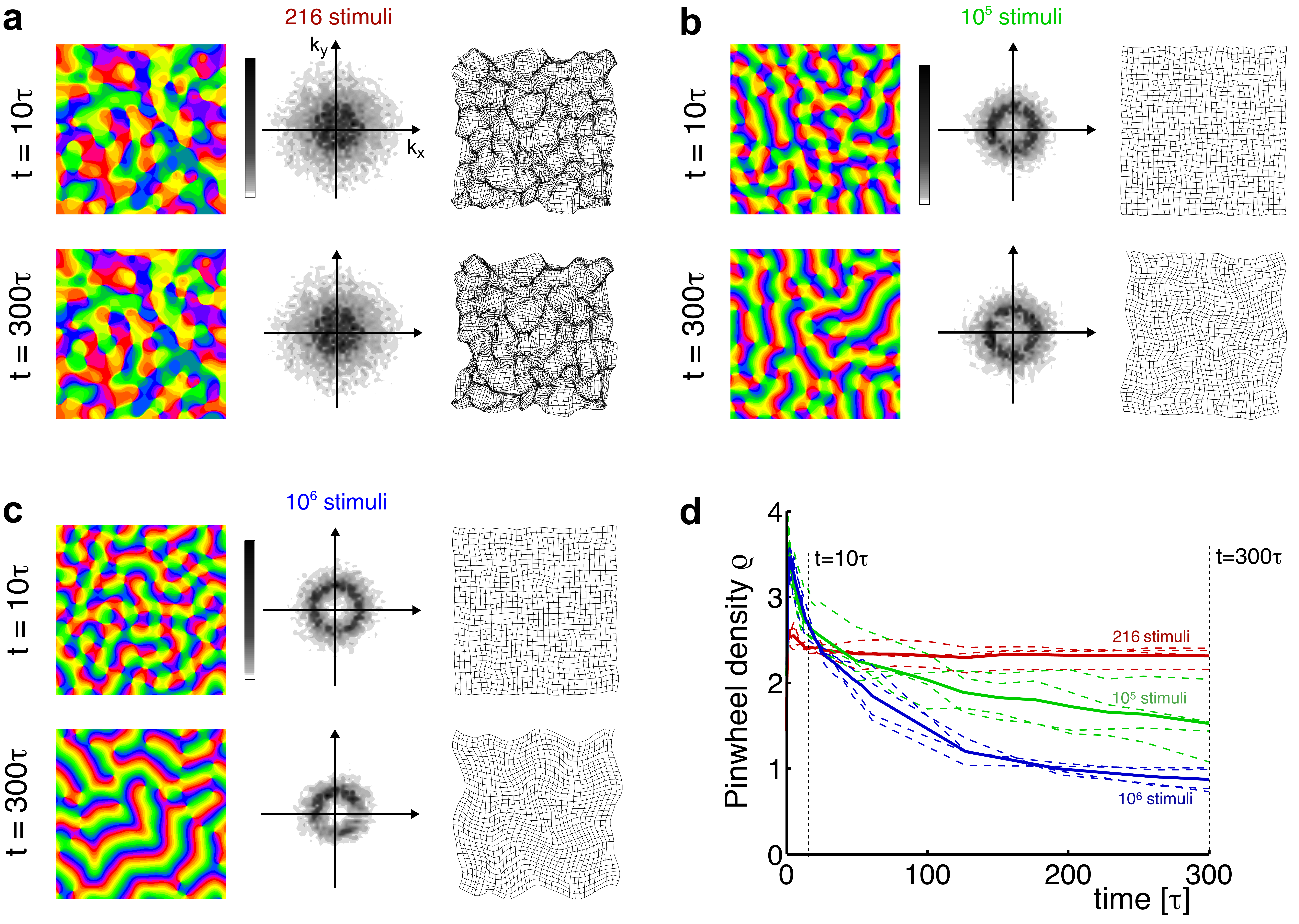}
\caption{\textbf{Development of OPM and retinotopic distortions in EN simulations with
fixed stimulus sets of different sizes.}
(\textbf{a}) OPMs (left), their power spectra (middle)
and retinotopic maps (right) for $t=10\tau$ (upper row) and $t=300\tau$
(lower row) obtained in simulations with fixed stimulus set ($\eta=0.028$, $\sigma/\Lambda=0.3$, $s4=4/3$,
$216$ stimuli). (\textbf{b}) $10^{5}$ stimuli (all other parameters
as in a). (\textbf{c}) $10^{6}$ stimuli (all other parameters as
in a). Large stripe-like OP domains are generated via pairwise pinwheel
annihilation for large simulation times. Retinotopic distortions are fairly weak.(\textbf{d})
Pinwheel density time course for EN simulations with fixed stimulus
sets of different sizes, including the simulations from a-c (red,
green, blue traces $216$, $10^{5}$, $10^{6}$ stimuli) (all other parameters
as in a). Dashed lines represent individual
simulations, solid lines an average over four simulations. Note, that
the pinwheel density rapidly decays below 2.0 in both cases, and in
particular for $10^{6}$ stimuli, the OPM pattern acquires large stripe-like
regions.
\label{Keil_Wolf_figure_17}}
\end{figure*}
Our reexamination of the EN model for the joint optimization of position and orientation selectivity has been so far carried out without addressing the apparently fundamental discrepancy between our results and the large majority of previous reports.
Since the seminal publication of Durbin and Mitchison in 1990 \cite{Durbin:1990p1196}, numerous studies have used
the EN model to simulate the development of visual cortical maps or
to examine the structure of optimal mappings by numerical simulation \cite{Goodhill_1990, Goodhill:2000p2141, Cimponeriu:2000p5167, CarreiraPerpinan:2004p6297, CarreiraPerpinan:2005p6295,Goodhill_2007,Simpson_2009,CarreiraPerpinan:2011}. These
studies have either used the circular or the uniform orientation stimulus
ensemble for which, to the best of our knowledge, the only two nontrivial
stationary solutions are square pinwheel crystals or orientation stripes. Furthermore, we found that the gradient descent dynamics seems to readily converge to the respective minima of the EN free energy. 
This indicates that other local minima and more complex intrinsically aperiodic states are not dominant in this model. In fact, we found that all aperiodic stationary solutions we could perturbative calculate analytically are unstable and thus represent hyperbolic saddle points and not local minima (see also Fig. \ref{Keil_Wolf_figure_21}a,b).
As these stable solutions barely resemble experimentally observed orientation
preference maps, it is not obvious how the EN model in all of these studies
could appear as a model well-suited to describe the complex layout of real cortical
orientation maps. Prior studies however often used computational methods different from our fixed parameter steepest descent simulations.

Two alternative approaches have been used predominantly to study dimension reducing mappings for cortical representations. These methods have been applied to both the EN model and the other widely used dimension reduction model, the Self-Organizing Feature Map (SOFM), originally introduced by Kohonen \cite{Kohonen1982p6163}. 
The simplest way to compute mappings from a high dimensional feature space onto the 2D model cortex is by iterating the following procedure for a large
number of randomly chosen stimuli (e.g. \cite{Bauer1992:570, Pawelzik_1996e, Swindale1998p827,Swindale:2000p6286, Swindale:2004p6287,Yu:2005p5850,Farley:2007p5840}): (i) Stimuli are chosen one at a time randomly from the complete feature space. (ii) The activation function for a particular stimulus is computed. In the case of the EN model, this activation function can acquire a rather complex form with multiple peaks (see Discussion). In the case of an SOFM, this activation function is a 2D-Gaussian. (iii) The preferred features of the cortical grid points are updated according to a discretized version of Eqs. (\ref{eq:continuous_z_dynamics}, \ref{eq:continous_r_dynamics}) or the corresponding equations for the SOFM model. Typically, this procedure is repeated on the order of 10$^{6}$ times. The resulting layout is then assumed to at least approximately solve the dimension reduction problem. In many studies, small stimulus sets have been chosen presumably for computational efficiency and not assuming specifically that the cortex is optimized for a discrete finite set of stimuli. In \cite{Durbin:1990p1196} for instance, a set of 216 stimuli was used, that was likely already at the limit of computing power available at this time.

In a more refined approach, the EN model as well as Kohonen's SOFM model have been trained with a finite set of stimuli (typically with on the order of 10$^{3}$-10$^4$) and the final layout of the model map has been obtained by deterministic annealing \cite{Rose:1998p6301}, i.e. by gradually reducing the numerical value of $\sigma$ in a numerical minimization procedure for the energy functional $\mathcal{F}$ at each value of $\sigma$ (see e.g. \cite{Durbin:1990p1196,CarreiraPerpinan:2004p6297,CarreiraPerpinan:2005p6295,CarreiraPerpinan:2011} and Methods). In such simulations, often non-periodic boundary conditions were used.
One might suspect in particular  the second approach to converge to OPM layouts deviating from our results. It is conceivable in principle, that deterministic annealing might track stationary solutions across parameter space that are systematically missed by both, our continuum limit analytical calculations as well as our descent numerical simulations. 

To assess the potential biases of the different approaches, we implemented (i) finite stimulus sampling in our gradient descent simulations and (ii) studied the results of deterministic annealing simulations varying both the size of the stimulus set as well as the type of boundary conditions applied.

We simulated the dynamics Eqs. (\ref{eq:continuous_z_dynamics},
\ref{eq:continous_r_dynamics}) with finite sets of stimuli of different
sizes (see Methods), drawn from the circular stimulus ensemble. As
e.g. in $ $\cite{Durbin:1990p1196,Wolf:1998p1199}, $\eta$ was set
to a small value ($\eta=0.025$) such that
the optimal configuration for the joint mapping of visual space and
orientation preference is the coupled 1-ECP (see Fig. \ref{Keil_Wolf_figure_10}), i.e.
a pattern of parallel orientation stripes without any retinotopic distortion (see
Fig. \ref{Keil_Wolf_figure_9}). Fig. \ref{Keil_Wolf_figure_17}
displays representative simulations for stimulus sets of size N = 216 (as used in \cite{Durbin:1990p1196}) (a), N = 10$^{5}$ (b), N
= 10$^{6}$ (c)  stimuli. Simulation time $t$ is measured in units of
the intrinsic time scale $\tau$ (see Methods). For N = 216 stimuli, RM and
OPM quickly reach an apparently stationary configuration with a large
number of pinwheels at around $t = 20\tau$. Power is distributed roughly isotropically around
the origin of Fourier space ($\mathbf{k}=0$). The stable OPM lacks a typical length scale and, expressing
the same fact, the power spectrum lacks the characteristic ring of enhanced Fourier amplitude. Retinotopic distortions are fairly pronounced.
Both obtained maps resemble the configurations reported in \cite{Durbin:1990p1196}. 

For N = 10$^{5}$  stimuli, we find that OPMs exhibit a characteristic scale (see dark shaded
ring in the power spectrum) and a dynamic rearrangement of the maps persists at least until $t=200\tau$. Stripe-like OP domains are rapidly generated
via pairwise pinwheel annihilation for $t>10\tau$. Retinotopic distortions
are fairly weak. 
For N = 10$^{6}$  stimuli, again OPMs exhibit a characteristic
scale (see dark shaded ring in the power spectrum) and the map dynamics persists beyond $t=200\tau$.
A larger fraction of the pinwheels annihilate pairwisely compared to N = 10$^{5}$  stimuli,
leading progressively to a pattern with large stripe-like domains.
Retinotopic distortions are fairly weak. 
For both cases with massive stimulus sampling ($N=10^5$, $N=10^6$), the pinwheel density rapidly drops below the range observed in tree shrews, galagos and ferrets and than further decreases during subsequent map rearrangement. 

In summary, the more stimuli are chosen for the optimization procedure, the less pinwheels are
preserved in the pattern of orientation preference and the more the
resulting map resembles the analytically obtained optimal solution. 

Deterministic annealing approaches which change parameters of the energy functional during the computational minimization process differ more fundamentally from our gradient descent simulations than the iterative schemes used with fixed parameters. Studies using deterministic annealing in addition frequently used non-periodic boundary conditions (e.g. \cite{CarreiraPerpinan:2004p6297,CarreiraPerpinan:2005p6295,CarreiraPerpinan:2011}). To study all potential sources of deviating results, we implemented deterministic annealing for the EN energy function (see Methods, Eq. \ref{eq:EN_energy_discrete}) for periodic boundaries, non-periodic boundary conditions as well as random and grid-like finite stimulus ensembles (see Methods).
We closely follow the refined methods used in \cite{CarreiraPerpinan:2004p6297,CarreiraPerpinan:2005p6295,CarreiraPerpinan:2011} and performed deterministic annealing simulations for the EN model with retinotopic distortions and stimuli drawn from the circular stimulus ensemble. 

Figs. \ref{Keil_Wolf_figure_18}a and \ref{Keil_Wolf_figure_19}a display representative simulations for random stimulus sets of size $N = 10^3$, $N = 10^4$ and $N = 10^5$ for periodic boundary conditions (Figs. \ref{Keil_Wolf_figure_18}a) and non-periodic boundary conditions (Figs. \ref{Keil_Wolf_figure_19}a).
Furthermore depicted are the pinwheel densities of stationary solutions as well as their energies, relative to the energy of a pinwheel-free stripe solution (see Methods) for different annealing rates $\xi$ (Figs. \ref{Keil_Wolf_figure_18}b-d, Figs. \ref{Keil_Wolf_figure_19}b-d). 
Figs. \ref{Keil_Wolf_figure_18}e-g and \ref{Keil_Wolf_figure_19}e-g additionally show the statistics of nearest neighbor (NN) pinwheel distances as well as the standard deviation (SD) of the pinwheel densities for randomly selected subregions in the OPM as introduced in \cite{Kaschube_2010}, averaged over four simulations with $N = 10^5$. To facilitate comparison, we superimposed fits to the experimentally observed statistics \cite{Kaschube_2010} for orientation maps in tree shrews, ferrets and galagos. 

When annealing with periodic boundary conditions, the maps found with deterministic annealing essentially resemble our gradient descent dynamics simulations. The larger the set of stimuli, the more stripe-like are the orientation preference maps obtained (Fig. \ref{Keil_Wolf_figure_18}a,b). Furthermore, the more carefully we annealed, the lower the pinwheel density of the obtained layouts (Fig. \ref{Keil_Wolf_figure_18}c). For $N=10^5$, the pinwheel density averaged over four simulations with annealing rate 0.999 was $\rho = 2.04$
As expected, the energy of the final layouts decreased with slower annealing rates (Fig. \ref{Keil_Wolf_figure_18}d). However, when starting from random initial conditions, the energy of the final layouts found was always higher compared to the energy of a pinwheel-free stripe solution (see Methods for details), which is the predicted optimum for the circular stimulus ensemble. NN-pinwheel distance histograms are concentrated around half the typical column spacing and in particular pinwheel pairs with short distances are lacking completely (Fig. \ref{Keil_Wolf_figure_18}e,f).

For non-periodic boundary conditions and random stimuli, we found that retinotopic distortions are more pronounced than for periodic boundary conditions. They however decreased with increasing number of stimuli. For large the stimulus numbers, we observed stripe-like orientation preference domains which are interspersed with lattice-like pinwheel arrangements (see Fig. \ref{Keil_Wolf_figure_19}c), lower row, upper left corner of the OPM). For $N=10^5$, the pinwheel density averaged over four simulations with annealing rate 0.999 was $\rho = 2.71$. 

Similarly to the results for periodic boundary conditions, short distance pinwheel pairs occur less frequently than in the experimentally observed maps, indicating an increased regularity in the pinwheel distances compared to real OPMs (Fig. \ref{Keil_Wolf_figure_19}e,f). This regularity is further indicated by a smaller exponent of the SD compared to the Poisson process (Fig. \ref{Keil_Wolf_figure_19}g). 
The perfect stripe-like solution is not the optimum for non-periodic boundaries. The energy of the map layouts found with very slow annealing rates is slightly lower than the energy of the pinwheel-free OPM layout (Figure \ref{Keil_Wolf_figure_19}d).
We note that the layout of the OPM at the boundaries does not differ substantially from the layout inside the simulated domain, suggesting that boundary effects affect the entire simulated domain for the relatively small region treated.

Finally, we performed simulations with grid-like stimulus patterns as  e.g. used in  \cite{CarreiraPerpinan:2004p6297,CarreiraPerpinan:2005p6295}. These simulations displayed a strong tendency towards rhombic pinwheel arrangements, i.e. the second stable stationary solution found for the circular stimulus ensemble. We refer to Appendix II for further details.

In summary, our results for the discrete EN model with deterministic annealing largely agree with the analytical results. Irrespective of the numerical methodology, the emerging map structure for large numbers of stimuli is confined to the states predicted by our analytical treatment of the continuum formulation of the EN. 
This behavior is expected because the energies underlying the deterministic annealing and the steepest descent simulations are mathematically equivalent (see Methods). In any kind of deterministic annealing simulation we tested, resulting patterns were patchworks of the two fundamental stable solutions identified by the analytical treatment: pinwheel free stripes and square lattices of pinwheels. Such patchworks are spatially more complicated than perfect stripes or crystals. Nevertheless, they qualitatively differ in numerous respects from the experimentally observed spatial arrangements (see Figs. \ref{Keil_Wolf_figure_18}, \ref{Keil_Wolf_figure_19} and \ref{Appendix_Keil_Wolf_figure_6} in Appendix II). 
How the fundamental stable solutions are stitched together somewhat differs between the different kinds of simulations. For instance, using a grid-like stimulus ensemble with non-periodic boundary conditions apparently energetically favors the rhombic pinwheel crystal compared to the pinwheel-free stripe regions (see Fig. \ref{Appendix_Keil_Wolf_figure_5} in Appendix II). In summary, while some of the patterns obtained by deterministic annealing might be called ``good-looking" maps, all of them substantially deviate from the characteristics of experimentally observed pinwheel arrangements.

We conclude, that the differences between our results and those of previous studies are most likely due to the small finite stimulus samples used largely for reasons of computational tractability. Deterministic annealing using stimulus samples that fill the feature space converges to the same types of patterns found by perturbation theory. We further conclude, that our methods do not systematically miss biologically relevant local minima of the classical EN energy function. 
\begin{figure*}
\centering
\includegraphics[width=14cm]{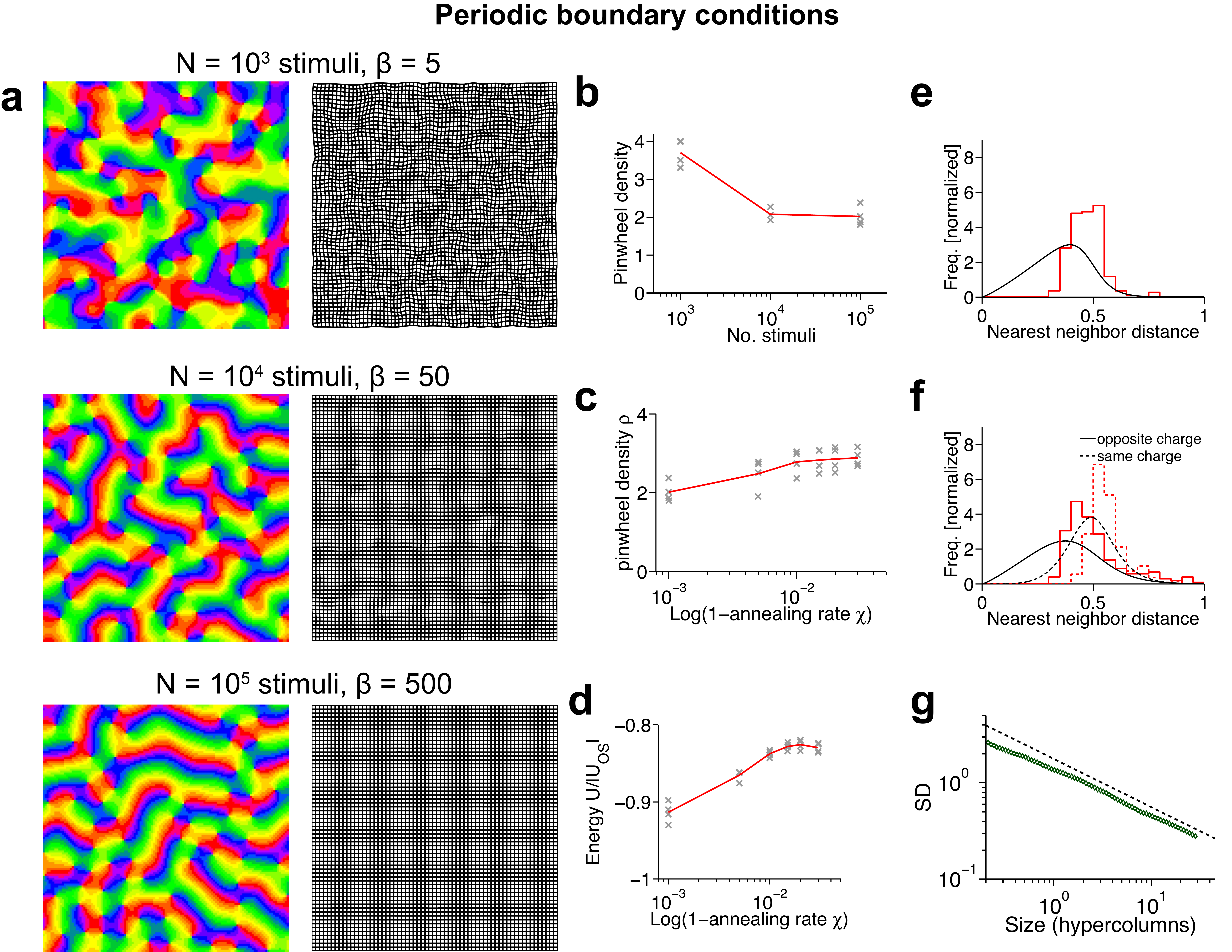}
\caption{\textbf{The EN model with periodic boundary conditions, solved with deterministic annealing}
(\textbf{a}) OPMs (left) and retinotopic maps (right) for $N=10^3$ (upper row), $N=10^4$(middle row) and $N=10^5$ (lower row) random stimuli and periodic boundary conditions (annealing rate  $\chi = 0.999$, see Methods). $\beta$ is the continuity parameter in the conventional definition of the EN model (see Methods, Eq. \ref{eq:EN_energy_discrete}) and is scaled, such that a comparable number of columns is emerging in the simulations for each size of the stimulus set.
(\textbf{b}) Pinwheel densities of EN solutions for different numbers of stimuli, $\chi=0.999$. 
(\textbf{c}) Pinwheel densities of EN solutions for $10^5$ stimuli and different annealing rates. 
(\textbf{d}) Energies of solutions for $10^5$ stimuli, relative to the energy of a pinwheel-free stripe solution (see Methods) for different annealing rates.
(b-d) Crosses mark individual simulations, red line indicates average values.
(\textbf{e}-\textbf{f}) Statistics of nearest neighbor pinwheel distances for pinwheels of (e) arbitrary and (f) opposite and equal charge for $10^{5}$ random stimuli and periodic boundary conditions, averaged over four simulations (red curves). Black curves represent fits to the experimental data from \cite{Kaschube_2010}. 
(\textbf{g}) Standard deviations (SD) of pinwheel densities estimated from randomly selected regions in the OPM. Black dashed curve indicates SD for a two-dimensional Poisson process of equal density. 
\label{Keil_Wolf_figure_18}}
\end{figure*}

\begin{figure*}
\centering
\includegraphics[width=14cm]{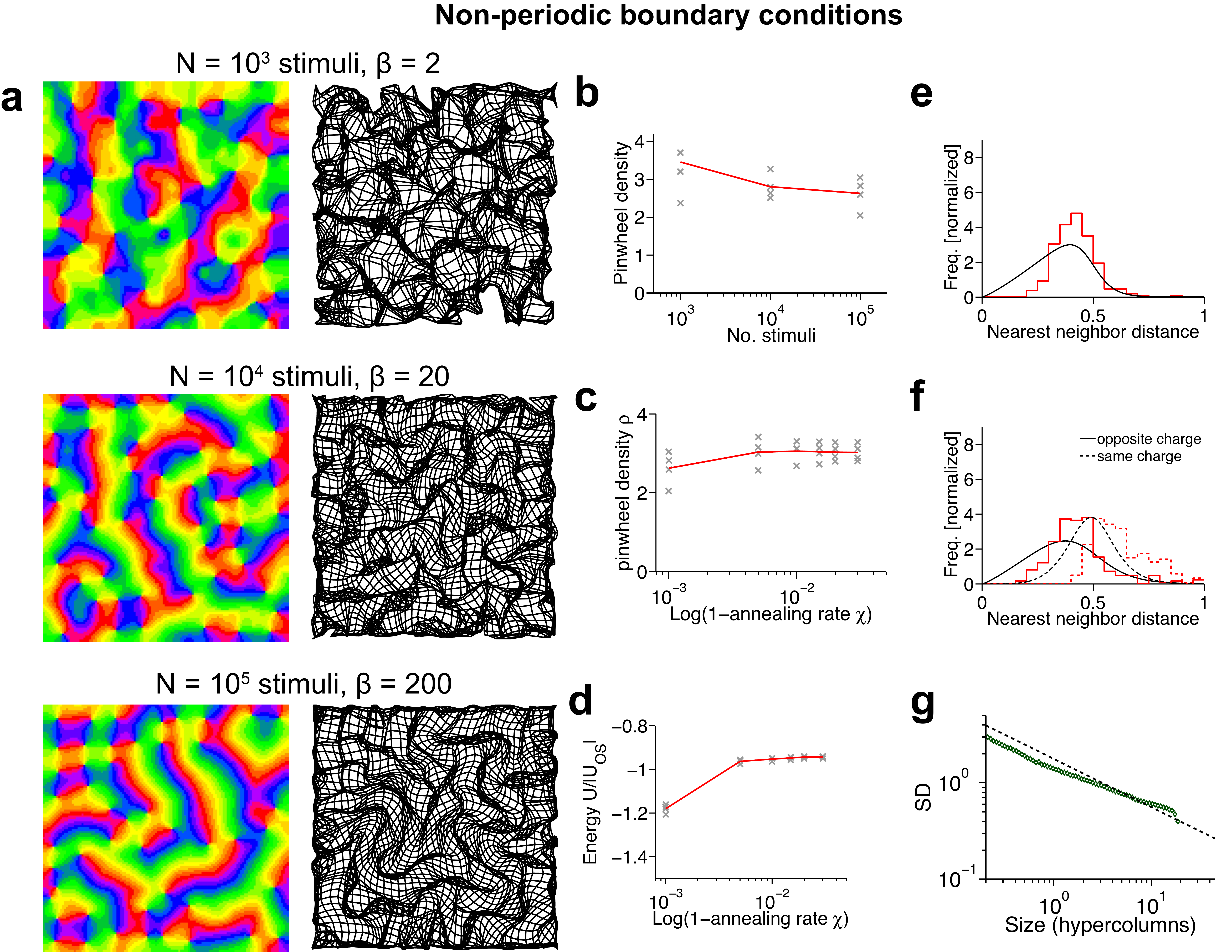}
\caption{\textbf{The EN model with non-periodic boundary conditions, solved with deterministic annealing}
(\textbf{a}-\textbf{g}) As Fig. \ref{Keil_Wolf_figure_18}, but for non-periodic boundary conditions.
\label{Keil_Wolf_figure_19}}
\end{figure*}

%
%
%
\section*{Discussion}
\subsection*{Summary}
In this study, we examined the solutions of what is perhaps the most prominent optimization model for the spatial layout of orientation and retinotopic maps in the primary visual cortex, the Elastic Network (EN) model. We presented an analytical framework that enables us to derive closed-form expressions for hyperbolic fixed points, local and global minima, and to analyze their stability properties for arbitrary optimization models for the spatial layout of OPMs and RMs. 
Using this framework, we systematically reexamined previously used instantiations of this model, dissecting the impact of stimulus ensembles and of interactions between the two maps on optimal map layouts.
To our surprise, the analysis yielded virtually identical results for all of these model instantiations that substantially deviate from previous numerical reports. Pinwheel-free orientation stripes and crystalline square lattices of pinwheels are the only optimal dimension-reducing OPM layouts of the EN model. Both states are generally stable but exchange their roles as optima and local minima at a phase border. Numerical simulations of the EN gradient descent dynamics as well as simulations utilizing deterministic annealing confirmed our analytical results.
For both processes, the initially spatially irregular layouts rapidly decayed into a patchwork of stripe-like or crystal-like local regions that then became globally more coherent on longer timescales.
Pinwheel-free solutions were approached after an initial phase of pattern emergence by pairwise pinwheel annihilation. Crystalline configurations were reached by the generation of additional pinwheels and pinwheel annihilation together with a coordinated rearrangement towards a square lattice. 
These results indicate that layouts which represent an optimal compromise of coverage and continuity for retinotopy and orientation do not per se reproduce the spatially aperiodic and complex structure of orientation maps in the visual cortex.

To clarify whether the EN model is in principle capable of reproducing the biological observations, we performed an unbiased comprehensive inspection of EN optima for arbitrary stimulus distributions possessing finite fourth moments. 
This analysis identified two key parameters determining pattern selection: (i) the effective intracortical interaction range and (ii) the fourth moment of the orientation stimulus distribution. We derived complete phase diagrams summarizing pattern selection in the EN model for fixed as well as variable retinotopy. 
Small interaction ranges together with low fourth moment values lead to either pinwheel-free orientation stripes, rhombic or hexagonal crystalline orientation map layouts as optimal states. Large interaction ranges together with orientation stimulus distributions with high fourth moment values lead to the stabilization of irregular aperiodic OPM layouts. These solutions belong to a class of solutions previously called n-ECPs. This solution class encompasses a large variety of OPM layouts and has been identified as optimal solutions of abstract variational models of OPM development \cite{Wolf:2005p190}. We showed that in the EN model due to a lack of a so-called permutation symmetry, among this family of solutions, states with low pinwheel densities are selected as global minima. In the extreme and previously unexplored parameter regime of very large effective interaction ranges and stimulus ensemble distributions with infinite fourth moment, permutation symmetry is restored and spatially aperiodic OPM layouts with higher pinwheel density are included in the repertoire of optimal solutions. Only in this limit, the repertoire of optima reproduces the recently described species-insensitive OPM design \cite{Kaschube_2010} and quantitatively matches experimentally observed orientation map layouts. None of these findings depend on whether the EN model is considered with variable or fixed retinotopy.
\subsection*{Comparison to previous studies}
It is an important and long-standing question, whether the structure of cortical maps of variables such as stimulus orientation or receptive field position can be explained by a simple general principle. The concept of dimension reduction is a prominent candidate for such a principle (see e.g. \cite{Swindale:1996p6540,Goodhill_2007} for reviews) and the qualitative agreement between experimental data and previous numerical results from dimension reduction models \cite{Durbin:1990p1196, Erwin:1995p1206, Farley:2007p5840,Swindale:1996p6540, Swindale:2004p6287, Obermayer:1991p505,Obermayer:1992pE75,  Goodhill:2000p2141, Cimponeriu:2000p5167, CarreiraPerpinan:2004p6297, CarreiraPerpinan:2005p6295,Goodhill_1994,Simpson_2009} can be viewed as evidence in favor of the dimension reduction hypothesis. Yet comprehensive analytical investigations of dimension reduction problems and in particular the determination of their optimal and nearly optimal solutions have been impeded by the mathematical complexity of these problems.
For the EN algorithm applied to the Traveling Salesman Problem, previous analytical results established the unselective fixed point above the first bifurcation point as well as the parameters at which this solution becomes unstable \cite{Durbin:1989p348}. Subsequent work extended these results to the EN model for cortical map formation. The periodicity of solutions depending on the model parameters has been obtained by computing the eigenvalues of the Hessian matrix of the energy function \cite{Goodhill:1990p348, Dayan:1993p392, Goodhill:2000p2141}. Hoffs\"ummer et al. confirmed these results, and computed the periodicity of the emerging patterns in the continuous EN model formulation by linear stability analysis of the EN gradient descent dynamics as used in the present study \cite{Hoffsummer95}. Our results extend these findings and for the first time provide analytical expressions for the precise layout of optimal and nearly optimal dimension-reducing maps.

In the light of the qualitative agreement between experimental data and numerical solutions of the EN model previously described, it is perhaps our most surprising result that the model's optimal dimension-reducing maps are regular periodic crystalline structures or pinwheel-free stripe patterns in large regions of parameter space. In particular, the species-insensitive pinwheel statistics observed experimentally \cite{Kaschube_2010} are not exhibited by optimal solutions of the classical EN in any of the previously considered parameter regimes. 

Our comparison of different numerical approaches indicates that the differences to previous studies are mainly attributable to differences in the sampling of the stimulus manifold in the numerical optimization procedures. In their seminal publication, Durbin \& Mitchison used sets of 216 stimuli from the circular stimulus ensemble and applied a Gauss-Seidel procedure to obtain stationary configurations \cite{Durbin:1990p1196}. A similar procedure was used in \cite{Goodhill_1994}. 
Quite frequently, the number of stimuli used for optimization is of the same order of magnitude as the number of model neurons or centroids in feature space. This provides a relatively sparse sampling of the stimulus manifold \cite{Goodhill:2000p2141, CarreiraPerpinan:2004p6297, CarreiraPerpinan:2005p6295}. Finite stimulus sampling effects are expected to worsen when feature spaces of higher dimension are considered.

The choice of small stimulus sets in previous dimension reduction studies was imposed mainly by the limitations of computing power. Using a parallelized implementation of the Cholesky-method for deterministic annealing \cite{Goodhill:2000p2141, Cimponeriu:2000p5167, CarreiraPerpinan:2004p6297, CarreiraPerpinan:2005p6295} on a multicore architecture with 2TB working memory, we explored the dependence of the obtained near optimal solutions on the sampling of the feature space manifold over two orders of magnitude. We find that, the more stimuli are sampled, the closer the numerically obtained configurations resemble our analytical predictions.
Our results on the classical EN model with deterministic annealing suggest that in the limit of large stimulus numbers, one would perfectly recover our analytical results both for periodic conditions or non-periodic boundary conditions with realistic system sizes. This dense stimulus sampling limit is also readily visible in our reproduction of the original Durbin \& Mitchison sampling and the modification of the predicted map structure with stimulus number (Fig. \ref{Keil_Wolf_figure_17}).

The finding that computational limitations prevented Durbin \& Mitchison from obtaining the genuine predictions of their dimension reduction model should not be viewed as diminishing the importance of their contribution. The dimension reduction approach has played a unique and extremely productive role in guiding the conceptualization of cortical functional architecture. It has established an abstract view on cortical representations without which most of our current theoretical knowledge about candidate theories for cortical architectures could not have been obtained.

Our results about optimal states of the EN for the circular and uniform stimulus ensembles however agree with some prior work.
In \cite{Wolf:1998p1199}, the gradient descent dynamics of the EN model Eqs. (\ref{eq:continuous_z_dynamics}, \ref{eq:continous_r_dynamics}) was used as a model for the emergence and refinement of cortical maps during development. Simulated visual stimulus features included retinotopy, orientation and eye dominance. The numerical procedures were similar to the one developed in the current study. Parameters were chosen such that $\left<|s_{z}|^{4}\right>=5.33$ and $\sigma/\Lambda\approx 0.366$. This study found that an initially large number of pinwheels decayed via pairwise annihilation of pinwheels with opposite topological charge. Our analysis predicts a stripe-like OP pattern as optimal solution in this regime, both in the case of a fixed uniform retinotopy as well as with variable retinotopy. In our simulations, this state is reached after an initial phase of symmetry breaking with the generation of numerous pinwheels via pairwise pinwheel annihilation. Our analytical and numerical results thus confirm, explain, and generalize these previous findings. 

The previous results also indicated that the inclusion of eye dominance in the EN model slightly slows down but does not stop the pinwheel annihilation process (see \cite{Wolf:1998p1199}, Fig. 3). This raises the possibility that the main features of our analysis of optimal solutions for the EN model may persist when additional feature dimensions are taken into account.    
Reichl et al. in fact observed that models with interacting OPM and ocular dominance maps (ODMs) exhibit a transition from pinwheel-free stripes to periodic pinwheel crystals similar to the transitions found in the EN \cite{Reichl2009:p208101} and demonstrated that this transition is a general feature of models with interacting OPM and ODMs \cite{reichl_2011}. A rigorous characterization of map structures predicted by the simultaneous optimization of multiple periodic feature representations such as orientation preference and ocular dominance constitutes an important goal for future studies. The recent work by Reichl and co-workers suggests that this issue can be successfully approached using concepts from the nonlinear dynamics of pattern formation \cite{Reichl2009:p208101}.
Finally, one recent study used the continuous formulation of the EN model to investigate the impact of postnatal cortical growth on the formation of ocular dominance columns in cat visual cortex \cite{Keil:2010p6536}. Consistent with our results, this study also observed perfectly regular stripe-like patterns as stationary states in gradient descent simulations. The dynamics of the convergence of the ODMs towards the stripes was modified by including cortical growth into the model. However, as soon as growth terminated, simulated ODM layouts readily converged towards regular stripes. How cortical growth interacts with the formation of orientation columns is currently not understood and represents a further interesting topic for future studies.
\subsection*{Geometric relationship between retinotopic distortions and orientation preference maps}
Experimental results on the geometric relationships between the map of visual space and the map of orientation preference are ambiguous. Optical imaging experiments in cat V1 suggested a systematic covariation of inhomogeneities in the retinotopic map with singularities in the pattern of orientation columns in optical imaging experiments \cite{Das:1997p5979}. Regions of high gradient in the map of visual space preferentially appeared to overlap with regions of high gradient of the OPM. In ferret, however, it has been reported that high gradient regions of the map of visual space correspond to regions of low gradient in the OPM \cite{Yu:2005p5850}. In tree shrew V1, no local relationships between the mapping of stimulus orientation and position seem to exist and the map of visual space appears to be ordered up to very fine scales \cite{Bosking:2002p6290}. In line with this, single unit recordings in cat area 17 revealed no correlation between receptive-field position scatter and orientation scatter across local cell ensembles \cite{Hetherington:1999p637,Buzas2003:p957}. 

Our analysis of the EN model shows that its optimal states exhibit a negative correlation between the rates of change of orientation preference and retinotopic position, similar to what has been observed in the ferret \cite{Yu:2005p5850}. This is expected from the principle of dimension reduction and in agreement with the original numerical results by Durbin \& Mitchison  \cite{Durbin:1990p1196}. However, both in simulations of the gradient descent dynamics and in deterministic annealing simulations with periodic boundary conditions as well as in analytically obtained optimal solutions, deviations from a perfectly uniform mapping of visual space are surprisingly weak (see Figs. \ref{Keil_Wolf_figure_9}, \ref{Keil_Wolf_figure_10}, \ref{Keil_Wolf_figure_18}, \ref{Appendix_Keil_Wolf_figure_4} in Appendix II). 

Deterministic annealing simulations with open non-periodic boundary conditions showed a substantially increased magnitude of retinotopic distortions. This raises the possibility that different behaviors observed in different experiments might be at least partially related to the influence of boundary effects. The influence of boundary effects is expected to decline into the interior of an area, in particular for large areas as V1 (see \cite{Wolf1996:p306}). In the bulk of V1, we thus expect only a weak coupling of orientation map and retinotopic distortions according to the EN model. In this regime, the predictions from models with reduced rotational symmetry (so-called Shift-Twist symmetry \cite{bressloff_et_al_01}) about the coupling between retinotopic distortions and orientation preference maps \cite{Thomas:2004p6132} appear to be more promising than the weak effects resulting from the coverage-continuity compromise. Consistent with the measurements of Das and Gilbert \cite{Das:1997p5979}, such models predict small but significant positive correlations between the rates of change of orientation preference and retinotopic position \cite{Thomas:2004p6132}. Moreover, the form of the retinotopic distortions in such models is predicted to differ for pinwheels with positive and negative topological charge \cite{Thomas:2006p119}. This interesting prediction of OPM models with Shift-Twist symmetry deserves to be tested by measuring the receptive field center positions around the two types of pinwheels with single cell resolution \cite{Ohki:2006p7461}.
\subsection*{Aperiodic orientation preference maps reflect long-range intracortical suppression}
\begin{figure*}
\includegraphics[width=16cm]{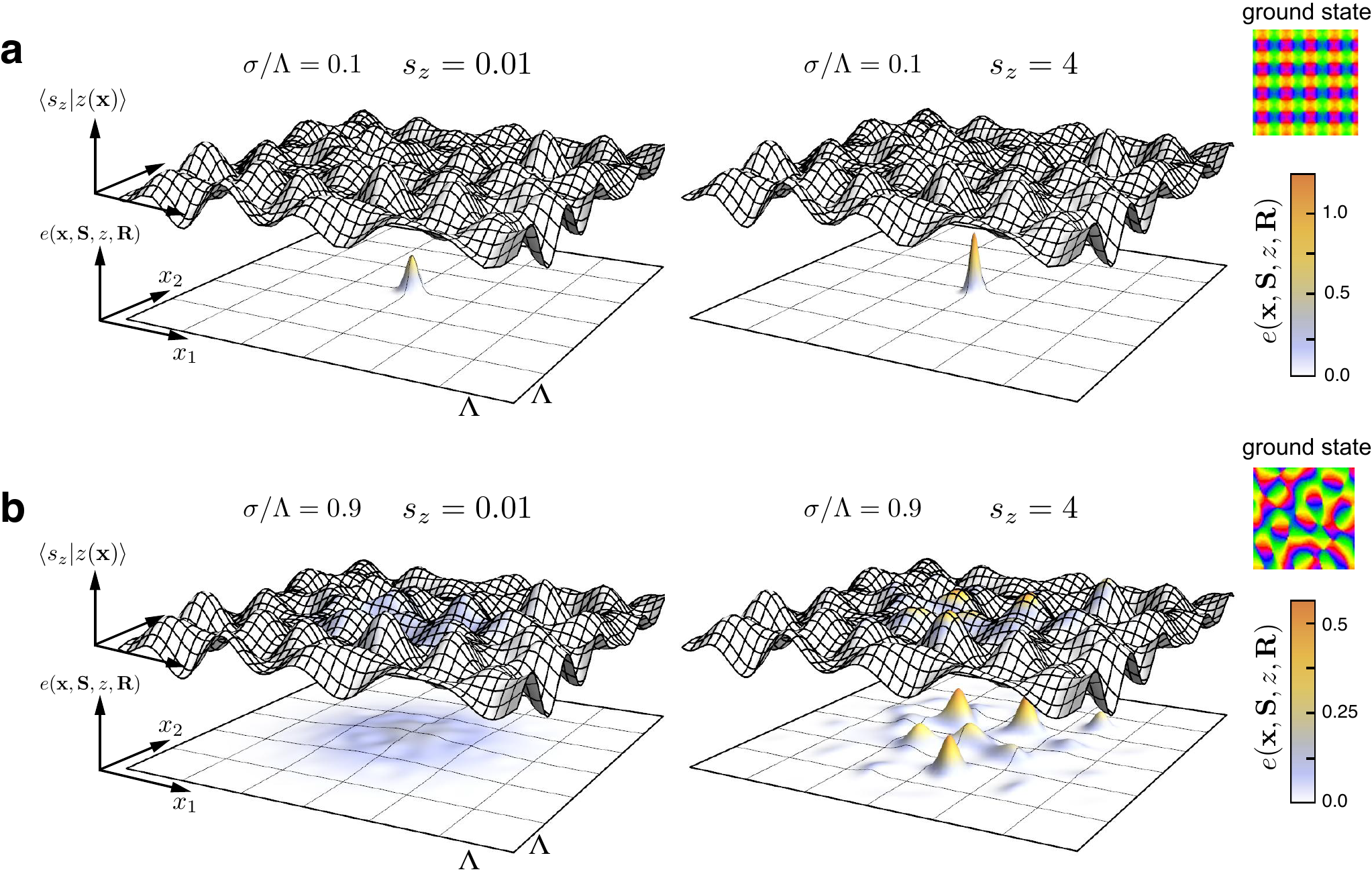}
\caption{\textbf{Different patterns of evoked activity for different effective ranges
of intracortical interaction in the EN model.}
The component $\left<s_{z}|z(\mathbf{x})\right>$
of the orientation map $z(\mathbf{x})$ in the direction of the stimulus
$s_{z}$ is plotted as a meshed 3D graph in a $6\Lambda\times6\Lambda$ patch. Color code and height of
the projection below indicate the strength of activation. The stimulus
is presented in the center of the displayed cortical subregion. (\textbf{a})
Evoked activity patterns $e(\mathbf{x},\mathbf{S},z,\mathbf{r})$
for small interaction range $\sigma/\Lambda=0.1$ and weakly oriented
stimulus with $s_{z}=0.01$ (left) and strongly oriented stimulus
with $s_{z}=4$ (right). rPWC (see upper right) are optimal in this regime. (\textbf{b}) Evoked activity patterns $e(\mathbf{x},\mathbf{S},z,\mathbf{r})$
for large interaction range $\sigma/\Lambda=0.9$ and weakly oriented
stimulus with $s_{z}=0.01$ (left) and strongly oriented stimulus
with $s_{z}=4$ (right). Spatially aperiodic 8-planforms (see upper right) are optimal in this regime. A uniform retinotopy was assumed in all cases for simplicity.
\label{Keil_Wolf_figure_20}}
\end{figure*}
Our unbiased search through the space of stimulus ensembles with finite fourth moment revealed the existence of spatially aperiodic optimal solutions in the EN. It is important to realize that the selection of these solutions is not easily viewed as resulting from an optimal compromise between coverage and continuity. In fact, the continuity parameter in the respective parameter regime is so small that solutions essentially maximize coverage (see Figs. \ref{Keil_Wolf_figure_12}, \ref{Keil_Wolf_figure_15}, and \ref{Keil_Wolf_figure_16}). Instead, this phenomenon reflects a different key factor in the stabilization of  pinwheel-rich aperiodic layouts, namely the dominance of long-ranged and effectively suppressive interactions. 
This is illustrated in Fig. \ref{Keil_Wolf_figure_20}  which depicts
different forms of stimulus-evoked activity patterns in the EN model.
For a short-range interaction (Fig. \ref{Keil_Wolf_figure_20}a),
the activity evoked by low as well as high orientation energy
stimuli is an almost Gaussian activity peak located near the stimulus position. The peak is shallow for low (left)
and sharp for high ``orientation energy" (right). In the corresponding parameter regime,
square pinwheel crystals are the optimal solution of the EN. For a longer
range of interaction where aperiodic OPM layouts are the optimal states, the
activity evoked by a single point-like stimulus is qualitatively different. Here, the activity pattern is extended and spans
several hypercolumns (Fig. \ref{Keil_Wolf_figure_20}b). It is weakly modulated for low orientation energy stimuli (left) and consists of several distinct peaks for high orientation energy stimuli (right). In this regime, neurons at a distance of several columns compete for activity through the normalization term in the EN which leads to a nonlocal and effectively suppressive intracortical interaction. 

It is presumably not a mere coincidence that recent studies of abstract variational models of OPM development \cite{Wolf:2005p190,Kaschube2008, Kaschube_2010} mathematically identified this type of interaction as a key mechanism for stabilizing realistic OPM layouts. It has been shown that all models for OPM development that share the basic symmetries (i) translational symmetry (ii) rotational symmetry (iii) shift symmetry and (iv) permutation symmetry and in addition are dominated by long-range suppressive interactions, form a universality class that generates maps exhibiting a universal and realistic pinwheel statistics. In such models, suppressive long-range interactions are key to stabilizing irregular arrangement of pinwheels, which otherwise largely disappear or crystalized during optimization. We have stressed that the EN model as considered here obeys the symmetries (i)-(iii). In the limit of infinite orientation stimulus ensemble fourth moment, permutation symmetry (iv) is restored. The EN can thus be tuned into the above universality class by sending the orientation stimulus distribution fourth moment to infinity and choosing an exponentially small continuity parameter to realize effective long-range coupling. Indeed, the phase diagrams for abstract variational models of OPM development \cite{Wolf:2005p190} and those of the EN model found here are structurally very similar. In both cases, a rather large orientation stripe phase is complemented by a cascade of phase transitions towards more complex, aperiodic and pinwheel-rich OPM layouts induced by long-range suppressive interactions. 
Using abstract variational models, it has been shown recently that the stabilization of regular crystalline pinwheel layouts can alternatively be achieved by a strong coupling between the map of orientation and the map of eye dominance \cite{Reichl2009:p208101,reichl_2011}. The structure of the phase diagrams of such models however appears fundamentally different from the structure of the EN phase diagrams. 

The parameter regime in which the EN model's optimal solutions exhibit the experimentally observed pinwheel statistics is not at all intuitive and in our opinion questions the conventional interpretation of the EN model to the formation of cortical feature maps. Firstly, the extremely small continuity parameter questions the fundamental role of a trade-off between coverage and continuity. We note that such a parameter regime is currently not accessible to numerical simulations. In addition, an apparently fundamental property for any adequate model for OPM optimization or development, namely a Turing-type finite wavelength instability of the unselective state \cite{Wolf:2003p210}, is lost in the limit $\eta\rightarrow 0$. 
At first sight the infinite fourth value required may appear reminiscent of the power-law distributions for orientation energy found in the statistics of natural images \cite{Field1987:p2379, Lee2003:p206}. 
However, as visualized in Fig. \ref{Keil_Wolf_figure_20}b, the essential property of the EN model in the infinite fourth moment regime is the occurrence of patterns of activity spatially extended beyond a single hypercolumn representing spatially localized point-like stimuli.  These activity patterns mediate the long-range interactions between distant orientation columns which in turn cause the stability of realistic pinwheel-rich aperiodic OPM layouts. It is obvious that spatially extended stimuli provide a much more plausible and realistic source of extended activity patterns in models for visual cortical development (for an extended discussion see \cite{Kaschube_2010_SOM}). Optimization models for cortical maps based on the representation of more complex spatially extended visual stimuli, such as natural scenes, rather than a model based on point-like stimuli with extreme statistics would then be a more appropriate basis for understanding visual cortical functional architecture.
\subsection*{Comparison to the SOFM model}
Several alternatives to the EN model have been proposed as optimization approaches that can account for the structure of visual cortical maps. One prominent alternative dimension reduction model is the so-called self-organizing feature map (SOFM), originally introduced by Kohonen \cite{Kohonen1982p6163}. It is widely believed that this model, albeit lacking an exact energy functional \cite{Erwin:1992p4656}, implements a competition between coverage and continuity very similar to the EN model \cite{Erwin:1992p4656, Swindale1998p827, Swindale:2004p6287, Swindale:2000p6286}. 
The SOFM has been reported to reproduce many of the experimentally observed geometric properties of visual cortical feature maps (e.g. \cite{Kohonen:1995p6543, Swindale1998p827, Swindale:2000p6286, Swindale:2004p6287, Yu:2005p5850, Farley:2007p5840}). The numerical procedures used in all of these studies were either the deterministic annealing procedure or the non-recurring application of a stimulus set without systematic assessment of pattern convergence. An analysis of the nontrivial stationary states of a dynamical systems formulation of the SOFM model is currently lacking. The main difference between the SOFM model and the EN model is that the activation function by definition has the form of a stereotypical Gaussian and competition is incorporated by a hard winner-takes-all mechanism. As a consequence, it is not obvious that a long-range suppressive interaction regime can be realized in this model. According to our analysis, one would thus expect orientation stripes and rhombic pinwheel crystals as nontrivial stationary states of the SOFM model. In a very recent study of the SOFM algorithm that used a numerical procedure similar to the gradient descent simulations developed in this paper, both pinwheel annihilation and rhombic pinwheel crystallization have been observed \cite{huang_08}. In addition, one study that examined the SOFM model for orientation and retinotopy found a fast convergence to pinwheel-free stripe-like solutions for a wide parameter range \cite{Wolf:1998p1199}. In view of these results, it seems worthwhile to also reexamine the SOFM model with respect to its stationary states.
%
%
%
%
\subsection*{Rugged or Smooth Energy landscape}
\begin{figure}
\centering
\includegraphics[width=16cm]{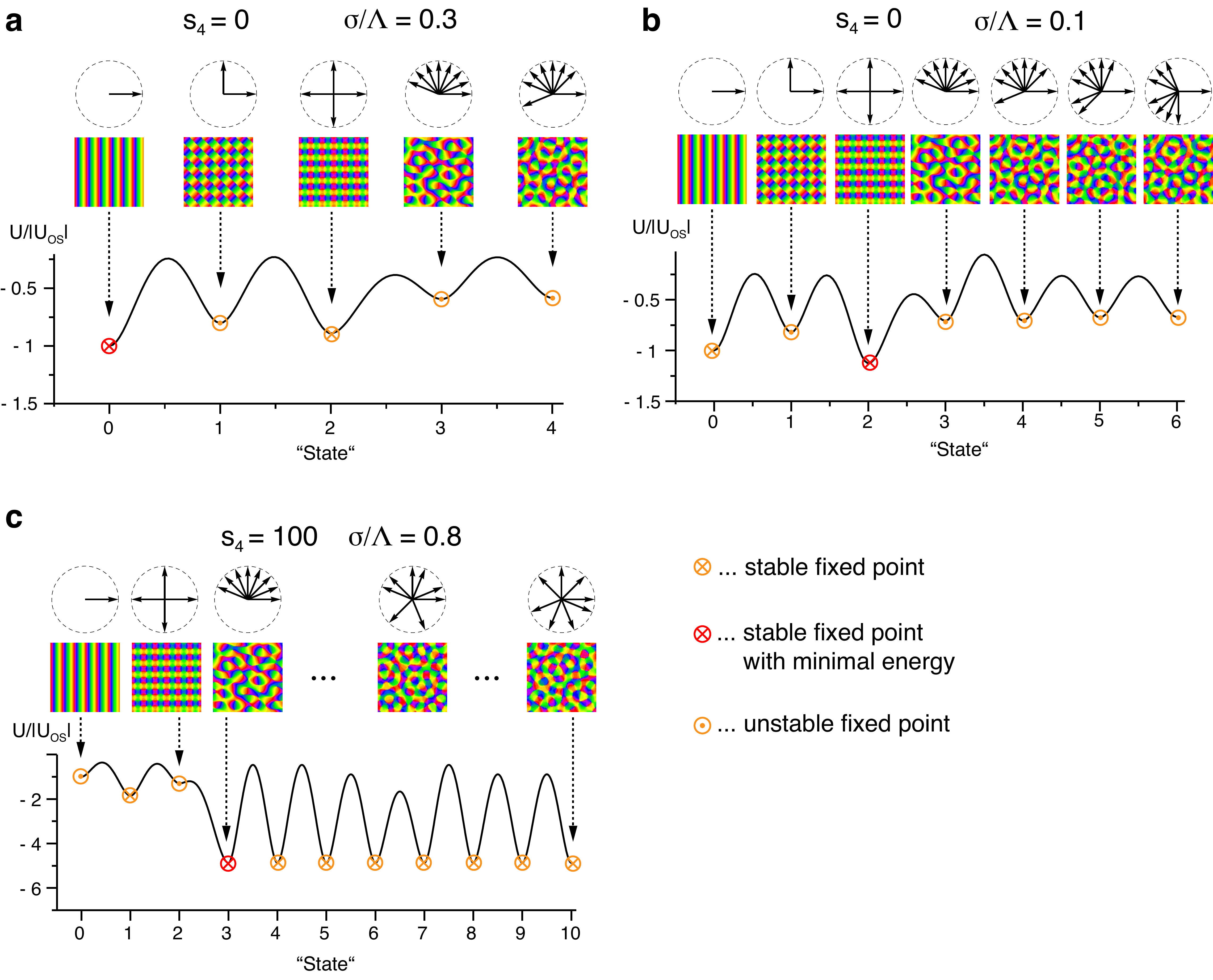}
\caption{\textbf{Illustration of the EN energy landscape close to pattern
formation threshold.}
The variation of the energy between states of
ideal orientation stripes (OS), the 2-ECP state, square pinwheel
crystals (sPWC) and possible mode configurations for 8-ECPs is shown
for the case that (\textbf{a}) the OS state has the lowest energy
($s_{4}=0$, $\sigma/\Lambda=0.3$), (\textbf{b}) the sPWC state has
the lowest energy ($s_{4}=0$, $\sigma/\Lambda=0.1$), and (\textbf{c})
the most anisotropic 8-ECP has lowest energy ($s_{4}=100$, $\sigma/\Lambda=0.8$). The energy values between
the state are computed from a state obtained by linear interpolation
between two neighboring states on the x-axis. Note that not all local
minima in $ $a-c correspond to a stable fixed point of the amplitude
dynamics (see text). 
\label{Keil_Wolf_figure_21}}
\end{figure}

As for many optimization problems in biology, the optimization of visual cortical functional architecture has been considered a problem characterized by a rugged energy landscape \cite{ritter_92}. In case of the EN model the expectation of a rugged energy landscape at first sight seems quite plausible. 
Originally, the Elastic Network algorithm was invented as a fast analogue method to approximately solve NP-hard problems in combinatorial optimization
such as the traveling salesman problem (TSP) \cite{durbin_87, Applegate2006}. In the TSP, the stimulus positions correspond to the locations of cities a salesman has to visit on the shortest possible tour. In problems such as the TSP, the energy functionals to be minimized are known to possess many local minima and the global minimization of these functionals generally represents an extremely difficult problem \cite{Applegate2006}.  Our analysis reveals that the trade-off between coverage and continuity for the mapping of a continuous feature space manifold leads to a much simpler structure of the energy landscape. This is also indicated by the fact that almost all of our gradient descent dynamics simulations readily converged to the predicted global minimum of the energy functional. Fig. \ref{Keil_Wolf_figure_21} illustrates the smooth structure of the EN energy landscape close to pattern formation threshold for different model parameters for a one dimensional path through the state space. In this landscape, the small set of stable planforms correspond to local minima of the EN energy functional, and unstable planforms to saddle points in the energy landscape. The optimal states correspond to global minima. Note that along the depicted state space path, unstable stationary solutions may appear as local minima if the unstable directions along which the energy decreases are orthogonal to the path.

What is the origin of this qualitative difference in the shape of the energy landscapes? In the traveling salesman problem, the finite repertoire of possible tours consists of all permutations of the N cities that the salesman has to visit. Via self-organized competition between the aim to visit all cities and the aim to minimize the path length, the elastic network algorithm converges to a specific ordering of the cities that eventually yields a very short tour. Most likely, the qualitative difference to the EN model for visual cortical map architecture originates from the transition from a finite number of cities to a continuum. When the elastic network algorithm is considered with an ensemble of cities (or stimuli) distributed according to a continuous probability density function, there is no discrete repertoire of tours. Both, the repertoire of tours as well as the path through the  landscape of cities or equivalently the space of visual stimulus features are determined by self-organization. The first is generated by the symmetry breaking mechanism that leads to the instability of the homogeneous state. The second corresponds to the selection of one of the many nontrivial stable steady states. 

An interesting property of the EN model dynamics that can be inferred from the energy landscape depicted in Fig. \ref{Keil_Wolf_figure_21} is the type of competition between two stable stationary states, where both are present in the system with a wall or a domain boundary
between them. The motion of the wall or domain boundary is predicted to proceed in the direction that increases the fraction of the pattern with lower energy. An example of such competition can be seen in Fig. \ref{Appendix_Keil_Wolf_figure_2}g. At $t=100\tau$, a small domain with an sPWC state is present. The area of this region is gradually reduced over the time course of the simulation until the pinwheel-free optimal state is reached.
\subsection*{Are simple OPM layouts an artifact of model simplicity?}
The perfectly periodic types of stationary solutions (stripes, crystals) that appear to dominate the classical EN model for retinotopy and orientation have been found in other models of visual cortical layouts that are relatively abstract. One might therefore suspect that they represent a mere artifact of model simplicity.  One conceptually appealing approach where perfectly periodic layouts have been found is wiring-length minimization \cite{Koulakov:2001p842}. According to this hypothesis, the structure of an OPM can be understood by minimizing the total length of dendritic and axonal processes. Maps obtained by stopping minimization of wire length exhibited qualitatively realistic layouts (see Fig. 6 in \cite{Koulakov:2001p842}). Complete optimization, however, leads to either stripe-like pinwheel-free patterns or rhombic pinwheel crystals, identical to the ones obtained in our investigation of optimal solutions of the EN model \cite{Koulakov:2001p842}.  Similarly, stripe-like and rhombic optima have been found in several abstract vector-field approaches for OPM development \cite{Lee:2003p5834, Thomas:2004p6132,Mayer2007:p150}.

It is ruled out by two observations, that the crystalline and perfectly periodic optima observed in all four optimization models, the EN model, the SOFM, the wiring-length minimization model, and the vector-field models are a mere artifact of the abstract order parameter field description of cortical selectivity patterns that is common to these approaches. Firstly, equally simplistic order parameter models for OPM development with long-range interactions have been shown to reproduce spatially irregular map layouts \cite{Wolf:2005p190,Kaschube2008, Kaschube_2010}. The occurrence of periodic optimal solutions is thus not a necessity in this model class. Secondly, pinwheel crystallization has also been observed in detailed network models for the development of orientation preference maps, notably in the first ever model for the self-organization of orientation selectivity by von der Malsburg in 1973 \cite{vonderMalsburg:1973p6457, GrabskaBarwinska:2008p6456}. Thus, on the one hand the phenomenon of pinwheel crystallization is thus not restricted to simple order parameter models and on the other hand abstract and mathematically relatively simple models can exhibit complex and biologically realistic optimal solutions.
\subsection*{Map rearrangement and layout optimization}
Irrespective of the optimization principle invoked to describe the structure of visual cortical maps, several common features of the resulting dynamics have been observed. The dynamics of optimization models usually starts with a phase of pattern emergence, where selectivity to visual features arises from an initially homogeneous unselective or weakly selective state. As we and others have shown, feature maps in these models continue to evolve after single cell selectivities reach mature levels. In fact, the phase of initial pattern emergence is typically followed by a prolonged phase of rearrangement of selectivities and preferences until a stable configuration is reached that represents a genuine optimum. This is not an exceptional type of dynamics but rather constitutes the generic expectation for a spatially extended system \cite{Cross:1993p922,manneville_90}.

What drives the second phase of map rearrangement? The initial emergence of feature selectivity is predominantly a local process in which merely neighboring units interact with each other to roughly match their selectivities. In the resulting spatial layout, selectivities are therefore far from being optimally arranged in space with respect to the global organization of selectivities on larger scales. Depending on the interactions incorporated in the model, local matching processes may (i) effectively propagate through space optimizing the pattern over gradually increasing spatial scales or (ii) distant sites may start to directly interact with each other to guide a rearrangement towards a globally optimized pattern after their initial emergence.

An illustrative example is provided by the emergence of pinwheel-sparse orientation stripes. Qualitatively, it is easy to see that a pattern of orientation stripes satisfies the continuity constraint very well. In a stripe pattern, preferred orientations are constant along one direction in space, realizing the absolute minimum of the orientation gradient in this direction. 
Reaching such a configuration obviously requires to select the preferred orientation at widely separated sites (along the stripe axis) to be identical. Because initially such sites develop independent preferred orientations, the optimized column layout can only emerge through a secondary rearrangement process. If the dominant low energy state has low pinwheel density, the later phase is governed by pinwheel motion and pairwise pinwheel annihilation. If this state is pinwheel-rich, e.g. a pinwheel crystal or an aperiodic pinwheel-rich state, both pinwheel annihilation and pinwheel creation together with a coordinated rearrangement of pinwheels are expected to occur. 

The local, essentially random processes during the initial emergence of a first pattern are in principle incapable of directly generating an optimized layout. In fact, it has been established that this initial so-called symmetry breaking phase will in general produce a random arrangement of selectivities of model-insensitive statistics \cite{schnabel_2007, Wolf:2003p210, Wolf:1998p1199}. The occurrence of some form of secondary reorganization is thus a qualitative prediction of any optimization model, provided that the optimal map is not seeded by an innate mechanism. The results presented in this study and many reports demonstrate that Hebbian plasticity is capable and often expected to achieve such rearrangements.

In gradient descent dynamics simulations of the EN model for retinotopy and stimulus orientation with conventional stimulus ensembles, pinwheel densities were found to be strongly time-dependent after the initial column formation (see e.g. Figs. \ref{Keil_Wolf_figure_6}, \ref{Keil_Wolf_figure_7}). In particular, the timescale for the establishment of full orientation selectivity and the time needed for either annihilation of a substantial fraction of pinwheels or their crystallization into periodic pinwheel crystals are in the same range of tens of tau. A similar time dependence of pinwheel density has also been observed in other models for OPM development with periodic optima \cite{Wolf:2005p190, Kaschube_2010}. Pinwheel annihilation in the EN can be slightly slowed down by additional features such as retinotopy Figs. \ref{Keil_Wolf_figure_10}, \ref{Keil_Wolf_figure_17} or ocular dominance \cite{Wolf:1998p1199,Reichl2009:p208101} but not by orders of magnitude. For this reason, signatures of the periodic optima of a developmental dynamics become visible at rather early simulation stages. Long-term minimization is apparently not essential to express the main layout features of the global minimum.

Because the main features of the dominant optimal solutions become apparent immediately after orientation selectivity saturates it appears not easy to reproduce the species-independent map layout in models with periodic crystalline optima by pattern freezing. In our simulations to match even only the pinwheel density, a very precise timing of the freezing point would be required. There is currently no evidence for such a freezing mechanism in early development.  In cats and ferrets, cortical maps for ocular dominance, orientation or direction arise on a timescale between hours and a few days (e.g. \cite{chapman_96, Crowley:2000p1199,White:2007p220}). The underlying circuits can be rapidly modified, e.g. by deprivation experiments, even on the timescale of hours \cite{Blakemore:1974p237, Antonini:1993p220} weeks after full selectivity has been established. 
Recently, evidence for long-term visual cortical circuit reorganization after the emergence of feature selectivity during normal development has emerged in diverse systems. In mouse, for example, activity-dependent changes induced by normal visual experience during the critical period, i.e. long after the primary emergence of orientation selectivity, have been shown to gradually match eye-specific inputs in the cortex \cite{wang_10}. Specifically, the data from mouse indicates that preferred orientations in the two eyes initially often emerge unmatched and subsequently change towards one binocularly matched orientation preference. Because preferred orientations in the two eyes initially are statistically independent, this suggests that neurons can rotate their orientation preferences up to at least 45${}^\circ$ during postnatal development. This is reminiscent of pairing experiments in kitten visual cortex in which Fr\'egnac and coworkers induced neurons to changed their preferred orientation by up to 90${}^\circ$ after pairing of a visual stimulus with intracortical stimulation \cite{Fregnac1988:p367,Fregnac1992:p1280} (see also \cite{Schuett2001:p325}). 
Also in the cat, visual cortical orientation columns in visual areas V1 and V2 have been found to undergo rearrangement during the late phase of the critical period \cite{Kaschube:2009p6537}. In this process, columns in mutually connected regions of areas V1 and V2 or in retinotopically matched regions in left and right hemisphere areas become progressively better matched in size. In the same species, a systematic reorganization of ocular dominance columns during postnatal development has been observed \cite{Keil:2010p6536}. Essential features of this columnar rearrangement are reproduced by the EN model for ocular dominance patterns simulated in a growing domain. 

In view of these observations, it seems unlikely that aperiodic orientation maps in the visual cortex represent frozen transient states of a developmental dynamics whose attracting layouts are pinwheel crystals or pinwheel free states. In fact, models for the activity-dependent development of OPMs with aperiodic optima predict only subtle changes of the OPM layout during the convergence after the establishment of selectivity \cite{Wolf:2005p190, Kaschube_2010}. This might also explain the apparent stability of cortical maps during normal development over short periods \cite{chapman_96}.
Further studies of the long-term rearrangement and stabilization of cortical functional architecture are needed to exhaustively characterize such processes. Given the fundamental role of map reorganization for any optimization theory of visual cortical development, chronic imaging experiments tracking the spatial arrangement of feature selectivities in individual animals beyond the emergence of selectivity and through later developmental stages are expected to be highly informative about fundamental principles of visual cortical optimization.

%
\section*{Conclusions}
Together with recent progress on the quantitative characterization of cortical functional architecture \cite{Kaschube2008, Kaschube_2010, Keil:2010p6536}, the current study lays the foundation for a mathematically rigorous and biologically informative search for optimization principles that successfully explain the architecture of columnar contour representations in the primary visual cortex.  A mathematically controlled and quantitatively precise determination of the predictions of candidate optimization principles is demanded by accumulating evidence indicating that geometrical features of visual cortical representations are biologically laid down with a precision in the range of a few percent \cite{Kaschube:2002p429, Kaschube:2003p430, Kaschube_2010}. Such data is expected to substantially reduce the range of candidate optimization principles that are consistent with biological observations. In particular, for the principle that cortical orientation maps are designed to optimally compromise stimulus coverage and feature continuity, our analysis demonstrates that the classical EN model for orientation preference and retinotopy essentially fails at explaining the biologically observed architecture. Our finding that the EN model exhibits biologically realistic optima only in a limit in which point-like stimuli are represented by complex spatially extended activity patterns corroborates that large-scale interactions are essential for the stabilization of OPM layouts with realistic geometry \cite{wolf_04, Wolf:2005p190, Kaschube2008, schnabel_2011}. In the light of these results, principles for the optimal representation of entire visual scenes by extended cortical activity patterns appear as promising candidates for future studies (see also \cite{Kaschube_2010_SOM}). In fact, there is recent evidence that visual cortical activity becomes progressively better matched to the statistics of natural stimuli but not to simplistic artificial stimulus ensembles \cite{Berkes2011:p83}.  We expect the methods developed here to facilitate a comprehensive characterization of such candidate principles.
%
%
%
\pagebreak
\section*{Methods}
\subsection*{Expansion of Elastic Net equation}
In order to analytically calculate the approximate optimal dimension-reducing mappings in the EN model with fixed retinotopy, an expansion of the nonlinear EN OPM dynamics Eq. \eqref{eq:continuous_z_dynamics} up to third order around the unselective fixed point has to be derived. This expansion is briefly sketched in the following.

Eq. \eqref{eq:continuous_z_dynamics} with $\mathbf{r}(\mathbf{x})=\mathbf{0}$ is of the form
\[
\partial_{t}z(\mathbf{x},t)=\mathcal{N}_{\mathbf{x}}[z]+\eta\Delta z(\mathbf{x},t)\,,
\]
where $\mathcal{N}_{\mathbf{x}}[z] $ is a nonlinear functional of $z(\mathbf{\cdot},t)$, parametrized by the position $\mathbf{x}$.
Clearly, the diffusion term contains no nonlinear terms in $z(\mathbf{\cdot},t)$ and therefore third order terms of the dynamics $\partial_t z(\mathbf{x},t)$ exclusively stem from third order terms of the Volterra series expansion  of the functional $\mathcal{N}_{\mathbf{x}}[z]$ around the fixed point $z(\mathbf{x},t) \equiv 0$. By the symmetry Eq. \eqref{eq:Shift-Symmetry-1}, only third order contributions of the form $N_3[z,z,\bar{z}]$ are allowed, i.e.
\begin{equation*}
N_3[z,z,\bar{z}] =  \frac{1}{2} \iiint d^2 y \, d^2 w\, d^2 v\, \left. \frac{\delta ^3 \mathcal{N}_{\mathbf{x}}[z]}{\delta z(\mathbf{y})\delta z(\mathbf{w})\delta \bar{z} (\mathbf{v})}\right|_{z \equiv 0} z(\mathbf{y})z(\mathbf{w})\bar{z}(\mathbf{v})\,.
\end{equation*}
Collecting all the terms yields 
\begin{equation}
N_{3}[z,z,\bar{z}] =\sum_{j=1}^{11}a_{j}N_{3}^{j}[z]\,,
\label{eq:third-order-EN-z-dynamics}
\end{equation}
where \begin{eqnarray}
N_{3}^{1}[z] & = & \left|z(\mathbf{x})\right|^{2}z(\mathbf{x})\label{eq:cubic_z_nonlinearities}\\
N_{3}^{2}[z] & = & \left|z(\mathbf{x})\right|^{2}\int d^{2}y\, K_{2}(\mathbf{y}-\mathbf{x})z(\mathbf{y})\nonumber \\
N_{3}^{3}[z] & = & z(\mathbf{x})^{2}\int d^{2}y\, K_{2}(\mathbf{y}-\mathbf{x})\bar{z}(\mathbf{y})\nonumber \\
N_{3}^{4}[z] & = & z(\mathbf{x})\int d^{2}y\, K_{2}(\mathbf{y}-\mathbf{x})\left|z(\mathbf{y})\right|^{2}\nonumber \\
N_{3}^{5}[z] & = & \bar{z}(\mathbf{x})\int d^{2}y\, K_{2}(\mathbf{y}-\mathbf{x})z(\mathbf{y})^{2}\nonumber \\
N_{3}^{6}[z] & = & \int d^{2}y\, K_{2}(\mathbf{y}-\mathbf{x})\left|z(\mathbf{y})\right|^{2}z(\mathbf{y})\nonumber \\
N_{3}^{7}[z] & = & z(\mathbf{x})\iint d^{2}y\, d^{2}w\, K_{3}(\mathbf{y}-\mathbf{x},\,\mathbf{w}-\mathbf{x},\,\mathbf{y}-\mathbf{w})\bar{z}(\mathbf{w})z(\mathbf{y})\nonumber \\
N_{3}^{8}[z] & = & \bar{z}(\mathbf{x})\iint d^{2}y\, d^{2}w\, K_{3}(\mathbf{y}-\mathbf{x},\,\mathbf{w}-\mathbf{x},\,\mathbf{y}-\mathbf{w})z(\mathbf{w})z(\mathbf{y})\nonumber \\
N_{3}^{9}[z] & = & \iiint d^{2}y\, d^{2}w\, d^{2}v\, K_{4}(\mathbf{y}-\mathbf{x},\,\mathbf{w}-\mathbf{x},\,\mathbf{v}-\mathbf{x},\,\mathbf{y}-\mathbf{w},\,\mathbf{v}-\mathbf{w},\,\mathbf{y}-\mathbf{v})\bar{z}(\mathbf{v})z(\mathbf{w})z(\mathbf{y})\nonumber \\
N_{3}^{10}[z] & = & \iint d^{2}y\, d^{2}w\, K_{3}(\mathbf{y}-\mathbf{x},\,\mathbf{w}-\mathbf{x},\,\mathbf{y}-\mathbf{w})\left|z(\mathbf{w})\right|^{2}z(\mathbf{y})\nonumber \\
N_{3}^{11}[z] & = & \iint d^{2}y\, d^{2}w\, K_{3}(\mathbf{y}-\mathbf{x},\,\mathbf{w}-\mathbf{x},\,\mathbf{y}-\mathbf{w})z(\mathbf{w})^{2}\bar{z}(\mathbf{y})\nonumber 
\end{eqnarray}
 and
 \begin{eqnarray*}
 K_{2}(\mathbf{x}) & =&  e^{-\mathbf{x}^{2}/(4\sigma^{2})} \\
 K_{3}(\mathbf{x}_1,\mathbf{x}_2,\mathbf{x}_3)  & =  & e^{-(\mathbf{x}_1^{2}+\mathbf{x}_2^{2}+\mathbf{x}_3^{2})/(6\sigma^{2})} \\
 K_{4}(\mathbf{x}_1, \mathbf{x}_2, \mathbf{x}_3,\mathbf{x}_4,\mathbf{x}_5,\mathbf{x}_6)  & =  & e^{-(\mathbf{x}_1^{2}+\mathbf{x}_2^{2}+\mathbf{x}_3^{2}+\mathbf{x}_4^{2}+\mathbf{x}_5^{2}+\mathbf{x}_6^{2})/(8\sigma^{2})} \,.
 \end{eqnarray*}
The coefficients $a_{j}$ for various orientation stimulus ensembles are given in the Results section. 

\subsection*{Adiabatic Elimination of $\mathbf{r}(\mathbf{x},t)$}
In order to analytically calculate the approximate optimal dimension-reducing mappings in the EN model with variable retinotopy, an expansion of the nonlinear EN retinotopy and orientation map dynamics Eqs. (\ref{eq:continuous_z_dynamics}, \ref{eq:continous_r_dynamics}) up to third order around the nonselective fixed point has to be derived and retinotopic distortions have to be adiabatically eliminated. Both of these calculations are briefly sketched in the following.
Eq. \eqref{eq:continuous_z_dynamics} is of the form
\[
\partial_{t}z(\mathbf{x},t)=\mathcal{N}_{\mathbf{x}}[z, \mathbf{r}]+\eta\Delta z(\mathbf{x},t)\,,
\]
where $\mathcal{N}_{\mathbf{x}}[z, \mathbf{r}]$ is a nonlinear functional of $z(\mathbf{\cdot},t)$ and $\mathbf{r}(\mathbf{\cdot},t)$, parametrized by the position $\mathbf{x}$.
The diffusion term contains no nonlinear terms in $z(\mathbf{\cdot},t)$ and therefore third order terms of the dynamics of $z(\mathbf{x},t)$ exclusively stem from third order terms of the Volterra series expansion of the functional  
$\mathcal{N}_{\mathbf{x}}[z, \mathbf{r}]$ around the fixed point $\left\{z(\mathbf{x},t)\equiv 0, \,\mathbf{r}(\mathbf{x},t)\equiv 0 \right\}$. By the symmetry Eq. \eqref{eq:Shift-Symmetry-1}, only terms in form of a cubic operator $N_3[z,z,\bar{z}]$ and a quadratic operator $Q^z[\mathbf{r},z]$ are allowed when expanding up to third order. $N_3[z,z,\bar{z}]$ is given in Eq.  \eqref{eq:third-order-EN-z-dynamics}. $Q[z,\mathbf{r}]$ can be calculated via
\begin{eqnarray*}
Q^z[\mathbf{r},z]&= \iint d^2 y \, d^2 w\, \left(\left.  \frac{\delta ^2 }{\delta z(\mathbf{y})\delta r_1(\mathbf{w})}\mathcal{N}_{\mathbf{x}}[z,\mathbf{r}] \right|_{z \equiv 0, \mathbf{r} \equiv 0} r_1(\mathbf{w}) + \left. \frac{\delta ^2 }{\delta z(\mathbf{y})\delta r_2(\mathbf{w})}\mathcal{N}_{\mathbf{x}}[z,\mathbf{r}]\right|_{z \equiv 0, \mathbf{r} \equiv 0} r_2(\mathbf{w}) \right) z(\mathbf{y})
\end{eqnarray*}
and this yields
\begin{eqnarray*}
Q^{z}[\mathbf{r},z] & = & \frac{\left(\left<|s_{z}|^{2}\right>-2\sigma^{2}\right)}{16\pi\sigma^{6}}z(\mathbf{x})\int d^{2}y\,\left<\mathbf{r}(\mathbf{y}),\mathbf{K}_{2}^{r}(\mathbf{y}-\mathbf{x})\right>\\
 &  & -\frac{\left<|s_{z}|^{2}\right>}{16\pi\sigma^{6}}\int d^{2}y\,\left<\mathbf{r}(\mathbf{x}),\, z(\mathbf{y})\mathbf{K}_{2}^{r}(\mathbf{y}-\mathbf{x})\right>+\frac{\left<|s_{z}|^{2}\right>}{16\pi\sigma^{6}}\int d^{2}y\,\left<\mathbf{r}(\mathbf{y}),z(\mathbf{y})\mathbf{K}_{2}^{r}(\mathbf{y}-\mathbf{x})\right>\\
 &  & -\frac{\left<|s_{z}|^{2}\right>}{36\pi^{2}\sigma^{8}}\iint d^{2}y\, d^{2}w\, z(\mathbf{y})\left<\mathbf{r}(\mathbf{w}),\mathbf{K}_{3}^{r}(\mathbf{y}-\mathbf{x},\mathbf{w}-\mathbf{x},\mathbf{y}-\mathbf{w})\right>\,,
 \end{eqnarray*}
where $\left<\cdot,\cdot\right>$ denotes the scalar product between two vectors and 
\begin{eqnarray}
\mathbf{K}_{2}^{r}(\mathbf{x})&=& e^{-\mathbf{x}^{2}/4\sigma^{2}}\mathbf{x}\\
\mathbf{K}_{3}^{r}(\mathbf{x}_1,\mathbf{x}_2,\mathbf{x}_3)&=& e^{-\frac{\mathbf{x}_1^{2}+\mathbf{x}_2^{2}+\mathbf{x}_3^{2}}{6\sigma^{2}}}\left[\mathbf{x}_1+\mathbf{x}_3\right]\,.
\end{eqnarray}
In complete analogy, by expanding the right hand side of the dynamical equation for the retinotopic distortions (Eq. \eqref{eq:continous_r_dynamics}) up to second order, the vector-valued quadratic operator $\mathbf{Q}^{r}[z,\bar{z}]$ can be obtained as
\begin{eqnarray}
\label{eq:Quadratic_operator_r_equation}
\mathbf{Q}^{r}[z,\bar{z}]&=&-\Re\left(\frac{\left<|s_{z}|^{2}\right>}{16\pi\sigma^{6}}\bar{z}(\mathbf{x})\int d^{2}y\,\mathbf{K}_{2}^{r}(\mathbf{y}-\mathbf{x})z(\mathbf{y})\right)\nonumber\\
& & +\frac{2\sigma^{2}-\left<|s_{z}|^{2}\right>}{32\pi\sigma^{6}}\int d^{2}y\,\mathbf{K}_{2}^{r}(\mathbf{y}-\mathbf{x})|z(\mathbf{y})|^{2}\nonumber\\
& & +\frac{\left<|s_{z}|^{2}\right>}{72\pi^{2}\sigma^{8}}\iint d^{2}y\, d^{2}w\,\mathbf{K}_{3}^{r}(\mathbf{y}-\mathbf{x},\mathbf{y}-\mathbf{w},\mathbf{w}-\mathbf{x})z(\mathbf{w})\bar{z}(\mathbf{y})\,.
\end{eqnarray}
Inserting $\mathbf{r}(\mathbf{x})=-\mathbf{L}_{r}^{-1}[\mathbf{Q}^{r}[z,\bar{z}]]$
into $Q^{z}[\mathbf{r},z]$ and using the linearity of $\mathbf{L}_{r}^{-1}$
as well as the bilinearity of both, $\mathbf{Q}^{r}[z,\bar{z}]$ and
$Q^{z}[\mathbf{r},z]$, yields a sum $N_{3}^{r}[z,z,\bar{z}]=\sum_{j=1}^{12}a_{r}^{j}N_{r}^{j}$,
with
\begin{eqnarray*}
N_{r}^{1} & = & z(\mathbf{x})\int d^{2}y\,\left<\mathbf{L}_{r}^{-1}\left[\Re\left(\bar{z}(\mathbf{y})\int d^{2}w\, K_{2}(\mathbf{w}-\mathbf{y})z(\mathbf{w})\right)\right],\,\mathbf{K}_{2}^{r}(\mathbf{y}-\mathbf{x})\right>\\
N_{r}^{2} & = & z(\mathbf{x})\int d^{2}y\,\left<\mathbf{L}_{r}^{-1}\left[\int d^{2}w\, K_{2}(\mathbf{w}-\mathbf{y})|z(\mathbf{w})|^{2}\right],\mathbf{K}_{2}^{r}(\mathbf{y}-\mathbf{x})\right>\\
N_{r}^{3} & = & z(\mathbf{x})\int d^{2}y\,\left<\mathbf{L}_{r}^{-1}\left[\iint d^{2}w\, d^{2}v\,\mathbf{K}_{3}^{r}(\mathbf{w}-\mathbf{y},\mathbf{w}-\mathbf{v},\mathbf{v}-\mathbf{y})z(\mathbf{w})\bar{z}(\mathbf{v})\right],\mathbf{K}_{2}^{r}(\mathbf{y}-\mathbf{x})\right>\\
N_{r}^{4} & = & \left<\mathbf{L}_{r}^{-1}\left[\Re\left(\bar{z}(\mathbf{x})\int d^{2}y\,\mathbf{K}_{2}^{r}(\mathbf{y}-\mathbf{x})z(\mathbf{y})\right)\right],\,\int d^{2}y\, z(\mathbf{y})\mathbf{K}_{2}^{r}(\mathbf{y}-\mathbf{x})\right>\\
N_{r}^{5} & = & \left<\mathbf{L}_{r}^{-1}\left[\int d^{2}y\,\mathbf{K}_{2}^{r}(\mathbf{y}-\mathbf{x})|z(\mathbf{y})|^{2}\right],\,\int d^{2}y\, z(\mathbf{y})\mathbf{K}_{2}^{r}(\mathbf{y}-\mathbf{x})\right>\\
N_{r}^{6} & = & \left<\mathbf{L}_{r}^{-1}\left[\iint d^{2}y\, d^{2}w\,\mathbf{K}_{3}^{r}(\mathbf{y}-\mathbf{x},\mathbf{y}-\mathbf{w},\mathbf{w}-\mathbf{x})z(\mathbf{w})\bar{z}(\mathbf{y})\right],\,\int d^{2}y\, z(\mathbf{y})\mathbf{K}_{2}^{r}(\mathbf{y}-\mathbf{x})\right>\\
N_{r}^{7} & = & \int d^{2}y\, z(\mathbf{y})\left<\mathbf{L}_{r}^{-1}\left[\Re\left(\bar{z}(\mathbf{y})\int d^{2}w\, K_{2}(\mathbf{w}-\mathbf{y})z(\mathbf{w})\right)\right],\mathbf{K}_{2}^{r}(\mathbf{y}-\mathbf{x})\right>\\
N_{r}^{8} & = & \int d^{2}y\, z(\mathbf{y})\left<\mathbf{L}_{r}^{-1}\left[\int d^{2}w\,\mathbf{K}_{2}^{r}(\mathbf{w}-\mathbf{y})|z(\mathbf{w})|^{2}\right],\mathbf{K}_{2}^{r}(\mathbf{y}-\mathbf{x})\right>\\
N_{r}^{9} & = & \int d^{2}y\, z(\mathbf{y})\left<\mathbf{L}_{r}^{-1}\left[\iint d^{2}v\, d^{2}w\,\mathbf{K}_{3}^{r}(\mathbf{w}-\mathbf{y},\mathbf{w}-\mathbf{v},\mathbf{v}-\mathbf{y})z(\mathbf{w})\bar{z}(\mathbf{y})\right],\mathbf{K}_{2}^{r}(\mathbf{y}-\mathbf{x})\right>\\
N_{r}^{10} & = & \iint d^{2}y\, d^{2}w\, z(\mathbf{y})\left<\mathbf{L}_{r}^{-1}\left[\Re\left(\bar{z}(\mathbf{w})\int d^{2}v\,\mathbf{K}_{2}^{r}(\mathbf{v}-\mathbf{w})z(\mathbf{v})\right)\right],\mathbf{K}_{3}^{r}(\mathbf{y}-\mathbf{x},\mathbf{y}-\mathbf{w},\mathbf{w}-\mathbf{x})\right>\\
N_{r}^{11} & = & \iint d^{2}y\, d^{2}w\, z(\mathbf{y})\left<\mathbf{L}_{r}^{-1}\left[\int d^{2}v\,\mathbf{K}_{2}^{r}(\mathbf{v}-\mathbf{w})|z(\mathbf{v})|^{2}\right],\mathbf{K}_{3}^{r}(\mathbf{y}-\mathbf{x},\mathbf{y}-\mathbf{w},\mathbf{w}-\mathbf{x})\right>\\
N_{r}^{12} & = & \iint d^{2}y\, d^{2}w\, z(\mathbf{y})\left<\mathbf{L}_{r}^{-1}\left[\iint d^{2}v\, d^{2}u\,\mathbf{K}_{3}^{r}(\mathbf{v}-\mathbf{w},\mathbf{v}-\mathbf{u},\mathbf{u}-\mathbf{w})z(\mathbf{v})\bar{z}(\mathbf{u})\right],\mathbf{K}_{3}^{r}(\mathbf{y}-\mathbf{x},\mathbf{y}-\mathbf{w},\mathbf{w}-\mathbf{y})\right>\,.
\end{eqnarray*}
The coefficients $a_r^j$ are given by
\begin{eqnarray*}
a_{r}^{1} & = & -\frac{\left(\left<|s_{z}|^{2}\right>-2\sigma^{2}\right)\left<|s_{z}|^{2}\right>}{256\pi^{2}\sigma^{12}} = - \frac{\left(1- \sigma^{2}\right)}{64\pi^{2}\sigma^{12}} \\
a_{r}^{2} & = & \frac{\left(\left<|s_{z}|^{2}\right>-2\sigma^{2}\right)^{2}}{512\pi^{2}\sigma^{12}} = \frac{\left(1-\sigma^{2}\right)^{2}}{128\pi^{2}\sigma^{12}} \\
a_{r}^{3} & = & \frac{\left(\left<|s_{z}|^{2}\right>-2\sigma^{2}\right)\left<|s_{z}|^{2}\right>}{1152\pi^{3}\sigma^{14}} =  \frac{\left(1-\sigma^{2}\right)}{288\pi^{3}\sigma^{14}} \\
a_{r}^{4} & = & -\frac{\left<|s_{z}|^{2}\right>^{2}}{256\pi^{2}\sigma^{12}} = -\frac{1}{64\pi^{2}\sigma^{12}}\\
a_{r}^{5} & = & -\frac{\left(\left<|s_{z}|^{2}\right> - 2\sigma^{2}\right)\left<|s_{z}|^{2}\right>}{512\pi^{2}\sigma^{12}} = -\frac{\left(1- \sigma^{2}\right)}{128\pi^{2}\sigma^{12}}\\
a_{r}^{6} & = & \frac{\left<|s_{z}|^{2}\right>^{2}}{1152\pi^{3}\sigma^{14}} = \frac{1}{288\pi^{3}\sigma^{14}}\\
a_{r}^{7} & = & -\frac{\left<|s_{z}|^{2}\right>^{2}}{256\pi^{2}\sigma^{12}} = -\frac{1}{64\pi^{2}\sigma^{12}}\\
a_{r}^{8} & = & -\frac{\left(\left<|s_{z}|^{2}\right>-2\sigma^{2}\right)\left<|s_{z}|^{2}\right>}{512\pi^{2}\sigma^{12}} = -\frac{\left(1 - \sigma^{2}\right)}{128\pi^{2}\sigma^{12}}\\
a_{r}^{9} & = & \frac{\left<|s_{z}|^{2}\right>^{2}}{1152\pi^{3}\sigma^{14}} = \frac{1}{288\pi^{3}\sigma^{14}}\\
a_{r}^{10} & = & \frac{\left<|s_{z}|^{2}\right>^{2}}{576\pi^{3}\sigma^{14}} = \frac{1}{144\pi^{3}\sigma^{14}}\\
a_{r}^{11} & = & \frac{\left(\left<|s_{z}|^{2}\right>-2\sigma^{2}\right)\left<|s_{z}|^{2}\right>}{1152\pi^{3}\sigma^{14}} = \frac{\left(1-\sigma^{2}\right)}{288\pi^{3}\sigma^{14}}\\
a_{r}^{12} & = & -\frac{\left<|s_{z}|^{2}\right>^{2}}{2592\pi^{4}\sigma^{16}} =  -\frac{1}{648\pi^{4}\sigma^{16}}\,,
\end{eqnarray*}
where the second equal sign is valid for $\left<|s_{z}|^{2}\right>=2$. 
\subsection*{Amplitude Equations from $N_{3}^{z}[z,z,\bar{z}]$}
We catalog the numerous stationary solutions
of Eq. \eqref{eq:third-order-approx-z-dynamics} following \cite{Wolf:2005p190}, by considering
planforms
$$z(\mathbf{x},t)=\sum_{j=1}^{N}A_{j}(t)e^{i\mathbf{k}_{j}\mathbf{x}}
$$
with an even number $N$ of modes with amplitudes $A_{j}$ and $\mathbf{k}_{j}=k_{c}(\cos(2\pi j/N),\sin(2\pi j/N))$.
In the vicinity of a finite wavelength instability - where the nonselective state $z(\mathbf{x})=0$
becomes unstable with respect to a band of Fourier modes around a finite wave number $k_{c}$ - by symmetry, the dynamics of the amplitudes $A_{j}$ at threshold
has the form
\begin{equation}
\dot{A}_{i}=A_{i}-A_{i}\sum_{j=1}^{N}g_{ij}|A_{j}|^{2}-\sum_{j=1}^{N}f_{ij}A_{j}A_{j^{-}}\bar{A}_{i^{-}}\,,\label{eq:amplitude_equations}
\end{equation}
where $j^{-}$ denotes the index of the mode antiparallel to the mode
$j$, $\mathbf{k}_{j}=-\mathbf{k}_{j^{-}}$, and the coefficients
$g_{ij}=(1-\frac{1}{2}\delta_{ij})g(|\alpha_{i}-\alpha_{j}|)$
and $f_{ij}=(1-\delta_{ij}-\delta_{i^{-}j})f(|\alpha_{i}-\alpha_{j}|)$
only depend on the angle $|\alpha_{i}-\alpha_{j}|$ between mode $i$
and $j$. The angle-dependent interaction functions $g(\alpha)$ and
$f(\alpha)$ are obtained from Eq. \eqref{eq:third-order-approx-z-dynamics}
by a multi scale expansion \cite{Cross:1993p922,manneville_90, wolf_04, Wolf:2005p190} as
\begin{eqnarray}
g(\alpha) & = & -e^{-i\mathbf{k}_{0}\mathbf{x}}\left[N_{3}^{z}(e^{i\mathbf{k}_{0}\mathbf{x}},e^{i\mathbf{h}(\alpha)\mathbf{x}},e^{-i\mathbf{h}(\alpha)\mathbf{x}})\right.\nonumber \\
 &  & + \left.N_{3}^{z}(e^{i\mathbf{h}(\alpha)\mathbf{x}},e^{-i\mathbf{h}(\alpha)\mathbf{x}},e^{i\mathbf{k}_{0}\mathbf{x}})\right]\label{eq:g_function_definition}\\
f(\alpha) & = & -\frac{1}{2}e^{-i\mathbf{k}_{0}\mathbf{x}}\left[N_{3}^{z}(e^{i\mathbf{h}(\alpha)\mathbf{x}},e^{-i\mathbf{h}(\alpha)\mathbf{x}},e^{i\mathbf{k}_{0}\mathbf{x}})\right.\nonumber \\
 &  & + \left.N_{3}^{z}(,e^{-i\mathbf{h}(\alpha)\mathbf{x}},e^{i\mathbf{h}(\alpha)\mathbf{x}},e^{i\mathbf{k}_{0}\mathbf{x}})\right]\,,\label{eq:f_function_definition}
 \end{eqnarray}
where $\mathbf{k}_{0}=k_{c}(1,0)$ and $\mathbf{h}(\alpha)=k_{c}(\cos\alpha,\sin\alpha)$.
$f(\alpha)$ is $\pi$-periodic, since the right hand side of Eq.
\eqref{eq:f_function_definition} is invariant with respect to the
transformation $\mathbf{h}(\alpha)\rightarrow\mathbf{h}(\alpha+\pi)=-\mathbf{h}(\alpha)$.
$g(\alpha)$ is $2\pi$-periodic in general. If, however, the nonlinearity
is permutation symmetric (Eq. \eqref{eq:permutation-symmetry}) it
can be seen from Eq. \eqref{eq:g_function_definition} that $g(\alpha)$ is $\pi$-periodic as well.
\subsection*{Stability of stationary planform solutions}
To determine the stability of fixed points of the amplitude equations Eq. \eqref{eq:amplitude_equations}, the eigenvalues
of their stability matrices have to be determined. In general, for any
fixed point $\mathbf{A}=\mathbf{A}^{0}$ of the dynamical system $\dot{\mathbf{A}}=\mathbf{F}(\mathbf{A})$ with complex-valued $\mathbf{A}$ and  
$\mathbf{F}$, we have to compute the eigenvalues of the Hermitian $2N\times2N$
matrix\[
\mathbf{M}=\left.\left(\begin{array}{cc}
\frac{\partial\mathbf{F}}{\partial\mathbf{A}} & \frac{\partial\mathbf{F}}{\partial\mathbf{\bar{A}}}\\
\frac{\partial\bar{\mathbf{F}}}{\partial\mathbf{A}} & \bar{\frac{\partial\mathbf{F}}{\partial\mathbf{A}}}\end{array}\right)\right|_{\mathbf{A}=\mathbf{A}_{0}}\,.
\]
For the system of amplitude equations, we obtain
\begin{eqnarray*}
\frac{\partial F_{i}}{\partial A_{k}} & = & r \delta_{ik}-\delta_{ik}\left(\sum_{j}^{N}g_{ij}|A_{j}|^{2}\right)-A_{i}g_{ik}\bar{A}_{k}-\bar{A}_{i^{-}}f_{ik}(A_{k^{-}}+A_{k})\\
\frac{\partial F_{i}}{\partial\bar{A}_{k}} & = & -A_{i}g_{ik}A_{k}-\delta_{i^{-}k}\left(\sum_{j}^{N}f_{ij}A_{j}A_{j^{-}}\right)\,.
\end{eqnarray*}
Stability of a solution, or more precise intrinsic stability is given, if all eigenvalues
of $\mathbf{M}$ are negative definite. Extrinsic stability is given, if the
growth of additional Fourier modes is suppressed. To test whether
a planform solution is extrinsically stable, we introduce a test mode
$T$ such that
\[
z(\mathbf{x})=Te^{i\mathbf{k}_{\beta}\mathbf{x}}+\sum_{j}^{N}A_{j}e^{i\mathbf{k}_{j}\mathbf{x}}\,,
\]
with $\mathbf{k}_{\beta}=(\cos\beta,\sin\beta)k_{c}$. We insert
this ansatz into Eq. \eqref{eq:general-amplitude-equations} and obtain
\[
\partial_{t}T= r T-\sum_{j}^{N}g(\beta-\beta_{j})|A_{j}|^{2}T+\mathcal{O}(T^{2})
\]
as the dynamics of the test mode $T$, where $g(\beta)$ is the angle-dependent
interaction function corresponding to $N_{3}[z,z,\bar{z}]$. For the
solution $T=0$ to be stable, we therefore obtain the condition
\[
r-\sum_{j}^{N}g(\beta-\beta_{j})|A_{j}|^2<0,\,\hspace{1em}\forall\alpha\in[0,2\pi]\,,
\]
where we assumed $\mathbf{k}_{\beta}\neq\mathbf{k}_{j},\mathbf{k}_{j}^{-}$.
These conditions for intrinsic and extrinsic stability were numerically evaluated to study the stability of n-ECPs and rPWCs.
\subsection*{Coupled Essentially Complex Planforms}
In the Results section, we presented a closed form expression
for the retinotopic distortions associated via Eq. \eqref{eq:slaved_retinotopy}
with stationary planform solutions of Eq. \eqref{eq:full_z_eq_insert_r}. Here,
we sketch how to explicitly calculate this representation. We start with the ansatz
\begin{equation}
z(\mathbf{x})=\sum_{j}^{n}A_{j}e^{i\mathbf{k}_{j}\mathbf{x}}\hspace{1em}\hspace{1em}|\mathbf{k}_{j}|=k_{c}\label{eq:planform_ansatz}
\end{equation}
for the orientation preference map $z(\mathbf{x})$. Note that this
general ansatz includes essentially complex planforms as well
as rhombic pinwheel crystals. To simplify notation, we denote the
individual terms in Eq. \eqref{eq:Quadratic_operator_r_equation}
\begin{eqnarray*}
\mathbf{Q}_{1}[z,\bar{z}] & = & -\Re\left(\frac{\left<|s_{z}|^{2}\right>}{16\pi\sigma^{6}}\bar{z}(\mathbf{x})\int d^{2}y\,\mathbf{K}_{2}^{r}(\mathbf{y}-\mathbf{x})z(\mathbf{y})\right)\\
\mathbf{Q}_{2}[z,\bar{z}] & = & \frac{2\sigma^{2}-\left<|s_{z}|^{2}\right>}{32\pi\sigma^{6}}\int d^{2}y\,\mathbf{K}_{2}^{r}(\mathbf{y}-\mathbf{x})|z(\mathbf{y})|^{2}\\
\mathbf{Q}_{3}[z,\bar{z}] & = & \frac{\left<|s_{z}|^{2}\right>}{72\pi^{2}\sigma^{8}}\iint d^{2}y\, d^{2}w\,\mathbf{K}_{3}^{r}(\mathbf{y}-\mathbf{x},\mathbf{y}-\mathbf{w},\mathbf{w}-\mathbf{x})z(\mathbf{w})\bar{z}(\mathbf{y})\,.\end{eqnarray*}
Each of the $\mathbf{Q}_{i}[z,\bar{z}],\,\, i=1,2,3$ can be evaluated for the ansatz Eq. \eqref{eq:planform_ansatz} and we
obtain
\begin{multline*}
\mathbf{Q}_{1}\left[\sum_{j}^{n}A_j e^{i\mathbf{k}_{j}\mathbf{x}}, \sum_{k}^{n} \bar{A}_k e^{-i\mathbf{k}_{k}\mathbf{x}}\right]  =  \frac{\left<|s_{z}|^{2}\right>e^{-k_{c}^{2}\sigma^{2}}}{2\sigma^{2}}\sum_{k, j<k}^{n}\left\lbrace\Re\left(A_{j}\bar{A}_{k}\right)(\mathbf{k}_{j}-\mathbf{k}_{k})\sin\left(\left(\mathbf{k}_{j}-\mathbf{k}_{k}\right)\mathbf{x}\right)\right.\\
 \left. +\Im\left(A_{j}\bar{A}_{k}\right)(\mathbf{k}_{j}-\mathbf{k}_{k})\cos\left(\left(\mathbf{k}_{j}-\mathbf{k}_{k}\right)\mathbf{x}\right)\right\rbrace
 \end{multline*}
\begin{multline*}
\mathbf{Q}_{2}\left[\sum_{j}^{n}A_j e^{i\mathbf{k}_{j}\mathbf{x}}, \sum_{k}^{n} \bar{A}_k e^{-i\mathbf{k}_{k}\mathbf{x}}\right] =  -\frac{(2\sigma^{2}-\left<|s_{z}|^{2}\right>)}{2\sigma^{2}}\sum_{j<k}^{n}e^{-(\mathbf{k}_{j}-\mathbf{k}_{k})^{2}\sigma^{2}}(\mathbf{k}_{j}-\mathbf{k}_{k})\left\lbrace \Re\left(A_{j}\bar{A}_{k}\right)\sin\left(\left(\mathbf{k}_{j}-\mathbf{k}_{k}\right)\mathbf{x}\right)\right.\\
 \left.+\Im\left(A_{j}\bar{A}_{k}\right)\cos\left(\left(\mathbf{k}_{j}-\mathbf{k}_{k}\right)\mathbf{x}\right)\right\rbrace 
\end{multline*}
\begin{multline*}
\mathbf{Q}_{3}\left[\sum_{j}^{n}A_j e^{i\mathbf{k}_{j}\mathbf{x}}, \sum_{k}^{n} \bar{A}_k e^{-i\mathbf{k}_{k}\mathbf{x}}\right]  =  -\frac{\left<|s_{z}|^{2}\right>e^{-\frac{k_{c}^{2}\sigma^{2}}{2}}}{\sigma^{2}}\sum_{j<k}^{n}(\mathbf{k}_{j}-\mathbf{k}_{k})e^{-\frac{(\mathbf{k}_{j}-\mathbf{k}_{k})^{2}\sigma^{2}}{2}}\left\lbrace \Re\left(A_{j}\bar{A}_{k}\right)\sin\left(\left(\mathbf{k}_{j}-\mathbf{k}_{k}\right)\mathbf{x}\right)\right.\\
  \left.+\Im\left(A_{j}\bar{A}_{k}\right)\cos\left(\left(\mathbf{k}_{j}-\mathbf{k}_{k}\right)\mathbf{x}\right)\right\rbrace \,.
 \end{multline*}
All resulting terms are proportional to either $(\mathbf{k}_{i}-\mathbf{k}_{j})\sin((\mathbf{k}_{i}-\mathbf{k}_{j})\mathbf{x})$
or $(\mathbf{k}_{i}-\mathbf{k}_{j})\cos((\mathbf{k}_{i}-\mathbf{k}_{j})\mathbf{x})\,,\,\, i\neq j$.
These functions are longitudinal modes (see Fig. \ref{Keil_Wolf_figure_3}b)
which have been identified as eigenfunctions of the linearized dynamics of retinotopic distortions $\mathbf{L}_{r}[\mathbf{r}]$
with eigenvalue \[
\lambda_{L}^{r}(|\mathbf{k}_{i}-\mathbf{k}_{j}|)=-|\mathbf{k}_{i}-\mathbf{k}_{j}|^{2}(\eta_{r}+e^{-\sigma^{2}|\mathbf{k}_{i}-\mathbf{k}_{j}|^{2}}\sigma^{2})\,.\]
Hence, they are eigenfunctions of $\mathbf{L}_{r}^{-1}[\mathbf{r}]$
with eigenvalue $1/\lambda_{L}^{r}(|\mathbf{k}_{i}-\mathbf{k}_{j}|)$.
Using this when inserting in Eq. \eqref{eq:slaved_retinotopy} and setting $\left<|s_{z}|^{2}\right >= 2$,
we obtain expression \eqref{eq:retinotopy_for_planforms} for the
retinotopic distortions belonging to an arbitrary planform.
\subsection*{Phase diagrams}
To compute the regions of minimal energy shown in Figs. \ref{Keil_Wolf_figure_6}, \ref{Keil_Wolf_figure_10}, \ref{Appendix_Keil_Wolf_figure_2},
\ref{Appendix_Keil_Wolf_figure_4} (Appendix), \ref{Keil_Wolf_figure_12}, \ref{Keil_Wolf_figure_15}, and \ref{Keil_Wolf_figure_16}, 
we first computed the fixed points of Eq. \eqref{eq:amplitude_equations} at each point
in parameter space. For n-ECPs, we constructed the coupling matrix
$\mathbf{g}$ in Eq. \eqref{eq:amplitudes_for_ECPs} for all mode configurations not related by any combination of the symmetry
operations: (i) Translation: $A_j\rightarrow A_j e^{i \mathbf{k}_j \mathbf{y}}$, (ii) Rotation: $A_j\rightarrow A_{j+\Delta j} $, (iii) Parity: $A_j\rightarrow \bar{A}_{(N-j)^-} $. 
Via Eq. \eqref{eq:amplitude-of-n-ECP}, we then computed
the absolute values of the corresponding amplitudes. If $\sum_{j=1}^{n}\left(\mathbf{g}^{-1}\right){}_{ij}\geq0$
for all $i$, a valid n-ECP fixed point of Eq. \eqref{eq:amplitude_equations}
was identified. Its energy was then computed via
Eq. \eqref{eq:energy-of-an-n-ECP}. For orientation stripes and rhombic pinwheel crystals, the derived analytical expressions for their energy
Eqs. (\ref{eq:energy-of-orientation-stripes}, \ref{eq:energy-of-rPWC}) were evaluated.  To analyze the stability of the fixed points, the conditions for intrinsic and extrinsic stability (see above) were numerically evaluated.
\subsection*{Numerical Procedures - Gradient descent simulations}
To test our analytical calculations and explore their range of validity, we simulated Eqs. \eqref{eq:continuous_z_dynamics}
and \eqref{eq:continous_r_dynamics} on a 64$\times$64 grid with
periodic boundary conditions. Simulated systems were spatially discretized
with typically 8 grid points per expected column spacing $\Lambda_{\textnormal{max}}$ of the orientation preference pattern (see Results section) to achieve
sufficient resolution in space. Test simulations with finer discretization
(16 and 32 grid points per $\Lambda_{\textnormal{max}}$) did not lead to
different results. Progression of time was measured in units of the
intrinsic timescale $\tau$ (see Results section) of the pattern formation process. The integration time step $\delta t$
was bounded by the relevant decay time constant of the Laplacian in
Eq. (\ref{eq:continuous_z_dynamics}) around $k_{c}$ and by the intrinsic
timescale $\tau$ of the system, using $\delta t=\min\left\{ 1/(20\eta k_{c}^{2}),\tau/10\right\} $.
This ensured good approximation to the temporally continuous changes
of the patterns. We used an Adams-Bashforth scheme for the first terms
on the respective r.h.s. of Eqs. (\ref{eq:continuous_z_dynamics},\ref{eq:continous_r_dynamics}). The second
terms (diffusion) were treated by spectral integration, exhibiting unconditional
numerical stability. The stimulus positions $\mathbf{s}_{r}$ were
chosen to be uniformly distributed in retinal coordinates.
The stimulus averages in Eqs. (\ref{eq:continuous_z_dynamics}, \ref{eq:continous_r_dynamics})
were approximated by choosing a random representative sample of $N_{s}$
stimuli at each integration time step, with
\[
N_{s}=\max\left\{10^5, \frac{N_{0}\Gamma^{2}}{(\varepsilon_{s})^{n}}\frac{\delta t}{\tau}\right\}\,,
\]
where $n$ corresponds to the dimensions of the feature space in addition
to the two retinal positions (in our case, $n=2$), $\Gamma^{2}=(L/\Lambda_{\textnormal{max}})^{2}$
the squared aspect ratio of the simulated system in units of $\Lambda^{2}$, $\varepsilon_{s}$
the resolution in feature space, $N_{0}$ the number of stimuli we
required to approximate the cumulative effect of the
ensemble of stimuli within each feature space voxel $\varepsilon^{n+2}$. With $N_{0}=100$
and $\varepsilon_{s}=0.05$, we ensured a high signal to noise ratio
for all the simulations. Typical values for $N_{s}$ were between
2.5x10$^{5}$ and 4x10$^{6}$. All simulations were initialized with
$z(\mathbf{x},t=0)=10^{-6}e^{i2\pi\xi(\mathbf{x})}$ and $\mathbf{r}(\mathbf{x},t=0)=\mathbf{0}$,
where the $\xi(\mathbf{x})$ are independent identically distributed random numbers uniformly in $[0,1]$. Different realizations were
obtained by using different stimulus samples.

Stimuli were drawn from different distributions, each with $\left<|s_{z}|^{2}\right>=2$.
We considered (i) stimuli uniformly distributed on a ring with $|s_{z}|^{2}=\sqrt{2}$
(circular stimulus ensemble) (ii) stimuli uniformly distributed within a circle $\left\{ s_{z},|s_{z}|\leq2\right\} $
(uniform stimulus ensemble) and  (iii) a Gaussian stimuli ensemble with $\rho_{s_{z}}=1/(2\pi)\exp(-|s_{z}|^{2}/2)$.
In addition, we considered mixtures of a high-fourth moment Pearson type VII distribution and the circular
stimulus ensemble. The Pearson distribution is given by
\[
\rho_{s_{z}}=\frac{1}{\alpha B(m-\frac{1}{2},\frac{1}{2})}\left[1+\frac{|s_{z}|^{2}}{\alpha^{2}}\right]^{-m}\,,
\]
where $B(\cdot,\cdot)$ is the Beta function \cite{abramowitz_stegun_64} and $\alpha=\sqrt{2m-3}$, and $m=\frac{5}{2}+\frac{12}{\gamma-12}$
such that $\left<|s_{z}|^{2}\right>=2$, $\left<|s_{z}|^{4}\right>=\gamma$
or equivalently $s_{4}=\gamma-4$.

In addition to simulations in which independent sets of stimuli were used for evaluating the stimulus average in Eqs. (\ref{eq:continuous_z_dynamics}, \ref{eq:continous_r_dynamics}) for every time step, we also performed simulations in which the same fixed set of N stimuli was used (see Results). 
To determine the time step $\delta t$ in these simulations, we first calculated \[
N_{\tau}=\frac{N_{0}\Gamma^{2}}{(\varepsilon_{s})^{n}}\]
(parameters as in regular simulations) which yields the number of
stimuli presented to the model in one intrinsic time unit $\tau$ in
regular simulations. To subject the network to the same number of stimuli per intrinsic time scale $\tau$ in fixed stimulus set simulations,
the integration time step $\delta t$ was in this case chosen as \[
\delta t=\min\left\{ \frac{N}{N_{s}}\tau,\frac{1}{20\eta k_{c}^{2}},\frac{\tau}{10}\right\} \,.\]
For small $N$, this resulted in very small integration steps. For
very large $N$, time steps were identical to the regular simulations.
Different realizations were obtained by different but fixed stimulus
sets. In all simulations with fixed stimulus sets, stimuli were drawn from the circular stimulus
ensemble. All other numerical methods were chosen as in regular simulations. 
%
%
%
\subsection*{Numerical Procedures - Solving the EN model with deterministic annealing}
A large body of previous work has solved the EN models for various aspects of visual cortical architecture for discrete fixed sets of stimuli and using deterministic annealing. We therefore also used deterministic annealing with fixed discrete sets of stimuli to solve the EN model for the most frequently used stimulus distribution. This allowed us to better compare our analytical and numerical results based on the gradient descent dynamics for a continuous stimulus with prior results. For the discrete deterministic annealing approach, cortical maps are described by a collection of M centroids $\{\mathbf{y}_{m}\}_{m=1}^{M}\subset\mathbb{R}^{d}$
that can be represented as a $D\times M$ matrix $\mathbf{Y}=(\mathbf{y}_{1},\dots,\mathbf{y}_{M})$. Maps forming a compromise of coverage and continuity are obtained for $\{\mathbf{x}_{n}\}_{n=1}^{N}\subset\mathbb{R}^{d}$ represented as a $D\times N$ matrix $\mathbf{X}=(\mathbf{x}_{1},\dots,\mathbf{x}_{M})$. In our case, $d=4$. 
The trade-off between coverage and continuity is formulated by the energy function
\begin{equation}
E(\mathbf{Y},\sigma)=-\alpha\sigma\sum_{n=1}^{N}\log\sum_{m=1}^{M}e^{-\frac{1}{2}\left\Vert \frac{\mathbf{x}_{n}-\mathbf{y}_{m}}{\sigma}\right\Vert ^{2}}+\frac{\beta}{2}\textnormal{tr}\left(\mathbf{Y}^{T}\mathbf{Y}\mathbf{S}\right)\,.\label{eq:EN_energy_discrete}
\end{equation}
The matrix $\mathbf{S}$ determines the topology of the network as well as the boundary conditions and is typically derived from a discretized derivative
based on a finite difference scheme or stencil (see below). 
For large N and M, the energy function Eq. (\ref{eq:EN_energy_discrete}) is equivalent to the energy functional of the continuum formulation Eq. \eqref{eq:Elastic-Net-Energy}  for $\beta=\eta N$ and S implementing the discretized Laplacian operator in two dimensions.

Following \cite{Goodhill:2000p2141, Cimponeriu:2000p5167, CarreiraPerpinan:2004p6297, CarreiraPerpinan:2005p6295, Giacomantonio_2010} we minimized the EN energy function Eq. (\ref{eq:Elastic-Net-Energy}) by an iterative deterministic annealing algorithm, starting with a minimization for large $\sigma$ and tracking this minimum to a small value of $\sigma$. As in [4], we reduced $\sigma$ from $\sigma_{\textnormal{init}}= 0.2$ to the point at which the amplitude of the orientation maps saturate ($\sigma\approx 0.03$), following $\sigma= \sigma_{\textnormal{init}} \times  \chi^j$ where $j$ counts the annealing step. This choice tracks stationary solutions of the EN to parameters that are very far from threshold. For high precision tracking of solutions, we used an annealing rate of $\chi=0.999$. 

At each value of $\sigma$, setting the gradient of Eq. (\ref{eq:EN_energy_discrete}) to zero yields a nonlinear system of equations
\begin{equation}
\mathbf{YA}=\mathbf{XW}\hspace{1em}\hspace{1em}\textnormal{with}\hspace{1em}\hspace{1em}\mathbf{A}=\mathbf{G}+\sigma \beta \left(\frac{\mathbf{S}+\mathbf{S}^{T}}{2}\right)\,,
\label{eq:Basic_EN_equation}
\end{equation}
Here, the $N\times M$-matrix $W$ is given by 
\[
w_{nm}=\frac{e^{-\frac{1}{2}\left\Vert \frac{\mathbf{x}_{n}-\mathbf{y}_{m}}{\sigma}\right\Vert ^{2}}}{\sum_{m'=1}^{M}e^{-\frac{1}{2}\left\Vert \frac{\mathbf{x}_{n}-\mathbf{y}_{m'}}{\sigma}\right\Vert ^{2}}}
\]
and the $M\times M$-matrix $G$ is 
\[
g_{ij}=\delta_{ij}\sum_{n=1}^{N}w_{ni}\,.
\]
$\mathbf{A}$ is a symmetric positive-definite $M\times M$ matrix.
The $M\times M$ matrix A is symmetric and positive-definite. Since both, G and W depend on Y, this equation is nonlinear in Y and has to be solved iteratively. Following \cite{Goodhill:2000p2141, Cimponeriu:2000p5167, CarreiraPerpinan:2004p6297, CarreiraPerpinan:2005p6295, Giacomantonio_2010}, we solved eq. (\ref{eq:Basic_EN_equation}) at each value of $\sigma$ and for each iteration directly via Cholesky-factorization.

We implemented periodic and non-periodic boundary conditions by appropriate choice of the matrix $\mathbf{S}$. $\mathbf{S}$ must be positive (semi)definite for the energy to be bounded from below. We used the canonical 2D Laplacian stencil of order 2, to construct the $M\times M$ matrix
\[
S=\left(\begin{array}{ccccccccccccc}
-4+2a & 1 & 0 & 0 & \cdots & 1 & 0 & 0 & \cdots & 0 & 0 & 0 & 1\\
1 & -4+a & 1 & 0 & 0 & \cdots & 1 & 0 & 0 & \cdots & 0 & 0 & 0\\
0 & 1 & -4+a & 1 & 0 & 0 & \cdots & 1 & \cdots\\
\vdots &  &  & \vdots\\
 &  &  & 1 & -4+2a & 1\\
 &  &  &  & 1 & -4+a & 1\\
0 & \cdots &  &  &  & 1 & -4 & 1 & 0 & \cdots & 1\\
0 &  &  &  &  &  & 1 & -4 & 1 & 0 & \cdots & 1\\
 &  &  & \vdots\\
 &  &  &  &  &  &  & 0 & 1 & -4+a & 1 & 0\\
 &  &  &  &  &  &  &  & 0 & 1 & -4+a & 1 & 0\\
1 &  &  &  &  &  &  &  &  &  &  & 1 & -4+2a
\end{array}\right)\,,
\]
Here, $a=0$ for periodic boundary conditions and $a=1$ for non-periodic boundary conditions. In Appendix II, we also present simulation results for a 4th derivative stencil, in which $S^2$ was used for the continuity term.
We used random stimulus positions and orientations as well as stimuli arranged on a grid in feature space. For random stimuli, positions were drawn from a uniform distribution in $[0,1] \times [0,1]$. Orientations $s_z$ were drawn from the circular stimulus ensemble, with $|s_z| = 0.08$ as in \cite{CarreiraPerpinan:2005p6295}. Stimuli from grid-like ensembles were distributed evenly-spaced in $[0,1] \times [0,1]$ and contained $2^k$ evenly space orientations with $|s_z| = 0.08$.

To compute the energy of pinwheel-free configuration, we initiated simulations with a stripe-like orientation preference pattern with the same typical spacing as the observed orientation maps and annealed from $\sigma = 0.035$ to $\sigma = 0.03$. 

To enable comparison between simulations with different numbers of stimuli, we scaled the continuity parameter proportionally to N such that the equivalent $\eta$ was approximately constant. The simulated domain then contained a comparable number of hyper columns for all stimulus numbers. %
\subsection*{Pinwheel density from simulations}
Pinwheels locations in models were identified by the crossings
of the zero contour lines of real and imaginary parts of the orientation
map. Estimation of local column spacing $\Lambda(\mathbf{x})$ was done using the wavelet analysis introduced in \cite{Kaschube:2002p429,Kaschube:2003p430}. 
In short, an overcomplete basis of complex Morlet wavelets at various scales and orientations was compared to
the OPM pattern at each location. $\Lambda(\mathbf{x})$ was
estimated by the scale of the best matching wavelet. The mean column
spacing $\left<\Lambda(\mathbf{x})\right>_{x}$ of a given map was then calculated
from the local column spacing by spatial averaging. For details we refer
to \cite{Kaschube:2002p429,Kaschube:2003p430, Kaschube_2010}. Given $N_{pw}$ pinwheels
in a simulated cortical area of size $L^{2}$, we defined the pinwheel
density \cite{Swindale:1996p6540,Wolf:1998p1199,Kaschube_2010} \[
\rho=N_{pw}\frac{\left<\Lambda\right>_{x}^{2}}{L^{2}}\,.\]
The pinwheel density $\rho$ is a dimensionless quantity and depends
only on the layout of orientation columns. The pinwheel density defined
in this way is large for patchy and small for more band-like columnar
layouts. 
%
%
%
%
%
%
\pagebreak
\section*{Appendix I}
\subsection*{The impact of non-oriented stimuli}
\begin{figure}[pb]
\centering
\includegraphics[width=9.5cm]{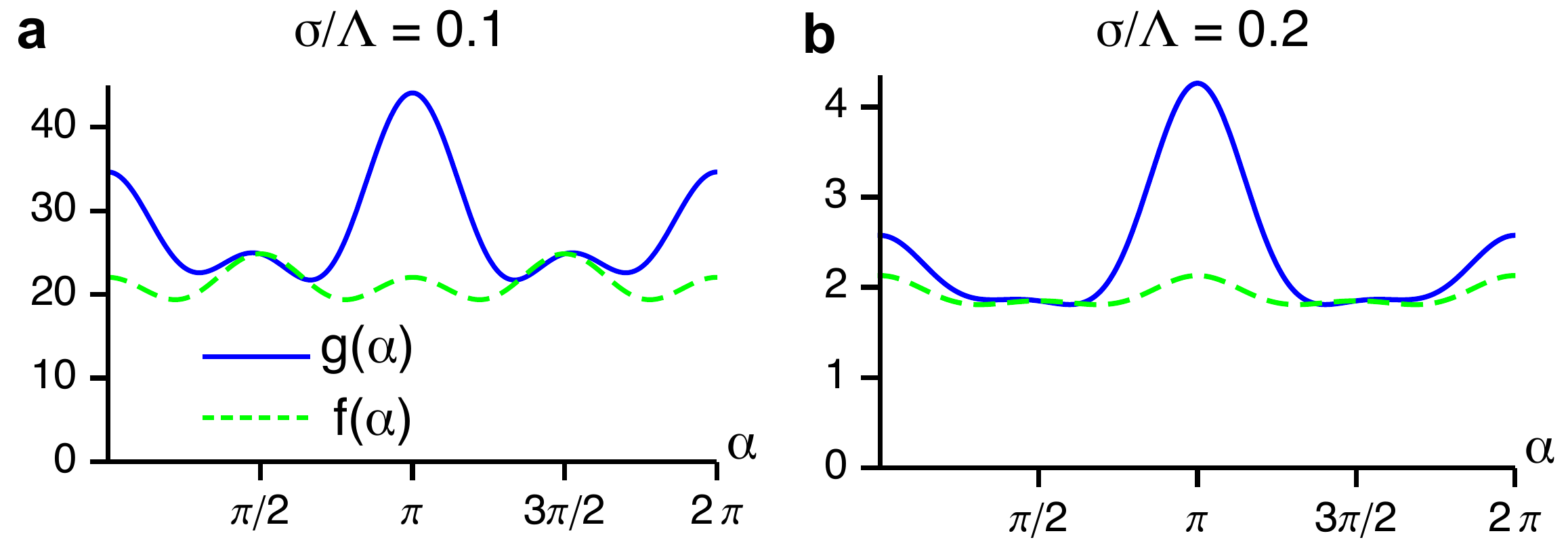}
\caption{\textbf{Angle-dependent interaction functions for the EN model with fixed
retinotopy and uniform orientation stimulus ensemble.}
(\textbf{a}, \textbf{b}) $g(\alpha)$ and $f(\alpha)$ for $\sigma/\Lambda=0.1$ (a) and $\sigma/\Lambda=0.2$ (b). 
\label{Appendix_Keil_Wolf_figure_1}}
\end{figure}
\begin{figure*}
\includegraphics[width=16cm]{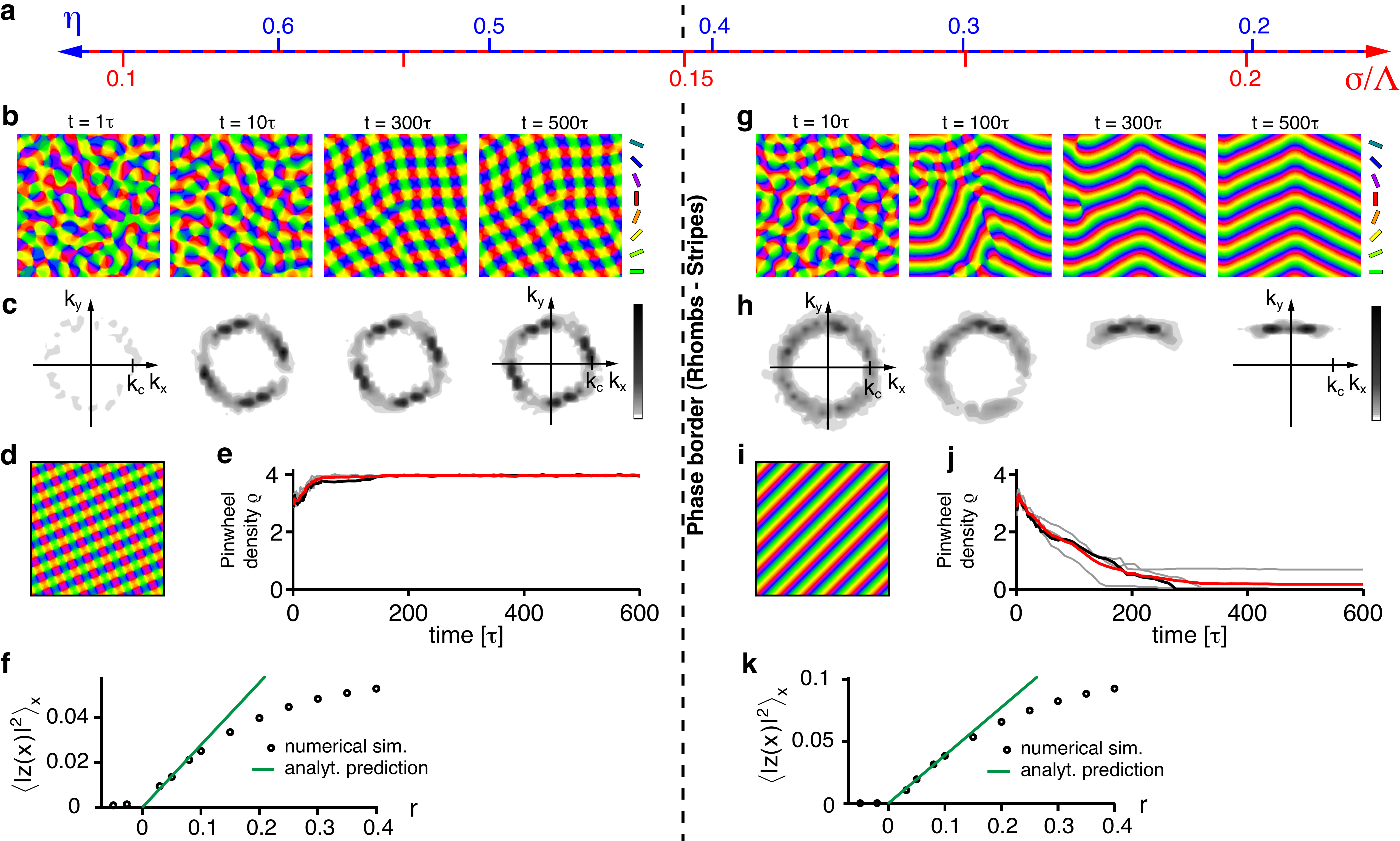}
\caption{\textbf{Optimal solutions of the EN model for uniform stimulus ensemble
and fixed representation of visual space.}
(\textbf{a}) At criticality, the phase space of this model is parameterized by either the continuity
parameter $\eta$ (blue labels) or the effective interaction range
$\sigma/\Lambda$ (red labels, see text).
(\textbf{b}-\textbf{c}) OPMs (b) and their power spectra (c)
in a simulation of Eq. \eqref{eq:continuous_z_dynamics} with $\mathbf{r}(\mathbf{x})=\mathbf{0}$, $r=0.1$,
$\sigma/\Lambda=0.12\,(\eta=0.57)$ and uniform stimulus ensemble.
(\textbf{d}) Analytically predicted optimum for $\sigma/\Lambda\lesssim0.15$
(rhombic pinwheel crystal).
(\textbf{e}) Pinwheel density time courses for four different simulations
(parameters as in b; gray traces, individual realizations; black trace,
simulation in b; red trace, mean value).
(\textbf{f}) Mean squared amplitude of
the stationary pattern in simulations (parameters as in b) for different
values of the control parameter $r$ (black circles) and analytically
predicted value (solid green line).
(\textbf{g}-\textbf{h}) OPMs (g) and their power spectra (h) obtained in
a simulation of Eq. \eqref{eq:continuous_z_dynamics} with $\mathbf{r}(\mathbf{x})=\mathbf{0}$, $r=0.1$,
$\sigma/\Lambda=0.15\,(\eta=0.41)$ and uniform stimulus ensemble.
(\textbf{i}) Analytically predicted optimum for $\sigma/\Lambda\gtrsim0.15$
(orientation stripes).
(\textbf{j}) Pinwheel density time courses for four different simulations
(parameters as in g; gray traces, individual realizations; black trace,
simulation in g; red trace, mean value).
(\textbf{k}) Mean squared amplitude of the
stationary pattern in simulations (parameters as in g) for different
values of the control parameter $r$ (black circles) and analytically
predicted value (solid green line). 
\label{Appendix_Keil_Wolf_figure_2}}
\end{figure*}
The main text of the present paper contains a complete analysis of optimal dimension-reducing mappings of the EN model with a circular ensemble
of orientation stimuli. These optima are simple regular orientation
stripes or square pinwheel crystals. The circular orientation
stimulus ensemble, however, contains only stimuli with a fixed and finite {}``orientation
energy'' or elongation $|s_{z}|$. This raises the question of whether the simple nature
of the circular stimulus ensemble might restrain the realm
of complex dynamics in the EN model. The EN dynamics are expected to
depend on the characteristics of the activity patterns evoked by
the stimuli and these will be more diverse and complex
with ensembles containing a greater diversity of stimuli. Therefore,
we also examined the EN model in detail for a richer
ensemble of stimuli. In this ensemble, called a \textit{uniform} stimulus ensemble in the following, orientation stimuli
are uniformly distributed on the disk $\{s_{z,}\,|s_{z}|\leq2\}$, a choice common to many previous studies, e.g. \cite{Obermayer:1990p1202,Obermayer:1992p1200,Obermayer:1992p4756,Wolf:1998p1199}. 
The uniform ensemble in particular contains unoriented stimuli with $|s_{z}|=0$.
Intuitively, the presence of these unoriented stimuli might be expected to fundamentally change the importance of pinwheels in the optimal OPM layout. Pinwheels' population activity is untuned for orientation. Pinwheel centers may therefore acquire a key role for the representation of unoriented stimuli. As such an effect should be independent of retinotopic distortions and to aid comparison with our previous results, we will again start with a fixed uniform retinotopy $\mathbf{r}(\mathbf{x})=\mathbf{0}$.

The linear stability properties of the unselective fixed point are independent of the ensemble of orientation stimuli ($\left<|s_z|^2\right>=2$ throughout this paper). The coefficients in Eq. \eqref{eq:sum-of-cubic-operators}, however, of course depend on the fourth moment of the stimulus distribution. Inserting $\left<|s_{z}|^{4}\right> = 16/3$ into Eq. \ref{eq:general-nonlinear_coefficients}, we obtain
\begin{equation*}
\begin{array}{lll}
a_{1}=\frac{1}{3\sigma^{6}}-\frac{1}{\sigma^{4}}+\frac{1}{2\sigma^{2}} & a_{2}=\frac{1}{4\pi\sigma^{6}}-\frac{1}{6\pi\sigma^{8}}\hspace{1em} & a_{3}=-\frac{1}{12\pi\sigma^{8}}+\frac{1}{8\pi\sigma^{6}}\\
a_{4}=-\frac{1}{6\pi\sigma^{8}}+\frac{1}{4\pi\sigma^{6}}-\frac{1}{8\pi\sigma^{4}}\hspace{1em} & a_{5}=-\frac{1}{12\pi\sigma^{8}} & a_{6}=\frac{1}{8\pi\sigma^{6}}-\frac{1}{12\pi\sigma^{8}}\\
a_{7}=\frac{1}{9\pi^{2}\sigma^{10}}-\frac{1}{12\pi^{2}\sigma^{8}} & a_{8}=\frac{1}{18\pi^{2}\sigma^{10}} & a_{9}=-\frac{1}{16\pi^{3}\sigma^{12}}\\
a_{10}=\frac{1}{9\pi^{2}\sigma^{10}}-\frac{1}{12\pi^{2}\sigma^{8}} & a_{11}=\frac{1}{18\pi^{2}\sigma^{10}}\,.
\end{array}
\end{equation*}
The angle-dependent interaction functions are then given by
\begin{eqnarray*}
g(\alpha) & = & \frac{1}{\sigma^{4}}\left(1-2e^{-k_{c}^{2}\sigma^{2}}-e^{2k_{c}^{2}\sigma^{2}(\cos\alpha-1)}\left(1-2e^{-k_{c}^{2}\sigma^{2}\cos\alpha}\right)\right)\nonumber \\
 &  & +\frac{1}{2\sigma^{2}}\left(e^{2k_{c}^{2}\sigma^{2}(\cos\alpha-1)}-1\right)+\frac{32}{3\sigma^{6}}e^{-2k_{c}^{2}\sigma^{2}}\sinh^{4}\left(1/2k_{c}^{2}\sigma^{2}\cos\alpha\right)\nonumber \\
f(\alpha) & = & \frac{1}{\sigma^{4}}\left(1-e^{-2k_{c}^{2}\sigma^{2}}\left(\cosh(2k_{c}^{2}\sigma^{2}\cos\alpha)+2\cosh(k_{c}^{2}\sigma^{2}\cos\alpha)\right)+2e^{-k_{c}^{2}\sigma^{2}}\right)\nonumber \\
 &  & +\frac{1}{2\sigma^{2}}\left(e^{-2k_{c}^{2}\sigma^{2}}\cosh(2k_{c}^{2}\sigma^{2}\cos\alpha)-1\right)+\frac{16}{3\sigma^{6}}e^{-2k_{c}^{2}\sigma^{2}}\sinh^{4}\left(1/2k_{c}^{2}\sigma^{2}\cos\alpha\right)\,.
\end{eqnarray*}
Both functions are depicted in Fig. \ref{Appendix_Keil_Wolf_figure_1}
for two different values of the effective intracortical interaction
range $\sigma/\Lambda$. They qualitatively resemble the functions depicted in Fig. \ref{Keil_Wolf_figure_5}.
Fig. \ref{Appendix_Keil_Wolf_figure_2} displays the phase diagram of the EN model with uniform stimulus ensemble. As summarized in the main part of the paper, it is almost identical to that obtained for the circular stimulus ensemble (Fig. \ref{Keil_Wolf_figure_6}). Two different optimal states are found, square pinwheel crystals (sPWC) and orientation
stripes (OS) separated by a phase border at $\sigma/\Lambda\simeq0.15$. 
Both fixed points are stable for all $\sigma/\Lambda$. Fig. \ref{Appendix_Keil_Wolf_figure_2}b-k demonstrates, that these analytical results are confirmed by direct numerical simulations of Eq. \eqref{eq:continuous_z_dynamics} with $\mathbf{r}(\mathbf{x})=\mathbf{0}$.
As for the circular stimulus ensemble, we also tested the stability
of stationary n-ECP solutions with $2\leq n\leq20$ by numerical evaluation of
the criteria for intrinsic and extrinsic stability (Methods). We find
all n-ECPs with $2\leq n\leq20$ intrinsically unstable for
all interaction ranges $\sigma/\Lambda$. The simple phase space structure furthermore apparently
remains unchanged if we consider the model far from pattern formation
threshold as shown in Fig. \ref{Appendix_Keil_Wolf_figure_3}. Simulations
bear a close resemblance to the simulations with circular orientation
stimulus ensemble (Fig. \ref{Keil_Wolf_figure_7}).
Either convergence to sPWC-like patterns or patterns with large orientation
stripe domains is observed. Again, pinwheel annihilation in the
case of large $\sigma/\Lambda$ is less rapid than close to threshold
(Fig. \ref{Appendix_Keil_Wolf_figure_3}a,b). The linear pinwheel-free
zones increase their size over the time course of the simulations, eventually
leading to a stripe pattern. For
smaller interaction ranges $\sigma/\Lambda$, the OPM layout rapidly converges towards
a crystal-like rhombic arrangement of pinwheels with dislocations and pinwheel density close to 4.
\begin{figure}
\centering
\includegraphics[width=14cm]{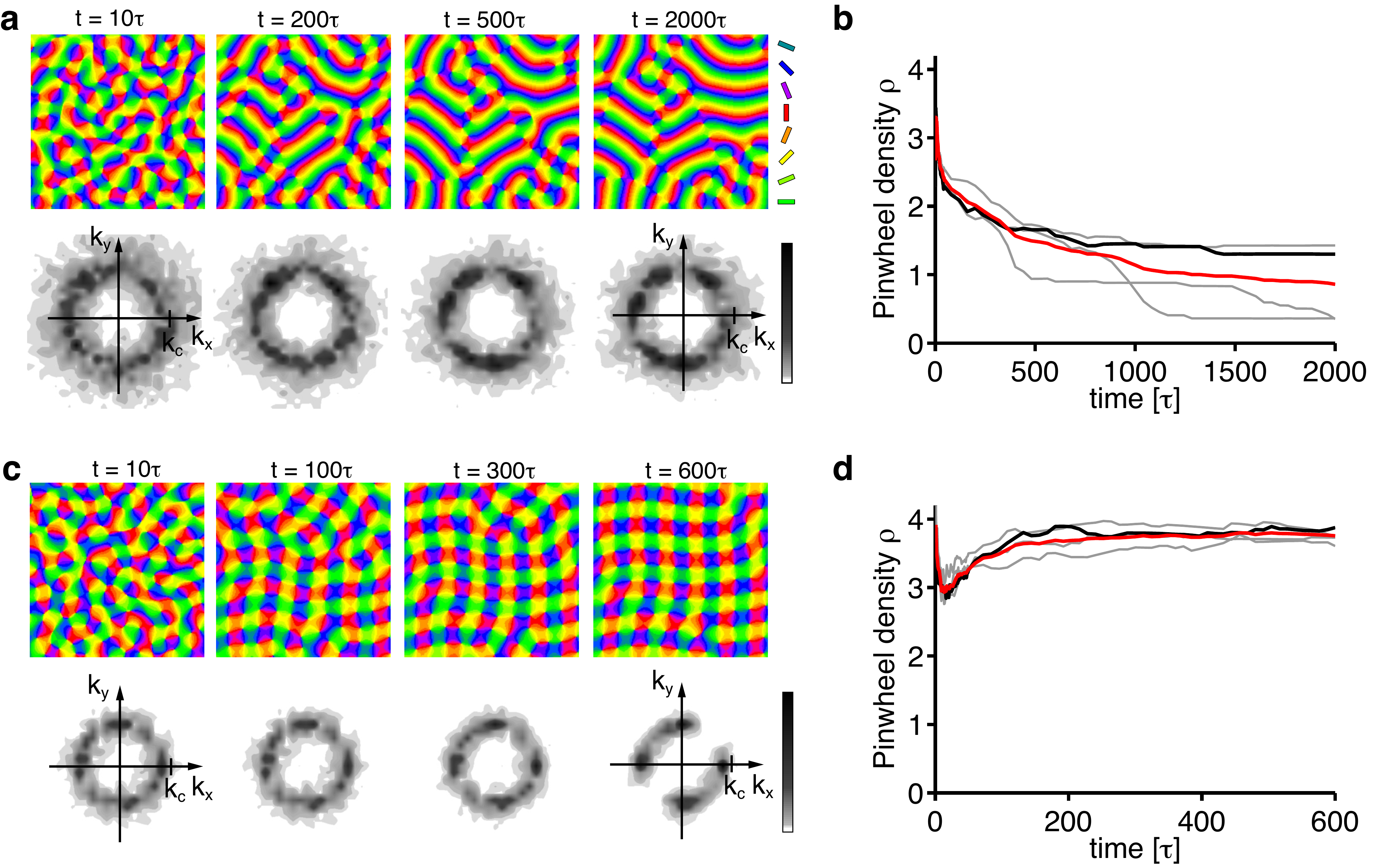}
\caption{\textbf{Numerical analysis of the EN dynamics with uniform orientation stimulus
ensemble and fixed representation of visual space far from pattern formation threshold.}
(\textbf{a}) OPMs and their power spectra in a representative simulation of Eq. \eqref{eq:continuous_z_dynamics} with $\mathbf{r}(\mathbf{x})=\mathbf{0}$, $r=0.8$, $\sigma/\Lambda=0.3\,(\eta=0.028)$ and uniform stimulus
ensemble.
(\textbf{b}) Pinwheel density time courses for four different simulations
(parameters as in a; gray traces, individual realizations; black trace,
simulation in a; red trace, mean value) (\textbf{c}) OPMs and their power spectra in a representative simulation
of Eq. \eqref{eq:continuous_z_dynamics} with $\mathbf{r}(\mathbf{x})=\mathbf{0}$ and $\sigma/\Lambda=0.12\,(\eta=0.57)$,
other parameters as in a. (\textbf{d}) Pinwheel density time courses
for four different simulations (parameters as in c; gray traces, individual
realizations; black trace, simulation in c; red trace, mean value)
\label{Appendix_Keil_Wolf_figure_3}}
\end{figure}

Fig. \ref{Appendix_Keil_Wolf_figure_4}
shows that taking retinotopic distortions into account yields an almost identical
picture compared to the circular stimulus ensemble. For small interaction range $\sigma/\Lambda$,
the analytically predicted optimum is a quadratic pinwheel crystal
with pinwheel density $\rho=4$. For larger $\sigma/\Lambda$, the
analytically predicted optimum is an orientation stripe pattern
with pinwheel density $\rho=0$. Our results are confirmed
by direct simulations of Eq. (\ref{eq:continuous_z_dynamics}, \ref{eq:continous_r_dynamics})
(Fig. \ref{Appendix_Keil_Wolf_figure_4}b-e). The simulation
results are virtually indistinguishable from the circular
stimulus ensemble.

All together, the EN Eq. (\ref{eq:continuous_z_dynamics},
\ref{eq:continous_r_dynamics}) and in particular the set of ground
states of the EN model and their stability regions appear almost identical when considering either a circular or a uniform
orientation stimulus ensemble. We found two different optima depending on the parameter regime,
orientation stripes for larger interaction ranges and quadratic pinwheel crystals for shorter interaction ranges.
In addition, the EN dynamics appears to be unchanged by the presence of unoriented stimuli.
\begin{figure}
\centering
\includegraphics[width=16cm]{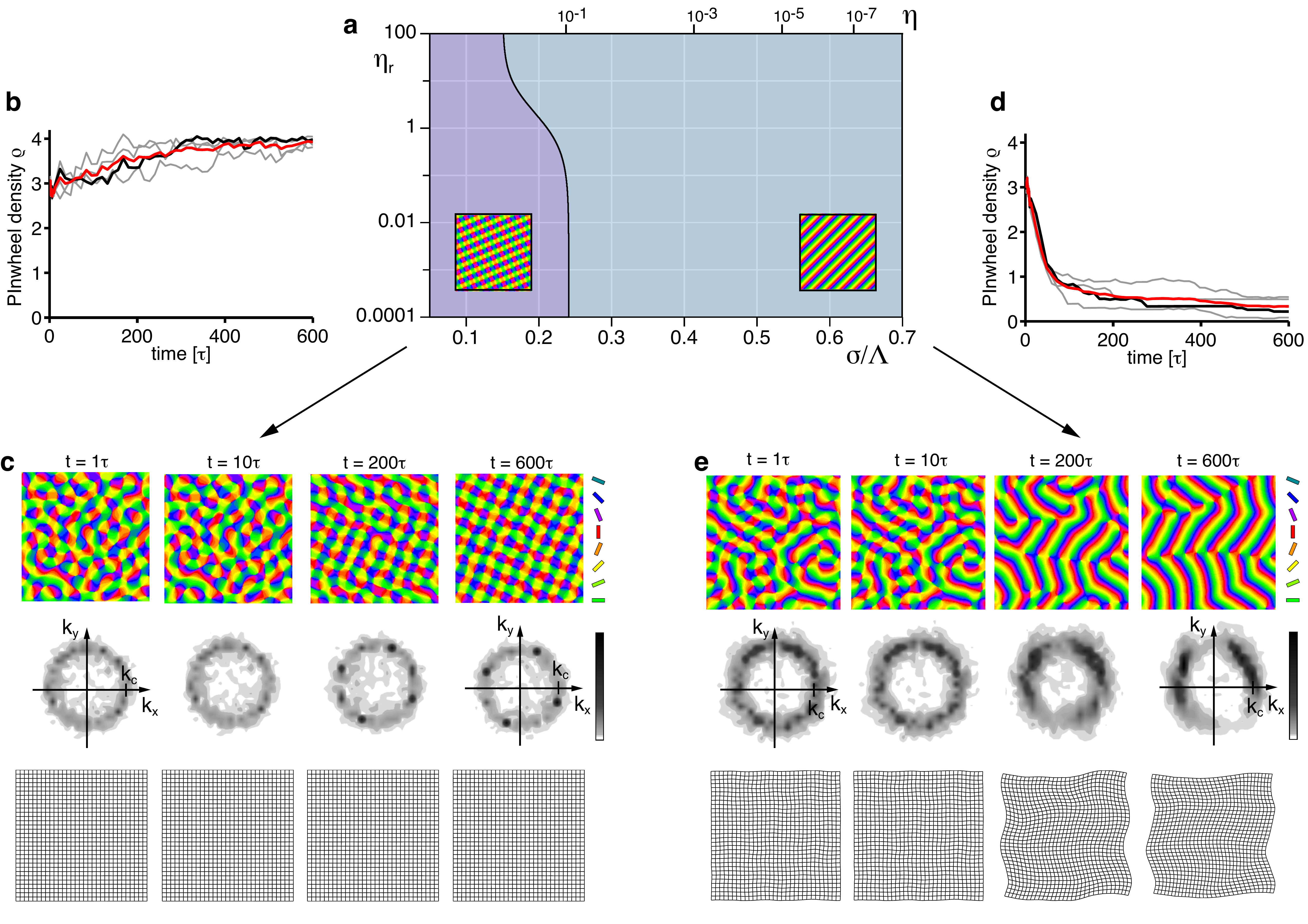}
\caption{\textbf{Phase diagram of the EN model for the joint mapping of visual space
and orientation with a uniform orientation stimulus ensemble.}
(\textbf{a}) Regions of the $\eta_{r}$-$\sigma/\Lambda$-plane in
which n-ECPs or rPWCs have minimal energy. 
(\textbf{b}) Pinwheel density time courses for four different simulations of Eqs. (\ref{eq:continuous_z_dynamics}, \ref{eq:continous_r_dynamics}) with $r=0.1$, $\sigma/\Lambda=0.13\:(\eta=0.51$),
$\eta_{r}=\eta$ (grey traces, individual realizations; red trace,
mean value; black trace, realization shown in c). 
(\textbf{c}) OPMs (upper row), their power spectra (middle row),
and retinotopic maps (lower row) in a simulation of Eqs. (\ref{eq:continuous_z_dynamics}, \ref{eq:continous_r_dynamics}); 
parameters as in b. (\textbf{d}) Pinwheel density time courses for four different simulations of 
Eqs. (\ref{eq:continuous_z_dynamics}, \ref{eq:continous_r_dynamics}) with $r=0.1$, $\sigma/\Lambda=0.3\:(\eta=0.03$),
$\eta_{r}=\eta$ (grey traces, individual realizations; red trace,
mean value; black trace, realization shown in e). 
(\textbf{e}) OPMs (upper row), their power spectra (middle row), and retinotopic maps
(lower row) in a simulation of Eqs. (\ref{eq:continuous_z_dynamics}, \ref{eq:continous_r_dynamics}); parameters as in d.
\label{Appendix_Keil_Wolf_figure_4}}
\end{figure}

\section*{Appendix II}
\subsection*{Grid-like stimulus ensembles}
Refs. \cite{CarreiraPerpinan:2004p6297,CarreiraPerpinan:2005p6295}) performed simulations with stimuli distributed in regular intervals in feature space, called grid-like ensemble. For comparison, we also performed deterministic annealing simulations with grid-like stimulus sets of varying size with non-periodic boundary conditions (see Methods). For these grid-like stimulus patterns, a competition between stripes and rhombs is observed. Notably, these are the only two stable states identified by our analysis for the circular stimulus ensemble. For non-periodic boundary conditions, rhombic pinwheel arrangements seem energetically favored for grid-like stimuli, almost independently of the size of the stimulus set. The average pinwheel density for $N=100\times100\times8$ stimuli was $\rho = 2.8$. As expected from the predominantly rhombic arrangement of pinwheels, NN-pinwheel distances concentrate around half the typical column spacing and pinwheel pairs at short distances are not observed (Fig. \ref{Appendix_Keil_Wolf_figure_5}b). With these features, the maps obtained substantially differ from the experimentally observed pinwheel statistics \cite{Kaschube_2010}.
\begin{figure}
\centering
\includegraphics[width=14cm]{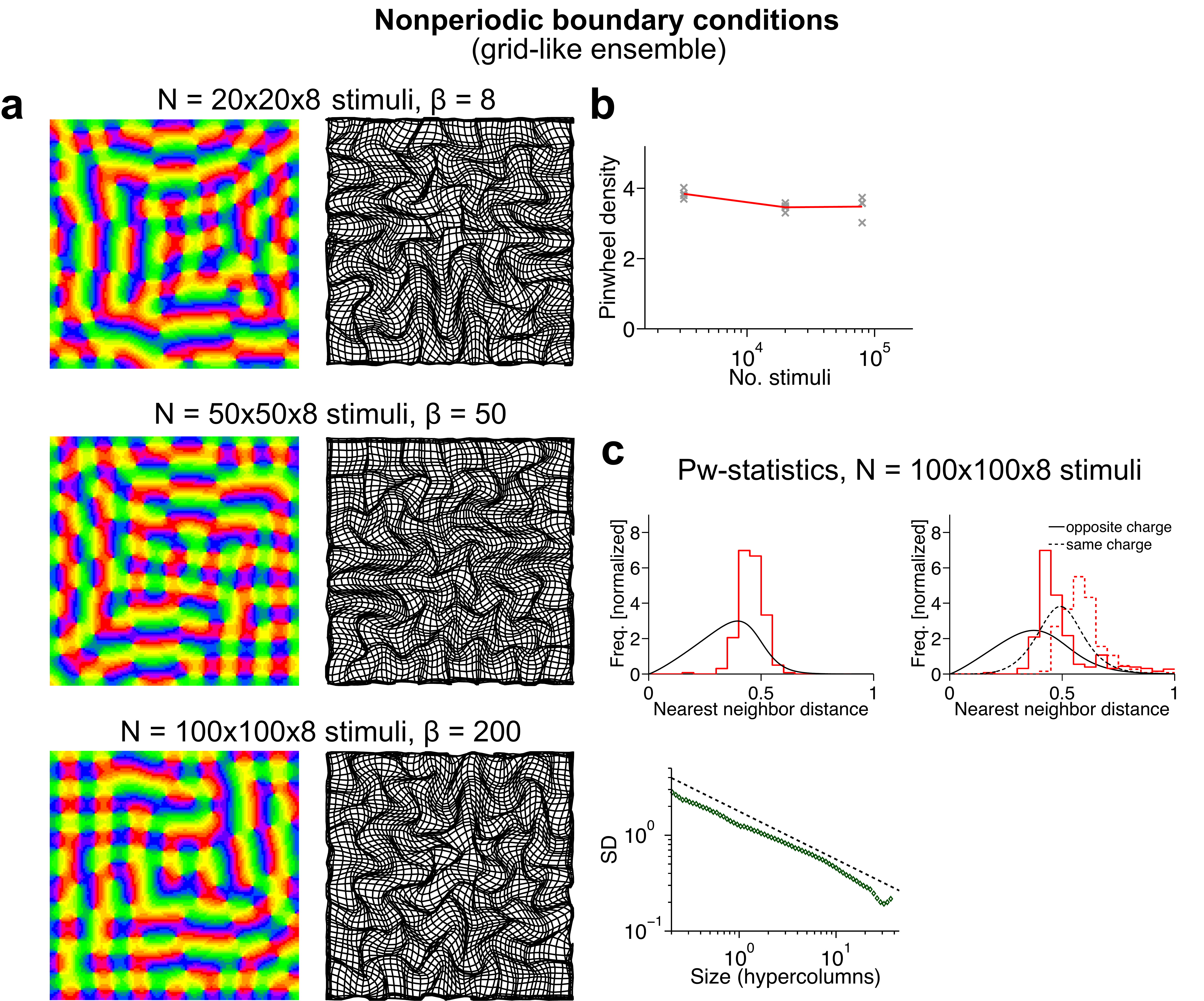}
\caption{\textbf{The EN model with deterministic annealing and stimuli, distributed on a grid in feature space.}
(\textbf{a}) OPMs (left) and retinotopic maps (right) for $N=20\times20\times8$ (upper row), $N=50\times50\times8$ (middle row) and $N=100\times100\times8$ (lower row) stimuli and non-periodic boundary conditions (annealing rate $\chi = 0.999$). $\beta$ is the continuity parameter in the conventional definition of the EN model (see Methods, Eq. (\ref{eq:EN_energy_discrete})) and is scaled, such that a comparable number of columns emerges in all simulation for each N.
(\textbf{b}) Pinwheel densities of EN solutions for different numbers of stimuli, (annealing rate $\chi = 0.999$). Crosses mark individual simulations, red line indicates average values.
(\textbf{c}) Statistics of nearest neighbor pinwheel distances for pinwheels of (upper left) arbitrary and (upper right) opposite and equal charge for $100\times100\times8$ stimuli and non-periodic boundary conditions, averaged over four simulations (red curves). Black curves represent fits to the experimental data from \cite{Kaschube_2010}. 
Lower left: Standard deviations (SD) of pinwheel densities estimated from randomly selected regions in the OPM. Black dashed curve indicates SD for a two-dimensional Poisson process of equal density. 
\label{Appendix_Keil_Wolf_figure_5}}
\end{figure}

\subsection*{The discrete EN model with 4th derivative}
In previous studies of the EN model, alternative definitions of the continuity term in the EN model have been explored \cite{CarreiraPerpinan:2004p6297}. A general continuity term for the spatially continuous formulation of the EN for OPM and retinotopy is a linear differential operator which will suppress the emergence of high-frequencies during the EN dynamics. 
A finite-wavelength instability is expected in this case, although the precise expressions for the critical $\sigma$ and the typical wavelength will differ. As linear terms do not enter in the higher-order derivatives of the EN functional, changing the continuity term is not expected to alter the stability results obtained in the present study.

To numerically test this expected robustness of our results for the EN model with discrete fixed sets of stimuli (see Figs. \ref{Keil_Wolf_figure_18} and \ref{Keil_Wolf_figure_19}), we also performed simulations  using deterministic annealing with a fourth derivative stencil (see Methods). Figure  \ref{Appendix_Keil_Wolf_figure_6} illustrates that the results almost perfectly match the ones for the 2nd order derivative, considered in the main part of our paper (Figs. \ref{Keil_Wolf_figure_18}, \ref{Keil_Wolf_figure_19} and \ref{Appendix_Keil_Wolf_figure_5} in the Appendix).  

When annealing with periodic boundary conditions, the solutions very much resemble our gradient descent dynamics simulations with Laplacian term. The larger the set of stimuli, the more stripe-like are the orientation preference maps obtained (Fig. \ref{Appendix_Keil_Wolf_figure_6}a) and consequently pinwheel densities decrease (Fig. \ref{Appendix_Keil_Wolf_figure_6}b, upper right). The exponent for the SD is considerably lower than for the Poisson process (Fig. \ref{Appendix_Keil_Wolf_figure_6}b, upper right).Typical NN-pinwheel distances concentrate around half the typical column spacing and in particular pinwheel pairs with short distances lack completely (Fig. \ref{Appendix_Keil_Wolf_figure_6}b, lower left and right). 

For non-periodic boundary conditions and random stimuli, we found that retinotopic distortions are much more pronounced. They however decreased with increasing number of stimuli. For large stimulus numbers, we observed stripe-like orientation preference domains which are interspersed with lattice-like pinwheel arrangements (see Fig. \ref{Appendix_Keil_Wolf_figure_6}c), lower row, upper left corner of the OPM). 
Similarly to the periodic boundary conditions, short distance pinwheel pairs occur much less frequently than in the experimentally observed maps, indicating an increased regularity in the pinwheel arrangements compared to realistic OPMs (Fig. \ref{Appendix_Keil_Wolf_figure_6}d, lower left and right). This regularity also manifests itself in a smaller exponent of the SD compared to the Poisson process (Fig. \ref{Appendix_Keil_Wolf_figure_6}d).

Simulations with grid-like stimulus as  e.g. used in  \cite{CarreiraPerpinan:2004p6297,CarreiraPerpinan:2005p6295} displayed a strong tendency towards rhombic pinwheel arrangements analogous to the 2nd derivative case (Fig. \ref{Appendix_Keil_Wolf_figure_5}e,f)
\begin{figure}
\centering
\includegraphics[width=16cm]{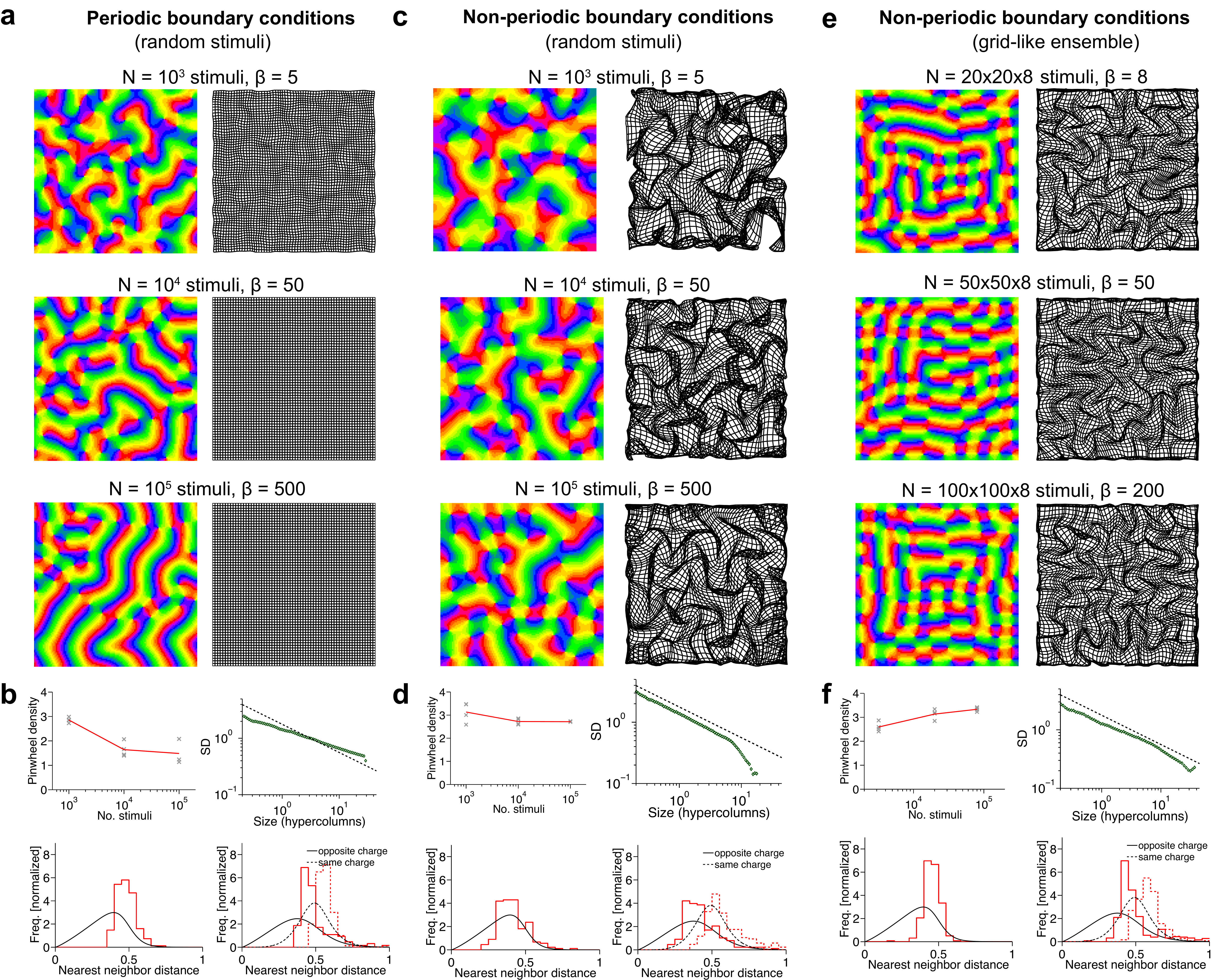}
\caption{\textbf{The EN model with deterministic annealing and fourth derivative stencil.}
(\textbf{a}) OPMs (left) and retinotopic maps (right) for $N=10^3$ (upper row), $N=10^4$ (middle row) and $N=10^5$ (lower row) stimuli, non-periodic boundary conditions, and annealing rate $\chi=0.999$. $\beta$ is the continuity parameter in the conventional definition of the EN model (see Methods, Eq. (\ref{eq:EN_energy_discrete})) and is scaled, such that a comparable number of columns is emerging in the simulations for each stimulus set.
(\textbf{b}) Pinwheel densities (upper left) of EN solutions, standard deviations (SD) of pinwheel densities estimated from randomly selected regions in the solutions (upper right). Crosses mark individual simulations, red line indicates average values.
Black dashed curve indicates SD for a two-dimensional Poisson process of equal density.
Statistics of nearest neighbor pinwheel distances for pinwheels of arbitrary (lower left) and (lower right) opposite and equal charge for $10^{5}$ random stimuli and periodic boundary conditions, averaged over four simulations (red curves). Black curves represent fits to the experimental data from \cite{Kaschube_2010}. 
(\textbf{c}) As a but for non-periodic boundary conditions.
(\textbf{d}) As b, but for non-periodic boundary conditions. 
(\textbf{e}) OPMs (left) and retinotopic maps (right) for $N=20\times20\times8$ (upper row), $N=50\times50\times8$ (middle row) and $N=100\times100\times8$ (lower row) stimuli, non-periodic boundary conditions, annealing rates $\chi=0.999$.
(\textbf{f}) As b, but for non-periodic boundary conditions and grid-like stimuli.
\label{Appendix_Keil_Wolf_figure_6}}
\end{figure}

\section*{Appendix III}
\subsection*{Strength of retinotopic coupling}
In our manuscript, we have shown that retinotopic distortions only have a weak influence on the optima of the EN model as well as its dynamics (see Figs. \ref{Keil_Wolf_figure_10} and \ref{Keil_Wolf_figure_12}). 
Here, we quantify the influence of retinotopic distortions on the pattern formation process by comparing the angle-dependent interaction function for retinotopic coupling $g_r(\alpha)$ with angle-dependent interaction function of the EN model with fixed retinotopy. We use the ratio 
\begin{equation*}
c = \frac{\|g_r(\cdot)\|_2 }{\|g(\cdot)\|_2}\,
\end{equation*}
as a measure to quantify the influence of retinotopic distortions. $\|\cdot\|_2$ denotes the 2-Norm,
$$
\|f(\cdot)\|_2=\int_0^{2\pi} f^2(\alpha)\, d\alpha\,.
$$ 
If $c$ is larger than one, $g_r(\alpha)$ dominates the total interaction function $g_r(\alpha) + g(\alpha)$ and retinotopic distortions may strongly influence the layout and stability of solutions of the EN model. On the other hand, if  c is small, the solutions and their stability properties are expected to not change substantially when including variable retinotopy into the EN model. Figure \ref{Appendix_Keil_Wolf_figure_7} displays the parameter $c$ in the $s_4$-$\sigma/\Lambda$-plane for the EN model at threshold for two different conditions, $\eta = \eta_r$ and $\eta_r = 0$. In the latter case, retinotopic distortions are expected to have the strongest impact. However, in both cases, $c \ll 1$, in almost all of the parameter space, implying little influence of retinotopic deviations. Only for small $\sigma/\Lambda$ and small $s_4$, $c$ is larger than one. As shown in Fig. \ref{Keil_Wolf_figure_12}, this leads to slight deformations of the stability regions for rhombs, and stripes in this region of parameter space but does not result in novel optimal solutions. 
\begin{figure}
\centering
\includegraphics[width=16cm]{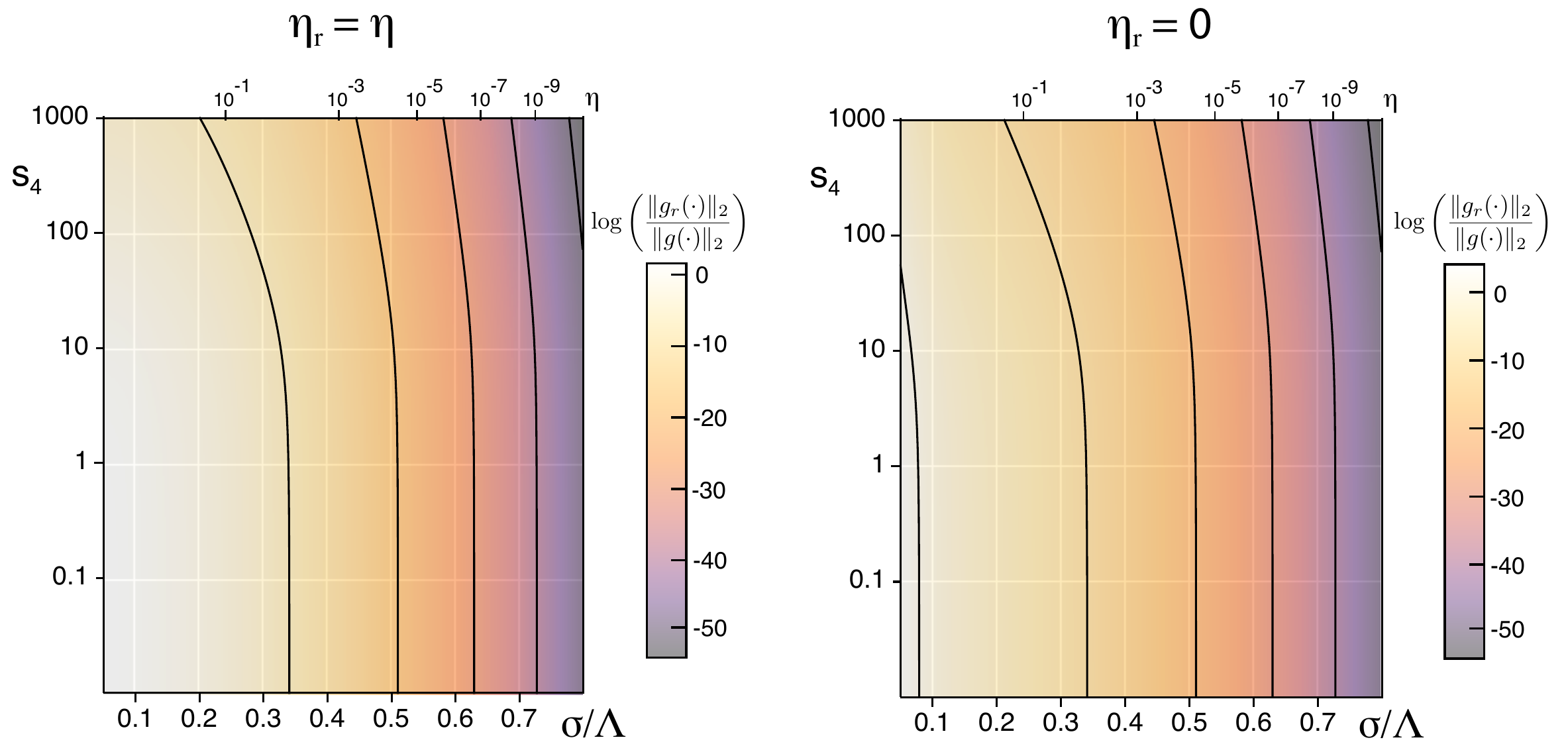}
\caption{\textbf{Strength of coupling between orientation map and retinotopy in the EN model.}
(\textbf{a}) Ratio of $\|g_r(\cdot)\|_2 / \|g(\cdot)\|_2$ in the $s_4$-$\sigma/\Lambda$-plane for the EN model at threshold and $\eta = \eta_r$. 
(\textbf{b}) As a, but for $\eta_r=0$, i.e. strongest coupling. Note the logarithmic scaling of the colormap.
\label{Appendix_Keil_Wolf_figure_7}}
\end{figure}
%
%
%
%
%
%
%
\section*{Authors contributions}
WK performed analytical calculations and numerical simulations. FW designed the study and performed analytical calculations. All authors read and approved the final manuscript.
\section*{Competing interest}
The authors declare that they have no competing interests.

\section*{Acknowledgements}
\ifthenelse{\boolean{publ}}{\small}{}
We thank E. Bodenschatz, D. Coppola, J. Crowley, A. Gail, T. Geisel, G. Goodhill, M. Huang, M. Kaschube, C. Kirst, S. L\"owel, J. Metzger, T. Mooser, L. Reichl, M. Schnabel, D. Tsigankov, and L. White for many inspiring discussions and the Kavli Institute for Theoretical Physics for its hospitality.

This work was supported by Grants No. 01GQ0922 (Bernstein Focus Learning) and No. 01GQ1005B (Bernstein Center for Computational Neuroscience, G\"ottingen) from the German Ministry for Education and Science (BMBF), Grant VWZN2632 from the Volkswagen Foundation, HFSP Grants Nos. RGP63/2003 and RGY0065/2007 as well as the NSF Grant No. NSF PHY05-51164.

%
%
{
\ifthenelse{\boolean{publ}}
{\footnotesize}{\small}
}     
\bibliographystyle{bmc_article}  


\bibliography{EN_article}
%
%
\ifthenelse{\boolean{publ}}{\end{multicols}}{}

\end{bmcformat}

\end{document}